\documentclass[
aps,prd,
nofootinbib,
superscriptaddress,
tightenlines,
amsmath,
amssymb
]{revtex4}
\usepackage{bm}
\usepackage{color,graphicx}
\usepackage{amsmath}
\usepackage{amssymb}
\usepackage{slashed}
\usepackage{float}
\usepackage[normalem]{ulem}
\usepackage{xcolor}
\usepackage{hyperref}
\usepackage{listings}
\usepackage{xcolor}
\usepackage[T1]{fontenc}

\newcommand{\xp}{x'}
\renewcommand{\d}{\delta}
\newcommand{\m}{\mathcal{M}}

\newcounter{comment}

\allowdisplaybreaks

\begin{document}
\title{Updated flexible global parametrization of generalized parton distributions from elastic and deep inelastic inclusive scattering data} 

\author{Zaki Panjsheeri}
\email{zap2nd@virginia.edu}
\affiliation{Department of Physics, University of Virginia, Charlottesville, VA 22904, USA.}

\author{Douglas Q. Adams}
\email{yax6jr@virginia.edu}
\affiliation{Department of Physics, University of Virginia, Charlottesville, VA 22904, USA.}

\author{Adil Khawaja} 
\email{taz8sp@virginia.edu}
\affiliation{Department of Physics, University of Virginia, Charlottesville, VA 22904, USA.}

\author{Saraswati Pandey}
\email{qbs8mh@virginia.edu}
\affiliation{Department of Physics, University of Virginia, Charlottesville, VA 22904, USA.}

\author{Kemal Tezgin}
\email{kemaltezgin@vt.edu}
\affiliation{Department of Physics, Virginia Tech, Blacksburg, VA, USA.}

\author{Simonetta Liuti} 
\email{sl4y@virginia.edu}
\affiliation{Department of Physics, University of Virginia, Charlottesville, VA 22904, USA.}


\begin{abstract}
An updated flexible parametrization of the generalized parton distributions in the quark, antiquark and gluon sectors is presented using constraints from high precision electron nucleon deep inelastic scattering data, as well as from the $u$, $d$ quark and gluonic contributions to the nucleon electromagnetic elastic form factors. The latter include recently updated lattice QCD moment calculations. 
The generalized parton distributions in the vector sector are $H$ and $E$. We rigorously constrain the partonic components, $H_{u_v}$, $H_{d_v}$, $H_{\bar{u}}$, $H_{\bar{d}}$, $H_{\bar{s}}$ and $H_{g}$, and the analogous quantities for $E$, with proper uncertainty quantification. These distributions obey leading order perturbative QCD evolution equations in $\alpha_S$. Parametric forms at the initial scale, $Q_o^2 = 0.58$ $\mathrm{GeV}^2$, for both quarks and gluon distributions are presented as a function of the relevant kinematic variables, namely, the parton momentum fraction, $x$, the skewness, $\xi$, and the invariant, $t$.
We also present the Compton form factors entering the deeply virtual Compton scattering process in the kinematic regimes for both fixed target and electron-ion collider settings.


\end{abstract}
\maketitle

\section{Introduction}
Generalized parton distributions (GPDs) parametrize the QCD quark and gluon correlation functions matrix elements where the initial and final proton states can have different momenta and helicity. 
At variance with the ordinary parton distribution functions (PDFs) measured in inclusive deep inelastic scattering (DIS), GPDs 
appear in processes that occur through a coherent emission by the proton of a photon or a meson, 
which is also deeply virtual, or characterized by a large momentum transfer, $Q^2$. 
The prototypes of these processes are deeply virtual Compton scattering (DVCS), where a photon is produced in the final state,  
or deeply virtual meson production (DVMP), where either a pseudoscalar or a vector meson is produced. GPDs are, however, also measurable in several crossed channels to DVCS, including {\it e.g.} timelike Compton scattering (TCS), where a high invariant mass lepton pair is produced coherently in real photon scattering off the proton, granting access to additional flavor and spin configurations (see reviews in \cite{Diehl:2003ny,Belitsky:2005qn,Kumericki:2016ehc}).  
GPDs carry a wealth of information on the 3D structure of the proton: the total angular momentum, $J$, is described in terms of the vector GPDs, $H$ and $E$, since their second Mellin moment defines the generalized form factors which parametrize the QCD energy momentum tensor \cite{Ji:1997nk}. Furthermore, through the Fourier transformation in the momentum transfer between the initial and final proton states,
one can access the spatial density distributions of specific partonic configurations in the transverse coordinate, or impact parameter, ${b}$.  This information allows us, in turn, to directly determine the orbital angular momentum of the partons, $L$ \cite{Polyakov:2002yz,Kiptily:2002nx, Rajan:2017cpx,Alkasassbeh:2024aws}. 
%
%

A known complication affecting GPDs analyses is in their extraction from the experimental observables, or the Compton form factors (CFFs) which parametrize the cross section for DVCS and related processes (see {\it e.g.} Refs.\cite{Kriesten:2019jep,Kriesten:2020apm,Adams:2024pxw}). 
The CFFs are defined in terms of GPDs as,
\begin{equation}
\label{eq:CFF_intro}
 \mathcal{F}(\xi,t; Q^2)   =   C^+ \otimes F   \equiv \int_{-1}^{1} dx  \,  C^+(x,\xi) F(x,\xi,t; Q^2)  . 
 \end{equation}
where ${\cal F} = {\cal H}, {\cal E}$, and $C^+$ is a complex kernel, calculable in perturbative QCD (PQCD).  
To extract GPDs from CFFs, therefore, involves an inverse problem, whose solution has been the focus of several  
recent analyses \cite{Moffat:2023svr,DallOlio:2024vjv,Riberdy:2023awf,Qiu:2023mrm,Almaeen:2024guo,Cuic:2020iwt}. 

In particular, as pointed out in previous work \cite{Kriesten:2021sqc,Almaeen:2022imx,Almaeen:2024guo}, the large uncertainties emerging in this, in principle, intractable problem can be mitigated by a careful consideration of physics constraints derived from the various symmetries observed by GPDs .
The latter can be taken into account by developing flexible parametrizations that are built on the symmetries of the theory, and constrained by the forward limit, $(\xi,t \rightarrow 0)$, and by specific sum rules dictated by the GPDs' polynomiality property \cite{Diehl:2001pm,Belitsky:2005qn,Kumericki:2016ehc}. 
Furthermore, flexible parametrizations aid the extraction of information from  lattice QCD (LQCD) calculations \cite{Lin:2020rxa,HadStruc:2024rix,Hannaford-Gunn:2024aix}. They provide also an essential tool for training neural networks in various machine learning-based approaches (see Refs.\cite{Cuic:2020iwt,Kumericki:2011zc,Kumericki:2015lhb}  and \cite{Almaeen:2022imx,Almaeen:2024guo,Dotson:2025omi,Dutrieux:2024rem}). 
Finally, flexible parametrizations with a proper uncertainty quantification are manifestly advantageous since they directly provide controlled physics-based extrapolations into unknown kinematic regions, thus allowing valuable insights that can help make informed predictions in the interpretation of various deeply virtual exclusive processes
as well as providing valuable input for event generators and Monte Carlo simulations in the low Bjorken $x$ ($x_{Bj}$) regime, dominated by gluonic components.

In what follows we present a physics-informed, flexible parametrization constrained by deep
inelastic scattering data, as well as from the $u$, $d$ quarks and gluonic contributions to the nucleon form factors. The latter include recently updated lattice QCD moments calculations. 
%
The physics background underlying our mathematical formulation is the reggeized spectator model (see \cite{Ahmad:2007vw,Ahmad:2006gn,Goldstein:2010gu,Kriesten:2019jep} and references therein) which, on one side, results in sufficiently simple mathematical forms for efficient calculations, and on the other, 
it provides a reference to a physical model that can be both explained and interpreted. The definition of our various parameters offers flexibility, in that for either $H$ or $E$,  we define 
three mass parameters and two Regge phenomenology-based parameters per flavor, in the forward limit (this number is consistent with the number of parameters used in PDF parametrizations, see {\it e.g.} Ref.\cite{Hou:2019efy,Thorne:2019mpt,Harland-Lang:2014zoa} and references therein); the elastic form factors constrain two extra parameters, consistently with the dipole parametrization  (Ref.\cite{Ye:2017gyb} and references therein). The number of parameters involved is large enough to provide a flexible GPD shape, while respecting the theoretical/symmetry constraints on GPDs.
%
We enhance the parametrization provided in Ref.\cite{Kriesten:2019jep} in that: 1) we provide more stringent constraints on both the antiquark and gluon distributions. The latter are fitted using updated LQCD form factor values evaluated at a pion mass value closer to the physical one, while the antiquark contribution for the GPD $E$ is entirely new in the present version  ; 2) we provide, for the first time, fully analytic forms in all kinematic variables, so that our parametrization can be easily implemented by users; 3) we define a common PQCD evolution initial scale for quarks and gluons, improving upon the version provided in Ref.\cite{Kriesten:2019jep}. 
Having this distribution, with a proper evaluation of GPDs at low $x_{Bj}$, we can make realistic estimates in a wide kinematic regime, from fixed target scenarios, dominated by the valence structure, to the electron ion collider (EIC) \cite{AbdulKhalek:2021gbh}, which is sensitive to the antiquark and gluon components.   
The present version of the parametrization is named  {\it UVA2}  as it supersedes the formulations and settings of \cite{Kriesten:2021sqc} (see also \cite{Ahmad:2007vw,Goldstein:2010gu,Goldstein:2012az} for the older versions and general settings of the model). The UVA2 version is readily available and usable in two different formats as we provide both analytic forms at the initial QCD evolution scale, and grids in the range of $Q^2$ at \textbf{\url{https://github.com/Exclaim-Collaboration/UVA2}}.

This paper  is organized as follows: in Section \ref{sec:formalism} we review the formalism, highlighting the constraints provided by the various symmetries of the theory and illustrating the working of PQCD evolution for GPDs; in Section \ref{sec:param} we present our parametrization for $H_{q,g}$ and $E_{q,g}$, in the quark sector for  $q=u_v,d_v,\bar{u},\bar{d},s$, and gluon sector, $g$, respectively; in Section \ref{sec:results} we present various results obtained using our parametrization, including the calculation of the Compton Form Factors for DVCS; we finally draw our conclusions and present an outlook on future work in Section \ref{sec:conclusions}. 


\section{General Formalism}
\label{sec:formalism}
\begin{figure}[H]
    \centering
    \includegraphics[width=0.5\linewidth]{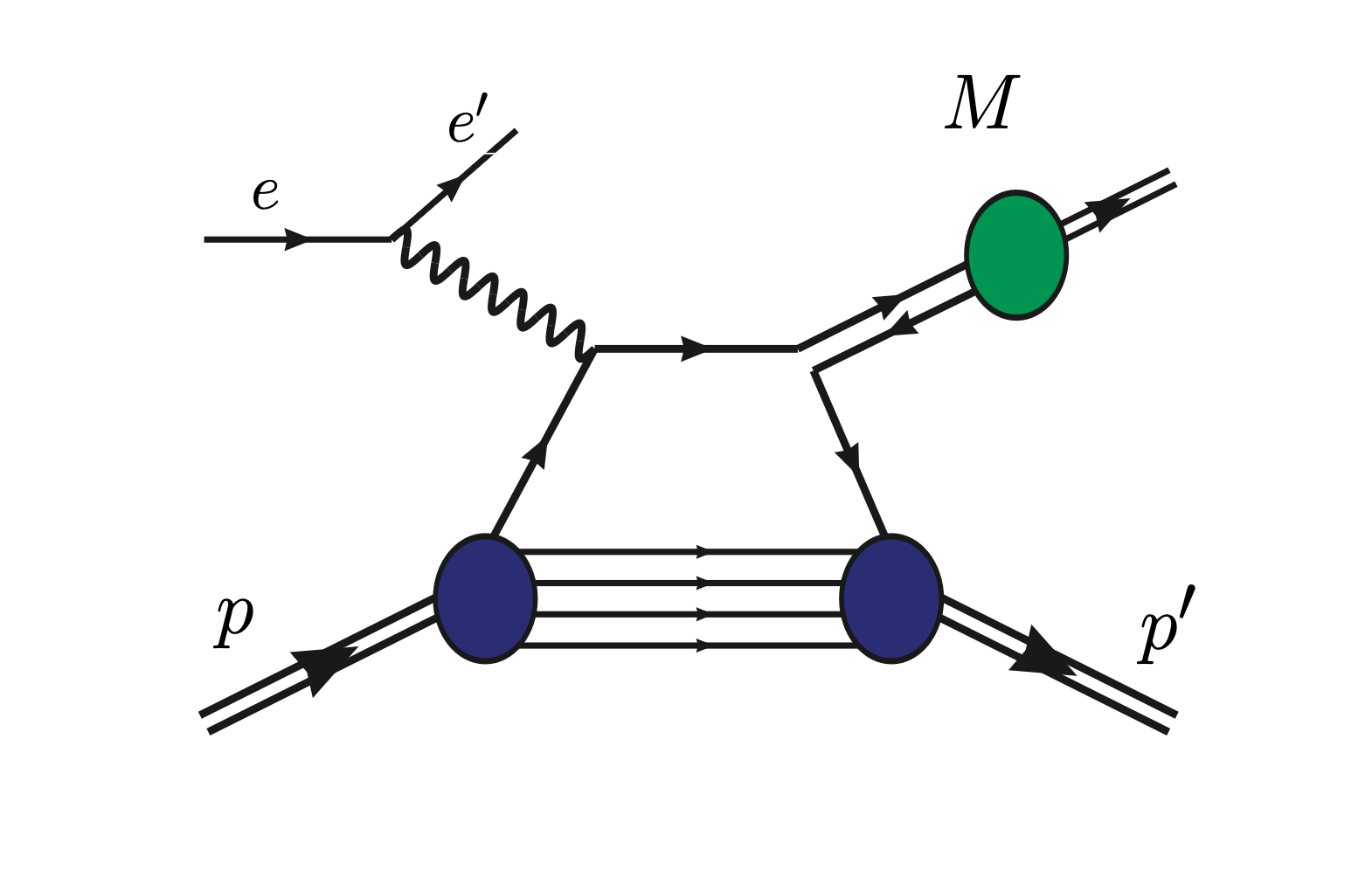}
    \caption{Schematic representation of a deeply virtual exclusive scattering graph within a QCD factorized scenario, from the reaction $e p \rightarrow e' p'M$, where $M$ can be either a meson or a photon.}
    \label{fig:Feynman}
\end{figure}

We present the main steps of the formalism for deeply virtual exclusive scattering (DVES) processes represented by the diagram in Figure \ref{fig:Feynman}, where the relevant four momenta for electroproduction from a nucleon target are: $p (p')$ for the initial (scattered) proton, $k_e (k_e')$ for the initial (scattered) electron, $q'$ for the photon/meson produced in the final state, $q=k_e - k_e'$, the four-momentum transfer between the initial and final electrons, and $\Delta=p-p'$, the four momentum transfer between the initial and final proton. The observables sensitive to GPDs, or the CFFs, written in terms of kinematic invariants obtained from the four-momenta, are chosen to be $(s, Q^2, x_{Bj}, t)$, where
\begin{itemize}
    \item $s=(k_e+p)^2$ is the electron-proton center of mass energy squared,
    \item $Q^2 = -(k_e-k_e')^2 = - q^2$ is the (positive) four-momentum transfer squared between the incoming and outgoing electrons,
    \item $x_{Bj} = Q^2 / 2(pq)$ is Bjorken-$x$. 
    \item $t = (p- p')^2 = \Delta^2$ is the four-momentum transfer squared between the initial and final protons 
\end{itemize}
Additional kinematic variables used to define GPDs are the light-cone momentum fractions,
\begin{itemize}
    \item $x = (k^+ + k^{\prime +})/(p^+ + p^{\prime +})$, $k$ and $k'$ being the initial and final parton four-momenta.
    \item $\xi = \Delta^+/(p^+ + p^{\prime +})$. In the asymptotic limit, disregarding $t/Q^2$ and $M^2/Q^2$ corrections, $x_{Bj}$ is written in terms the skewness parameter, $\xi = (\Delta q)/[(pq)+(p'q)]= x_{Bj}/(2- x_{Bj})$ 
    \end{itemize}
An alternative set of variables, which will be useful for QCD evolution is the so-called {\it asymmetric set} is defined through the transformation \cite{GolecBiernat:1998ja}, \begin{subequations}
\label{eq:asym}
 \begin{eqnarray}
X & = & \frac{x+\xi}{1+\xi} = \frac{k^+}{p^+}, \quad \quad X-\zeta= \frac{x-\xi}{1+\xi}  = \frac{k'^+}{p^+},  \quad \Rightarrow \quad  x = \frac{X-\zeta/2}{1-\zeta/2}. \\
 \zeta & = & \frac{2 \xi}{1+\xi} = \frac{\Delta^+}{p^+}, \quad \Rightarrow \quad \xi = \frac{\zeta}{2-\zeta}.
 \end{eqnarray}
 \end{subequations}
A detailed description of the connection between the symmetric and asymmetric frame variables is presented in Appendix \ref{app:asymmetric}. 
In what follows, we present the CFFs formulation, Section \ref{sec:CFFs}, the GPD flavor symmetries, Section \ref{sec:sym}, and their impact on the CFFs determination, Section \ref{sec:CFFsym}); the GPDs perturbative QCD (PQCD) evolution equations are summarized in Section \ref{sec:evol}; the Mellin moments and their symmetries are listed in Section \ref{subsec:Mellin}. 
 
\subsection{Compton Form Factors}
\label{sec:CFFs}
In a QCD factorized framework \cite{Ji:1998xh,Collins:1998be} the amplitude for the DVES process in Fig.\ref{fig:Feynman} is parametrized in terms of CFFs defined as convolutions of the GPDs for each quark flavor, $q$, with appropriate Wilson coefficient functions. 
\subsubsection{DVCS}
\label{sec:CFF_DVCS}
For DVCS off a proton target, $\gamma ^* p \rightarrow \gamma p$, at leading order, both  
the vector, ${\cal F}_q$ = (${\cal H}_q$, ${\cal E}_q$), and axial-vector $\widetilde{\cal F }_q$= ($\widetilde{\cal H}_q$ $\widetilde{\cal E}_q$), sectors contribute, yielding the following CFFs,
\begin{subequations}
\label{eq:CFFdef}
\begin{eqnarray}
\mathcal{F}_q(\xi,t)  & = &  C^+ \otimes F_q \equiv \int_{-1}^{1} dx  \,  C^+(x,\xi) F_q(x,\xi,t), 
 \\
\mathcal{\widetilde{F}}_q(\xi,t) & = & C^-  \otimes  \widetilde{F}_q  \equiv \int_{-1}^{1} dx  \,  C^-(x,\xi) \widetilde{F}_q(x,\xi,t) , 
\label{eq:CFF}
\end{eqnarray}
\end{subequations}
where $F_q= H_q, E_q$, $\widetilde{F}_q= \widetilde{H}_q, \widetilde{E}_q$ are the quark GPDs. 
The GPDs parametrize the correlation function which, for the quark-quark case, considering nucleon states, with $\mid p, \, \Lambda \rangle$ with momentum $p$ ($p'$) and helicity $\Lambda$ ($\Lambda'$), reads,
\begin{eqnarray}
W_{\Lambda' \Lambda}^{[\Gamma]} &=& \frac{1}{2} 
\int \frac{d^{4}z}{(2\pi)^{4}}e^{ik\cdot z} \left\langle p',\Lambda'\bigg| \bar{\psi}\left(-\frac{z}{2}\right) \Gamma {\cal W}\left(-\frac{z}{2},\frac{z}{2} \bigg| n\right) \psi\left(\frac{z}{2}\right)\bigg|p,\lambda\right\rangle
\label{eq:correlator}    
\end{eqnarray}
where $\Gamma$ is a Dirac operator, and ${\cal W}$ is the gauge link operator, or Wilson line, conserving color gauge invariance.
In this paper, we focus on the vector sector GPDs and CFFs, $F=H,E$. The total CFF for the proton is composed, in this case, of the contributions from all quark flavors, weighted by $e_q^2$, i.e., 
\begin{equation}
    \mathcal{H} = \sum_{q} e_q^2 \mathcal{H}_q, \quad\quad \mathcal{E} = \sum_{q} e_q^2 \mathcal{E}_q,.
\end{equation}
From Eqs.\eqref{eq:CFFdef}, one can see that to obtain the full $(x,\xi,t)$ dependence of GPDs from data requires addressing an inverse problem with singular kernels given by $C^{\pm}$.  
Physically motivated parametrizations can help address the ``deconvolution problem'' in DVCS by narrowing the range of possible solutions, which can provide information on the trends of GPDs in various kinematic regions, which in turn is essential to inform quantitative fits of the DVES data.     

The leading order coefficients functions are given by  (see {\it e.g.} \cite{Kriesten:2021sqc} and references therein),
 \begin{equation}
 \label{eq:CFFpm}
 { C^\pm(x,\xi) = \frac{1}{x-\xi + i \epsilon} \pm \frac{1}{x+\xi - i \epsilon} . } 
 \end{equation}
The real and imaginary parts of the CFFs can be separated out as,
\begin{subequations}
\label{eq:CFF}
\begin{eqnarray}
\mathcal{F}_q(\xi,t)  & = & \Re e \mathcal{F}_q + i \, \Im m \mathcal{F}_q = P.V. \int_{-1}^{1} dx  \left( \frac{1}{x-\xi} + \frac{1}{x+\xi} \right)  F_q(x,\xi,t) + i \pi \, F_q(\xi,\xi,t)  , 
 \\
\mathcal{\widetilde{F}}_q(\xi,t) & = &  \Re e \mathcal{\widetilde{F}}_q +  i \, \Im m \mathcal{\widetilde{F}}_q = P.V. \int_{-1}^{1} dx  \left( \frac{1}{x-\xi} - \frac{1}{x+\xi} \right)  \widetilde{F}_q(x,\xi,t) + i \pi \, \widetilde{F}_q(\xi,\xi,t) , 
\end{eqnarray}
\end{subequations}
The coefficient functions obey the following symmetry relations for $x \rightarrow -x$,
\begin{subequations}
\label{eq:C_coeff_app}
\begin{eqnarray}
C^-(-x,\xi) & = & C^-(x,\xi) \quad\quad symmetric \\
C^+(-x,\xi) & = & -C^+(x,\xi) \quad antisymmetric
\end{eqnarray}
\end{subequations}
where we notice that the $i\epsilon$ term also contributes to the symmetry as it appears with opposite signs in Eq.\eqref{eq:CFFpm}.

\subsection{Quark-Parton Symmetries}
\label{sec:sym}
GPDs observe crossing symmetry relations with respect to $x \rightarrow -x$, which allow us to introduce flavor non-singlet, valence and flavor singlet distributions (see also discussion in Refs.\cite{GolecBiernat:1998ja,Goldstein:2010gu}). In DVCS, the proton GPD is written in terms of the quark GPDs as,
\begin{eqnarray}
\label{eq:Hp}
H_p & = & \sum_q e_q^2 \, (H_q+H_{\bar{q}}) = \frac{4}{9} H_{u+\bar{u}}+ \frac{1}{9} H_{d+\bar{d}}+ \frac{2}{9} H_{s}   \nonumber \\
\end{eqnarray}
where $e_q$ is the quark charge, and we consider contributions up to the strange quark. 
%
We define the unpolarized anti-quark distribution, $\bar{q}(x)$, as,
\begin{eqnarray}
\bar{q}(x) = - q(-x)
\end{eqnarray}
Similarly, we define the antiquark GPD, $H_{\bar{q}}(x,\xi,t)$ as, 
\begin{eqnarray}
    H_{\bar{q}}(x,\xi,t) = - H_{{q}}(-x,\xi,t) 
\end{eqnarray}
The $+$ and $-$ contributions are defined analogously to PDFs as,
\begin{subequations}
\label{eq:Hplusminus}
\begin{eqnarray}
    H^+_q(x,\xi,t) = H_{{q}}(x,\xi,t)  + H_{\bar{q}}(x,\xi,t) \\
    H^-_q(x,\xi,t) = H_{{q}}(x,\xi,t)  - H_{\bar{q}}(x,\xi,t)
\end{eqnarray}
\end{subequations}
with symmetry properties,
\begin{eqnarray}
 H^+_q(x,\xi,t) = -  H^+_q(- x,\xi,t)  \\
 H^-_q(x,\xi,t) =  H^-_q(- x,\xi,t)
\end{eqnarray}

Focusing on the GPD $H$, the ``$+$" and ``$-$" components, Eqs.\eqref{eq:Hplusminus},
 obey  the following symmetry  properties for $x \rightarrow -x$ (we omit writing the dependence on the $\xi,t$ variables),
\begin{subequations}
\label{eq:Hplusminu}
\begin{eqnarray}
H^-_q(-x) &= & H^-_q(x) \quad\quad symmetric \; \\
H^+_q(-x) &=& -H^+_q(x) \quad antisymmetric 
 \end{eqnarray}
\end{subequations}
In the Kuti-Weisskopf ansatz \cite{Kuti:1971ph},
\begin{subequations}
\begin{eqnarray}
H_{q}(x) &= & H_{q_v}(x) + H_{q_{sea}}(x) \, \Rightarrow H_{q_v}(x) = H_{q}(x) - H_{\bar{q}}(x) \equiv H_q^-(x) \\
H_{\bar{q}} &=& H_{q_{sea}},
\end{eqnarray}
\end{subequations}
Using the $SU(N_f=3)$ flavor symmetry, we can write the flavor non-singlet ($NS$) contributions, respectively,  as,
\begin{subequations}
\label{eq:nonsinglet_H}
\begin{eqnarray}
H^-_q  & = &  H_{q_v} , \quad q=u,d\\
 H^{T3} & = & H_{u}^+ - H_{d}^+ \\
H^{T8} & = & H_{u}^+ + H_{d}^+ - 2 H_{s}^+  
\end{eqnarray}
\end{subequations}
while the singlet ($S$) term is given by,
\begin{eqnarray}
\label{eq:single_H}
H_S &  =  &  H_{u}^+ + H_{d}^+ + H_{s}^+  \nonumber \\
&=& H_{u_v}+ H_{d_v} + 2 (H_{\bar{u}} + H_{\bar{d}} + H_{\bar{s}} ) ,
\end{eqnarray}

\noindent The unpolarized/polarized gluon distributions are even/odd functions of $x$.
\begin{align}
\label{eq:gluon_symmetry}
    H_{g}(x,\xi) &= H_{g}(-x, \xi)\\
    \tilde{H}_{g}(x,\xi) &= -\tilde{H}_{g}(-x, \xi)
\end{align}
%
Analogous relations exist in the asymmetric frame, where the axis of symmetry corresponding to $x=0$, becomes $X=\zeta/2$. These relations are summarized in Appendix \ref{app:asymmetric}.
Notice that in the forward limit, $\zeta,t=0$, {the following correspondence holds between the unpolarized GPDs $H_q$ ($H_g$) and the unpolarized PDFs $q(x)$ ($g(x)$),}
\begin{eqnarray}
    H_q(x,0,0) = q(x), \quad\quad\quad  H_g(x) = x \, g(x) 
    \label{eq:forward}
\end{eqnarray}

\subsection{Compton Form Factor Symmetries}
\label{sec:CFFsym}
\begin{figure}[h]
    \centering
    \includegraphics[width=7.5cm]{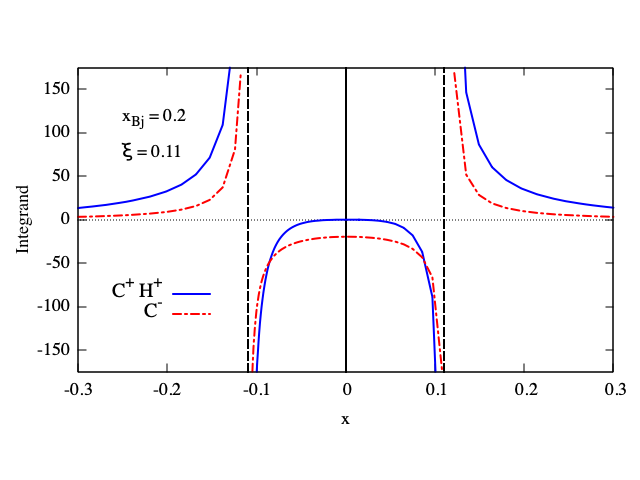}
    \includegraphics[width=7.5cm]{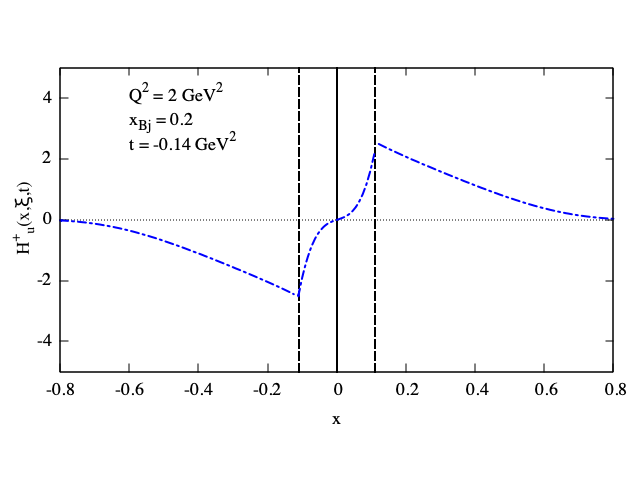}
    \caption{{\it Left}: Integrand, $H_q^+(x,\xi,t) C^+$, in Eq.\eqref{eq:ReH} compared to the coefficient $C^-$ which shows a similar symmetry; {\it Right}: GPD $H^+_u$ used to evaluate the integrand in Eq.(\ref{eq:ReH}) shown on the {\it l.h.s.}.}
    \label{fig:integrand_cff}
\end{figure}

%
Using the symmetry properties from Section \ref{sec:CFFs} and \ref{sec:sym}, one finds, 
 \begin{eqnarray}
 \label{eq:CFFsymm}
 {\cal H}_q   &= &   \int_{-1}^{1} dx \, C^+  \frac{H^+_q + H_q^-}{2} = \frac{1}{2} \int_{-1}^{1} dx \, C^+  H^+_q  = \int_{0}^{1} dx \, C^+  H^+_q \\
 {\cal H}_{\bar{q}} & = &  \int_{-1}^{1} dx \, C^+  \frac{H^+_q - H_q^-}{2} = \frac{1}{2} \int_{-1}^{1} dx \, C^+  H^+_q = \int_{0}^{1} dx \, C^+  H^+_q 
  \end{eqnarray}
  therefore, ${\cal H}_q = {\cal H}_{\bar{q}}$. For $u$ and $d$ quarks we define,
  \[ H_{u_v} = H_u^-, \quad\quad H_{d_v} = H_d^-   \] 
(analogous definitions are for the GPD $E$).
%
\begin{figure*}[h!]
    \centering
    \includegraphics[width=7.5cm]{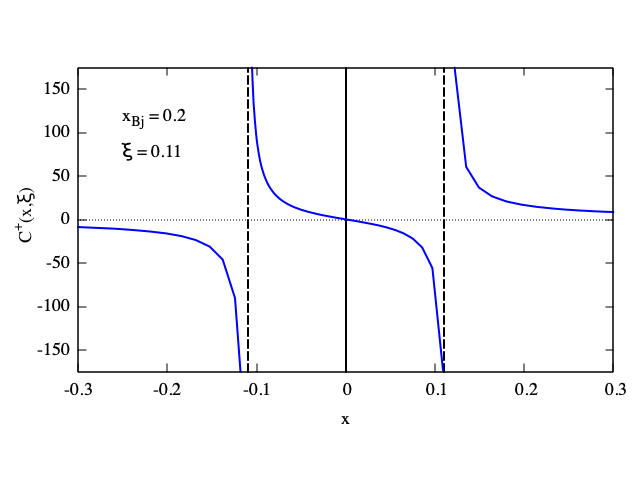}
    \hspace{1cm}
    \includegraphics[width=7.5cm]{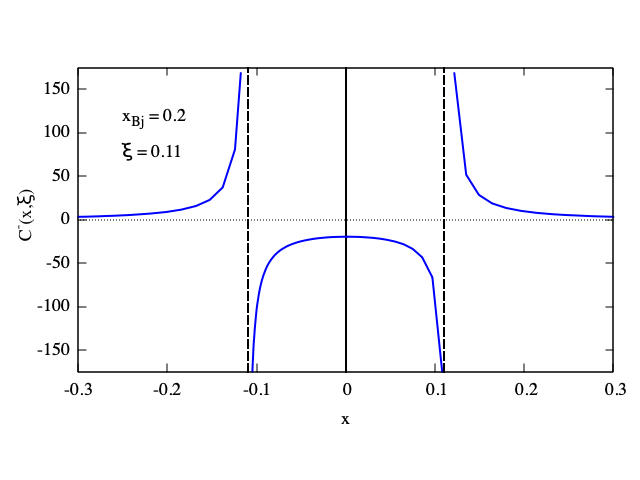}
    \caption{Coefficients of the Compton form factors, $C^+$ (left) and $C^-$ (right). The figure displays the symmetries around $x=0$, displayed in Eq.\eqref{eq:CFFpm}.}
    \label{fig:CFF}
\end{figure*}
Inserting $C^\pm$ in Eqs.\eqref{eq:CFF} we obtain for the real part,
\begin{eqnarray}
\label{eq:ReH}
{\Re} e {\cal H}_q & = & P.V. \int_{-1}^1 dx \,  H_q(x,\xi,t) \left( \frac{1}{x-\xi} + \frac{1}{x+\xi} \right) 
= P.V. \int_{0}^1 dx \, \frac{H_q^+(x,\xi,t)}{x-\xi} + \int_{0}^1 dx \,  \frac{H_q^+(x,\xi,t)  }{x+\xi} 
\end{eqnarray}
where we notice that the second term in each equation is not singular in the integration region of interest.   
The coefficient functions, $C^+$ and $C^-$ are plotted vs. $x$ at $\xi=0.11$ in Figure \ref{fig:CFF}, where one can see their respective symmetries around the $y$-axis (an equivalent symmetry is obtained for the $(X,\zeta)$ set of variables, around $X=\zeta/2$). 
The integrand in Eq.\eqref{eq:ReH}, $H_q^+(x,\xi,t) C^+$ is plotted in Figure \ref{fig:integrand_cff} ({\it l.h.s.}) along with the coefficient function, $C^-$ (Eq.\eqref{eq:CFFpm}, and Fig.\ref{fig:CFF}), to show the similarity in the respective symmetry properties of the two terms.  $H_q^+$,  was calculated for $q=u$ in the model of Ref.\cite{Kriesten:2021sqc} (Fig.\ref{fig:integrand_cff}, {\it r.h.s.}).
Similar values can be obtained for the $d$ quark. 

Eq.\eqref{eq:CFFsymm} can be written in terms of the  asymmetric set variables (Eqs.\eqref{eq:asym}) as, 
 \begin{eqnarray}
 \label{eq:CFFasymm}
 {\cal H}_q & = &   \frac{1}{2} \int_{-1+\zeta}^{1}  dX \, C^+  H^+_q = 
 \int_{\zeta/2}^{1} {dX}\, C^+  H^+_q .
 \end{eqnarray}
By examining Figures \ref{fig:CFF} and \ref{fig:integrand_cff}, one can see that kernels $C^{\pm}$ being highly singular in $x=\xi$, concentrate the GPD {\it information} content of $\Re e H_q$, in the region $x \approx  \xi$. Similar graphs can be rendered in terms of the asymmetric set of variables with a shift of the axis of symmetry from $x=0 \rightarrow X=\zeta/2$, and the asymptotes position, from $x = \xi \rightarrow X=\zeta$. 

To study the singular behavior, we evaluated the principal value integral through the modified Gaussian method (see also Appendix C in Ref.\cite{Goldstein:2010gu}). We use the asymmetric notation and add and subtract in the P.V. integral the term,
\[ \frac{H^+_q(\zeta,\zeta,t)}{X-\zeta} \quad \left( \equiv \frac{H^+_q(\xi,\xi,t)}{x-\xi} \right) , \] 
obtaining,
\begin{eqnarray}
\label{eq:ReH_Gaussian}
{\Re} e {\cal H}_q & = &  \int_{\zeta/2}^1 dX \, \frac{H^+_q(X,\zeta,t) - H^+_q(\zeta,\zeta,t)}{X-\zeta} + H^+_q(\zeta,\zeta,t) \, \ln \frac{(1-\zeta)}{\zeta/2}  +  \int_{\zeta/2}^1 dX  \,  \frac{ H_q^+(X,\zeta,t) }{X}
\end{eqnarray}
where we used,
\begin{eqnarray}
 P.V. \int_{\zeta/2}^1 dX  \, \frac{1}{X-\zeta} =  \ln \frac{(1-\zeta)}{\zeta/2}   .
\end{eqnarray}
\begin{figure}[h]
    \centering
    \includegraphics[width=10cm]{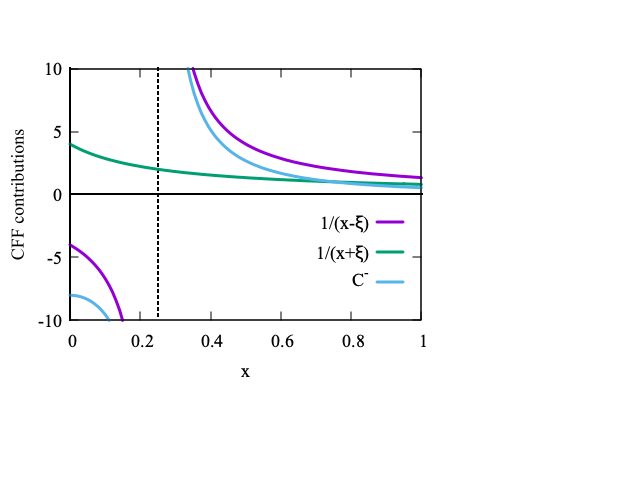}
    \caption{Contributions to the kernel of the CFF illustrating the role of the singular term, $1/(x-\xi)$, and the non singular term, $1/(x+\xi)$ in the integration, Eq.\eqref{eq:CFFasymm}, for $\xi=0.25$. The coefficient $C^-$ is shown because it has the same symmetry as $C^+H^+$.}
    \label{fig:2nd_integrand_cff}
\end{figure}
\begin{figure*}
    \centering
    \includegraphics[width=0.49\linewidth]{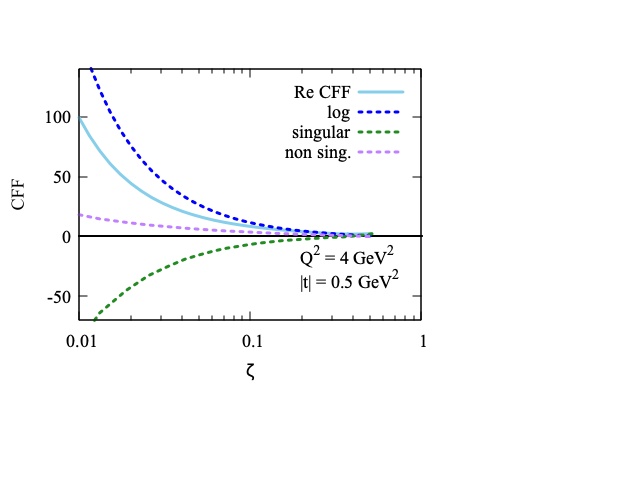}
    \caption{Contributions to the real CFF calculated according to Eq.(\ref{eq:ReH_Gaussian}) for the $u$-quark, plotted vs. $\zeta$, at the kinematics point, $Q^2 = 4$ GeV$^2$, $t=-0.5$ GeV$^2$. The allowed maximum value of $\zeta=x_{Bj}$ at this kinematic point is $\zeta_{max}=0.51$, which is where the curves subside. The curve denoted ``log" corresponds to ${\bf (1)}$ in the text, while``singular" and ``non-singular" correspond to terms ${\bf (2)}$  and  ${\bf (3)}$, respectively.} 
    \label{fig:Re_cff1}
\end{figure*}
We distinguish three terms in the equation,
\begin{subequations}
\label{eq:integral_components}
\begin{eqnarray}
\label{eq:term1}
& {\bf (1) } & \quad\quad  H_q^+(\zeta,\zeta,t) \, \ln\frac{1- \zeta}{\zeta/2} \\
\label{eq:term2}
& {\bf (2) } & \quad\quad  \int_{\zeta/2}^1 dX \, \frac{H_q^+(X,\zeta,t) - H_q^+(\zeta,\zeta,t)}{X-\zeta} \\
\label{eq:term3}
& {\bf (3) }& \quad\quad  \int_{\zeta/2}^1  dX \,  \frac{ H_q^+(X,\zeta,t)  }{X}. 
\end{eqnarray}
\end{subequations}
Term ${\bf (1)}$  dominates over the other two, which are given by integrals over smooth functions, partially canceling one another. To illustrate the behavior of the three different terms, in Figure \ref{fig:2nd_integrand_cff} we show at a given kinematic point, $x=0.25$, the $x$ behavior of the singular part of the kernel in $\xi$, $1/(x-\xi)$, the non singular part, $1/(x+\xi)$, and their algebraic sum corresponding to $C^-$ (this corresponds to an analogous behavior in $X$).
Next, in Figure \ref{fig:Re_cff1} we show the integrated terms corresponding to Eqs.\eqref{eq:integral_components}, plotted vs. $\zeta$, for the kinematic point, $t = -0.5$ GeV$^2$,  $Q^2 = 4 $ GeV$^2$, and corresponding range in $x_{Bj} \equiv \zeta = [0,0.53]$. In Figure \ref{fig:Re_cff1} we show on the {\it l.h.s.} the contributions of all three components to the $u$-quark CFF, $\Re e {\cal H}_q$, at the kinematic point, $Q^2= 4$ GeV$^2$, $| t | = 0.5$ GeV$^2$ , plotted as a function of $\zeta$, in the allowed kinematic range. Note that $H_u^+ = H_{u_v} + 2 H_{\bar{u}}$, where the antiquark component dominates the low $\zeta$ behavior. One can see that while the logarithmic ``endpoint" component dominates the CFF, the contribution of the integral parts, ${\bf (2)}$ and ${(\bf 3)}$ is still important and it cannot therefore be disregarded. On the {\it rhs} we plot ${\Re e H_u}$ and ${\Im m H_u}$: one can see that the curves track each other.

In summary, in this Section we examined the information on the GPD structure that can be obtained from the experimental extraction of the imaginary and real parts of the CFFs. The mathematical form of the coefficient functions, $C^\pm(x,\xi)$, observes symmetries around $x=0$ ($X=\zeta/2$) and $x=\pm \xi$ ($X=0,\zeta$), which, in turn, control the GPD content of the CFFs. In particular, the only non-vanishing contributions to the CFFs for DVCS under $x \rightarrow -x$ inversion correspond to the flavor singlet components, $F^+_q = F_{q_v} + 2 F_{\bar{q}}$. 
An important consequence of being sensitive to the singlet component is that they couple to the gluon GPDs in the PQCD evolution equations. Gluon GPDs can therefore be obtained from DVCS experiments. Quantitative evaluations are given in the results discussion in Section \ref{sec:results}. 

Most importantly, we can address the question of what type of information can be extracted from $\Re e H$ {\it and} $\Re e E$, in addition to $\Im m H$ and $\Im m E$.  A possible answer, expounded here, is that $\Re e H$,  $\Re e E$ carry essentially the same information as their imaginary counterparts in that they are only sensitive to values of the GPD which are concentrated around the crossover point $X=\zeta$ ($x=\xi$) between the DGLAP and ERBL regions. This is an important step towards deconstructing the "inverse problem" of extracting GPDs from the observables given by measurable CFF convolution integrals.

\subsection{PQCD Evolution}
\label{sec:evol}
PQCD evolution equations for GPDs have been derived from the inception in Refs.\cite{Ji:1997nk,Musatov:1999xp,Radyushkin:1998es,GolecBiernat:1998ja} in a LO framework, followed  by more studies at NLO (Ref.\cite{Freund:2001bf}, and references therein). 
A practical approach with directly usable expressions that are not specific to the GPD representation ({\it i.e.} whether they are given by double distributions, or in the treatment of their kinematic limits) was proposed in Ref.\cite{GolecBiernat:1998ja}, and more recently, in the comprehensive study of Ref.\cite{Bertone:2022frx}, where GPD evolution was reviewed, and several technical aspects were listed systematically.  
In our approach, we start from the DGLAP region, and we use the asymmetric system, therefore $X>\zeta$, and $X<0$. Note that, because of the symmetries described in Sections \ref{sec:sym} and \ref{sec:CFFsym}, we can restrict calculations to positive momentum fraction $X$, namely $X>\zeta/2$. We believe that the asymmetric system is the most appropriate choice to derive evolution equations in momentum space, as momentum fractions are taken directly from the parent proton, thereby obtaining simpler expressions. Then, the evaluation of evolution equations in the DGLAP region proceeds similarly to the PDFs, as a natural extension of the ones defined for DIS observables. 

For the flavor $NS$ case, one has ($Q^2=\mu$), 
\begin{eqnarray}
\label{eq:evolNS}
\mu \frac{\partial }{\partial \mu} H_{NS}(X,\zeta) &= &\frac{\alpha_S(\mu)}{2\pi}  \int_X^1 \frac{dZ}{Z}  \, 
P_{qq}\left(\frac{X}{Z},\frac{X-\zeta}{Z-\zeta} \right) \, H_{NS}(Z,\zeta)      
\end{eqnarray}
and for the singlet and gluon distributions, $H_{S}$ and $H_{g}$, respectively, 
\begin{subequations}
\label{eq:evolS}
\begin{eqnarray}
\mu \frac{\partial }{\partial \mu} H_{S}(X,\zeta) &= &\frac{\alpha_S(\mu)}{2\pi} \left[  \int_X^1 \frac{dZ}{Z}  \, 
P_{qq}\left(\frac{X}{Z},\frac{X-\zeta}{Z-\zeta} \right) \, H_{S}(Z,\zeta)  
+  2 N_f \int_X^1 \frac{dZ}{Z}  \, P_{qg}\left(\frac{X}{Z},\frac{X-\zeta}{Z-\zeta} \right) \, H_{g}(Z,\zeta)    \right] \\
\mu \frac{\partial }{\partial \mu} H_{g}(X,\zeta) &= &\frac{\alpha_S(\mu)}{2\pi} \left[  \int_X^1 \frac{dZ}{Z}  \, 
P_{gq}\left(\frac{X}{Z},\frac{X-\zeta}{Z-\zeta} \right) \, H_{S}(Z,\zeta)  +  \int_X^1 \frac{dZ}{Z}  \, 
P_{gg}\left(\frac{X}{Z},\frac{X-\zeta}{Z-\zeta} \right) \, H_{g}(Z,\zeta)    \right]
\end{eqnarray}
\end{subequations}
The LO off-forward splitting functions read,
\begin{subequations}
\begin{eqnarray}
    P_{qq}(Y,Y') & = & C_F \left[ \frac{1 + YY'}{(1-Y)_+} + \frac{3}{2} \delta(1-Y) \right]   ,   \\ 
    P_{qg}(Y,Y') & = & \frac{1}{2} \left[1- Y'(1-Y) - Y(1-Y')  \right]   ,   \\ 
    P_{gq}(Y,Y') & = & C_F \left[ \frac{(1-Y)(1-Y') + 1 }{Y}\right]    ,   \\ 
    P_{gg}(Y,Y') & = & 2 C_A \left[ \frac{(1-Y')(1+YY')}{Y} + \frac{1 + YY'}{(1-Y)_+} + \frac{3}{2} \delta(1-Y) \right]  
\end{eqnarray}
\label{eq:LOsplitting}    
\end{subequations}
where we set, $Y=X/Z$, $Y'=(X-\zeta)/(Z-\zeta)$, and $C_F=4/3$, $C_A=N_c=3$. The endpoint from the $+$ distribution plus the delta function contribution is,
\begin{equation}
 C_F \left(   \frac{3}{2} +  \frac{\log(1-X)^2 }{1-\zeta} \right) 
\end{equation}
These equations follow closely the ones presented in Ref.\cite{Golec-Biernat:1998zbo} (Appendix A), and they relate to Eq.(35) of Ref.\cite{Bertone:2022frx}, which are written in $(x,\xi)$ coordinates. Similar equations can be written for the GPD $E_{q,g}$.
%
The evolution equations are written in terms of the flavor NS and S combinations as in Eqs.\ref{eq:nonsinglet_H}, and \eqref{eq:single_H} formed from the individual quark and gluon contributions at the initial scale, $Q_o^2$; the final results for evolution are then obtained by inverting the system, as, 
\begin{subequations}
\label{eq:SU6}
\begin{eqnarray}
    	H_{\bar{u} } & = & \frac{1}{4} \left[ H_{T_3} + \frac{1}{3} H_{T_8} + \frac{1}{6} H_{T_{15}} +
        \frac{1}{10} H_{T_{24}} + \frac{1}{15} H_{T_{35}}  - 2 H_{u_v} + \frac{1}{3} H_S  \right] \\
  H_{\bar{d}} & = & \frac{1}{4}  \left[ -H_{T_3} + \frac{1}{3} H_{T_8}  + \frac{1}{6} H_{T_{15}} +
 \frac{1}{10} H_{T_{24}} + \frac{1}{15} H_{T_{35}}  - 2 H_{d_v}   + \frac{1}{3} H_S  \right] \\
	H_{\bar{s} }  & =  & - \frac{1}{6} H_{T_8}+   \frac{1}{24} H_{T_{15}} +
     \frac{1}{40} H_{T_{24}} +  \frac{1}{60} H_{T_{35}} + \frac{1}{12} H_S 
    \end{eqnarray}
\end{subequations}
We adopt the variable-flavor-number scheme with charm and bottom quarks thresholds: $m_c= 1.4$ GeV, and 
and $m_b = 4.5$ GeV, respectively.

\vspace{0.3cm}
Practical approaches to PQCD evolution in the ERBL region, which technically can be derived from a separate limit of a common kernel describing both the ERBL and DGLAP region, involve addressing two main physical constraints: on one side evolution has to be consistent with the symmetries around $X=\zeta/2$ ($x=0$), and, on the other,  the crossover at $X=\zeta$ ($x=\xi$) between the DGLAP and ERBL regions needs to be taken carefully, in order to avoid possible singularities and/or the emergence of unphysical discontinuities.
In other words, the PQCD evolution equations have to be solved simultaneously in the ERBL and in the DGLAP
regions while observing the basic symmetries of the theory and resolving any possible singularities \cite{Freund:2001bf,GolecBiernat:1998ja}.
In the present approach, we start with a symmetric/antisymmetric parametric form in the ERBL region at the initial scale, $\mu_o=Q_o^2$ (given in Section \ref{sec:results}) that matches the form in the DGLAP region. Symmetry at $Q^2>Q_o^2$, is then preserved by letting the ERBL parameters be $Q^2$ dependent, and imposing the constraint on the area rule subtended by GPDs, induced by their polynomiality property \cite{Belitsky:2001ns,Diehl:2003ny,Kumericki:2016ehc}. Our method is complementary to the approach of Ref.\cite{GolecBiernat:1998ja} where the symmetry constraint was accomplished by introducing an extra term directly in the kernel of their equivalent to Eqs.\eqref{eq:Hplusminu},\eqref{eq:gluon_symmetry}. We, however, find a similar issue in the presence of a cusp at $X=\zeta$, which arises from a discontinuity in the derivatives of distributions. As we explain in our Results section (Section \ref{sec:results}), this discontinuity is circumvented when calculating physical observables given by the CFFs.   

The latter are evaluated with their full $Q^2$ dependence by the convolution of the  GPDs evolved using Eqs.\eqref{eq:evolNS},\eqref{eq:evolS}, with the off-forward Wilson coefficient functions,
\begin{eqnarray}
\label{eq:NLO1}
{\cal F}_q(\zeta,t,\mu) & = & \int_{-1+\zeta}^{1} dX \left[ C_q(X,\zeta,\mu) H_q(X,\zeta,t;\mu) + C_g(X,\zeta,\mu) H_g(X,\zeta,t;\mu) \right]  .
\end{eqnarray}
The $C_q$ and $C_g$ coefficients in Eq.(\ref{eq:NLO1}), evaluated up to NLO read,
\begin{subequations}
\begin{eqnarray}
C_q(X,\zeta) & = & C_q^{LO}(X,\zeta) + \frac{\alpha_S(Q^2)}{2\pi} C_q^{NLO}(X,\zeta) \\
C_g(X,\zeta) & = &  \frac{\alpha_S(Q^2)}{2\pi} C_g^{NLO}(X,\zeta) 
\end{eqnarray}
\end{subequations}
In this paper we present results on CFFs using our parametrization evolved at LO, with $C_q^{LO}$ given by Eq.\eqref{eq:CFF}, consistent with present benchmarks \cite{Bertone:2022frx}. Note that using Eqs.\eqref{eq:SU6}, enables us to extract the gluon distributions from the coupling of $H_g$ to the singlet, component $H_S$, Eq.\eqref{eq:evolS}.


%


\subsection{Mellin Moments}
\label{subsec:Mellin}
The Mellin moments of the GPDs are defined as \cite{Ji:1998pc},
\begin{align}
\label{eq:Mellin}
    M_{n}^{q}(Q^2,\xi,t) = \int_{-1}^{1}dx x^{n-1} F_q(x, \xi,t) \equiv \int_{-1+\zeta}^{1} \frac{dX}{(1-\zeta/2)^n} \, X^{n-1}F_q(X, \zeta,t)
\end{align}
where $n=1, 2, ...$, and $F_{q,g}= H_{q,g}, E_{q,g}$.
One can show that, similarly to what was derived for the PDFs in the forward limit, Eq.\eqref{eq:Mellin} corresponds directly to the matrix elements of the operators in the Wilson operator product expansion (OPE) and their $Q^2$ dependence is thus determined by the anomalous dimensions. The property of ``polynomiality" written below, is obtained by extending the OPE to include the new kinematic variable, $\Delta_\mu$, 
\begin{eqnarray}
M_n^{q,H} (\xi,t,Q^2) & = & \sum_{i, even}^n (2 \xi)^i A_{n+1,i}(t,Q^2) + \mod(n,2) (2 \xi)^{n+1} C_{n+1,i}(t,Q^2) \\
M_n^{q,E}(\xi,t,Q^2) & = & \sum_{i, even}^n (2 \xi)^i B_{n+1,i}(t,Q^2) -\mod(n,2) (2 \xi)^{n+1} C_{n+1,i}(t,Q^2) 
\end{eqnarray}
For $n=1$, $A_{10}$, $B_{10}$ are the nucleon Dirac and Pauli form factors, respectively, 

\begin{eqnarray}
\label{eq:dirac}
F_1^q &=& A_{10} = \int_{-1+\zeta}^1 \frac{dX}{1-\zeta/2} H_q 
\equiv \frac{1}{2} \int_{-1+\zeta}^1 \frac{dX}{1-\zeta/2} \, H^-_q = \int_{\zeta/2}^1 \frac{d X}{1-\zeta/2} \, H_{q_v}(x)  
\\
\label{eq:pauli}
F_2^q &=& B_{10} = \int_{-1+\zeta}^1 \frac{dX}{1-\zeta/2} E_q 
\equiv \frac{1}{2} \int_{-1+\zeta}^1 \frac{dX}{1-\zeta/2} \, E^-_q = \int_{\zeta/2}^1 \frac{d X}{1-\zeta/2} \, E_{q_v}(x)  
\end{eqnarray}
For $n=2$, we obtain, 
\begin{eqnarray}
\label{eq:A20}
M_2^{q,H} =  A_{20} + (2\xi)^2 C_{20} & = & \int_{-1+\zeta}^1 \frac{dX}{1-\zeta/2} X \, H_q = \int_{-1+\zeta}^1 \frac{dX}{1-\zeta/2} X \,\frac{H^+ +H^-}{2}  = \int_{\zeta/2}^1 \frac{d X}{1-\zeta/2} \, X H_q^+  
\\
\label{eq:B20}
M_2^{q,E} =   B_{20} - (2\xi)^2 C_{20}  &=&  \int_{-1+\zeta}^1 \frac{dX}{1-\zeta/2} X \, E_q 
= \int_{-1+\zeta}^1 \frac{d X}{1-\zeta/2} \, X \frac{E^+ +E^-}{2}   =  \int_{\zeta/2}^1 \frac{dX}{1-\zeta/2} \, X E^+_q 
\end{eqnarray}
In the forward limit, $(\xi,t=0)$, we define: $A_{20} = \langle X \rangle_{q,g}$, the average momentum, and $(A_{20}+ B_{20})/2= J_{q,g}$, the average total angular momentum, respectively carried by a parton of flavor $q$ or a gluon, $g$.

\noindent 
We use the form factors for the $u$ and $d$ flavors  extracted from lepton-nucleon scattering experiments and also evaluated in LQCD, while for the higher $n$ moments, with $n \geq 2$,  we use the LQCD results of Ref.\cite{Bhattacharya:2023ays}.
Due to symmetry, the elastic form factors are completely transparent to the sea quark contribution with the exception of the $s-\bar{s}$ component, that is expected to be small and effectively treated like a valence component (see discussion in \cite{Sufian:2016pex}).
Using flavor symmetry, taking the strange form factor $F_1^s \approx 0$ \cite{Diehl:2007uc}, we obtain a constraint on both the normalization and $t$ dependence of the GPDs, $H_{u_v}, H_{d_v}, E_{u_v}$, and $E_{d_v}$,  from the flavor separated  elastic proton and neutron data \cite{Cates:2011pz,Qattan:2004ht}, using,
\begin{subequations}
\label{eq:ud_formf}
\begin{eqnarray}
F_{1,2}^u & = & 2 F_{1,2}^p + F_{1,2}^n \\
F_{1,2}^d & = &   F_{1,2}^p + 2 F_{1,2}^n ,
\end{eqnarray}
\end{subequations}
where it is a standard procedure to define the $u$ and $d$-quark form factors as the $u$ and $d$ distributions in the proton $F_1^{u,d} \equiv F_1^{(u,d)/p}$.

Using the symmetries from Section \ref{sec:GPDs}, we obtain the normalization and $t$ dependence of the antiquark GPDs, $H_{\bar{u}}, H_{\bar{d}}, E_{\bar{u}}$ and $E_{\bar{d}}$  from the $n=2$ moment constraint for $F=H,E$, Eqs.\eqref{eq:A20},\eqref{eq:B20}, as follows, 
\begin{eqnarray}
\label{eq:kutiW_moment}
    M_2^{q,F} & = & \frac{1}{2} \int_{-1+\zeta}^1 \frac{dX}{1-\zeta/2} X ( H^+ + H^-) = \int_{\zeta/2}^1 \frac{dX}{1-\zeta/2} X  H^+ 
    = M_2^{q_v} + 2 M_2^{\bar{q}} .
\end{eqnarray}
We, therefore, obtain a constraint on the antiquark $\bar{u}$ and $\bar{d}$ using results from the LQCD evaluation of  $M_2^{q,F}$ \cite{Bhattacharya:2023ays}, and $M_2^{q_v}$, using the parameters obtained from Eqs.\eqref{eq:ud_formf}. 
Notice that the strange quarks, as well as the heavier flavors, are taken equal to zero at our initial scale which is $\lesssim 2 m_s^2$. However, we generate strange quark GPDs through PQCD evolution. The treatment of the $s, \bar{s}$ components will be the subject of a future dedicated paper.

Finally, both the normalization and $t$-dependence of the gluon GPD, $H_g$, is constrained by LQCD results \cite{Shanahan:2018pib,Pefkou:2021fni}, through the moments,
For the gluon component one has,
\footnote{In Ref.\cite{Shanahan:2018pib}, the following definition is used for $D_{g}(t)$, which is just $4 C_{g}(t) = D_{g}(t)$.}
\begin{align}
    \int_{0}^{1} H_g(x, \xi, t; Q^{2}) dx = A_{g}(t) + (2\xi)^{2} C_{g}(t)\\
    \int_{0}^{1} E_g(x, \xi , t; Q^{2}) dx = B_{g}(t) - (2\xi)^{2} C_{g}(t), 
\end{align}
The gluon form factor can also be written in asymmetric notation as,
\begin{eqnarray}
    F^g &=& A_{10}^g = \int_{-1+\zeta}^1 \frac{dX}{1-\zeta/2} H_g
\end{eqnarray}
We conclude this Section by noting that, in order to merge consistently information from experimental data, theoretical constraints, and LQCD results, we need to address whether the quark, antiquark and gluon contributions are defined equivalently in all sectors. LQCD evaluations being path-integral-based,  can differ, in principle, from the Kuti-Weisskopf ansatz,  depending on how the sea quarks are introduced. In LQCD, the contribution of the connected sea quarks (CS) is obtained from a four point current-current
correlator for the nucleon through a ``connected" insertion of the current on a valence quark line, while the ``disconnected" sea quarks (DS) originate in a quark loop from vacuum polarization.  
As shown most recently in Ref.\cite{Alexandrou:2025vto}, one can determine the separate contribution of the CS and DS, to the nucleon form factors. The point that is of interest here and that will be explored in a dedicated paper, is the impact of considering both CS and DS in the evaluation of GPDs. Our analysis will extend the one already carried out in Ref.\cite{Hou:2019efy} in the forward case, for PDFs ($\zeta=t=0$). 


\section{Parametrization}
\label{sec:param}
In this section, we present analytic expressions for GPDs in the variables $X,\zeta,t$ that serve as an input at a given low scale, $Q^2=Q_o^2$. The expressions are obtained building on previous work: in \cite{Ahmad:2006gn,Ahmad:2007vw} a Reggeized spectator model of the nucleon GPDs was first introduced, using scalar and axial-vector diquark configurations that allowed one to distinguish between different $u,d$ flavor structure; in \cite{Goldstein:2010gu,Goldstein:2013gra} this model was developed into a fully fledged parametrization that was quantitatively constrained by elastic form factor data and by PDFs; finally, in \cite{Kriesten:2021sqc}, we started constructing a global parametrization that could be used to examine various quark and antiquark flavors and gluons. The latter was possible since the LQCD on the gluon form factor were made available \cite{Shanahan:2018pib}.

Here we present a readily usable data table generated from our parameterization available at \url{https://github.com/Exclaim-Collaboration/UVA2}. 
We have calculated GPD values covering a range of kinematic values in a grid with ranges: $(a < X < b) \otimes (b < \xi < c) \otimes (e < t < f) $. 
We have multiple versions with different spacing of input kinematics. 
We have produced different grids with linear, logarithmic, and hybrid spacings. 
The grids are stored as ``.csv" files, and can easily be loaded into memory using software.
They are also easily human-readable if loaded using a spreadsheet program such as Excel.
Basic linear interpolation between points in our table, is a reasonable way to use/visualize/test/compare our parameterization in future work.
As a concrete example we provide the following few lines of code which make use of a CSV file using Python, but the same procedure could be used from any other language.


\begin{lstlisting}[]
import numpy as np
import scipy.interpolate
table = np.loadtxt("UVA_grid_linear.csv", delimiter=',')
table_inputs, table_outputs = table[:, :4] , table[:, 4:]
query_input = [.5,.6,.7]  #<-- ask for GPD(X=0.5, xi=0.6, t=.7)
interpolated_GPD_value = scipy.interpolate.griddata(
        table_inputs, table_outputs, query_input, 
        method='linear')
\end{lstlisting}

Comparison with data can be made in a standard way, by evolving the parametric forms to the $Q^2$ values of interest through the evolution equations provided in Section \ref{sec:formalism}. 
The asymmetric set affords us an easier evaluation of the $Q^2$ evolution equations, although this choice of variables is equivalent to, and can be easily transformed into the symmetric variables set $x,\xi$, as shown in Appendix \ref{app:asymmetric}.  
While GPDs are defined in the entire support region in $X$ ($x$), namely, $X \in [-1 +\zeta,1]$ ($x \in [-1,1]$), we distinguish between two parametric forms for the kinematic regions defines as: DGLAP $X \in [-1 +\zeta, 0]$ and $[\zeta,1]$, ($x \in [-1,-\xi]$ and $x \in [\xi,1]$), and ERBL $X \in [0,\zeta]$, ($x \in [-\xi,\xi]$), based on the form perturbative QCD evolution takes in the two regions. 
\subsection{DGLAP region}
\label{sec:dglap}
\begin{figure}[H]
    \centering
\includegraphics[width=0.75\linewidth]{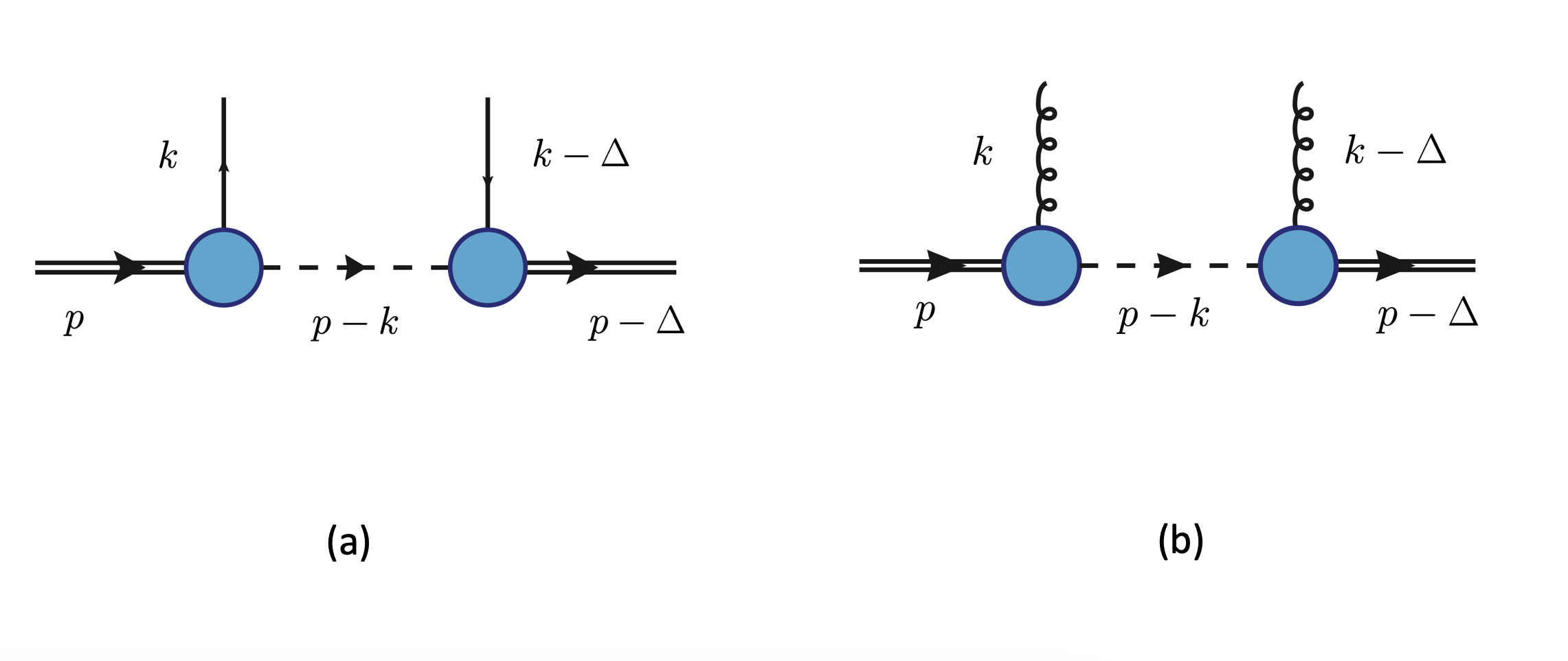}
    \caption{Spectator model configurations for quarks (a) and gluons (b). The intermediate state with four-momentum $p-k$ is a diquark (tetraquark) for the quark (antiquark) component, and an octet baryon for the gluon.}
    \label{fig:diquark_feynman}
\end{figure}

We parametrize the DGLAP region using the Reggeized spectator model for GPDs \cite{Ahmad:2006gn,Ahmad:2007vw,Goldstein:2010gu,GonzalezHernandez:2012jv,Kriesten:2021sqc}. In what follows we highlight the main characteristics and parameter definitions of the model. Starting from the QCD correlation function, we insert a complete set of spectator states, $|X (k_X, \Lambda_X )\rangle$, and we assume that, at tree level, the outgoing system ``$X$" 
 scatters coherently. 
The quark(antiquark)-proton helicity amplitude can then be defined as (Figure \ref{fig:diquark_feynman}a),
\begin{equation} 
\label{eq:As}
A^q_{\Lambda^\prime\lambda^\prime, \Lambda\lambda}
  =  \int \frac{d^2 k_\perp}{1-X} \, \phi^*_{\lambda^\prime \Lambda^\prime}(k^\prime,p^\prime) \phi_{\lambda \Lambda}(k,p),
\end{equation} 
with 
\begin{eqnarray}
\phi_{\Lambda,\lambda}(k,P) =  \Gamma(k) \frac{\bar{u}(k,\lambda) U(p,\Lambda)}{k^2-m^2}
\end{eqnarray}
defining the helicity structures at each vertex. 
The function $\Gamma$, is the spectator form factor, modeled as,
\begin{equation}
\Gamma(k) = g_s\frac{k^2 - m^2}{(k^2 - M_{\Lambda}^2)^2},
\end{equation}
where we introduce an extra mass parameter, $M_\Lambda$, the dipole mass.

For gluons, the spectator is a color-octet baryon, $p \rightarrow g + p_8$, and the proton-gluon-spectator vertex takes the form, 
\[ \Gamma_g(k) u(p-k) \gamma_\mu \varepsilon^\mu(k) U(p). \] 
The gluon-proton helicity amplitude is written as, 
\begin{eqnarray}
A^g_{\Lambda^{\prime}\lambda^{\prime},\Lambda \lambda} = \int \frac{d^2k_{\perp}}{1-X} \:
\sum_{\Lambda_X=\pm 1/2} \, \phi^{\ast \, \Lambda_X}_{ \lambda' \Lambda'}(k', p') \, \phi^{\Lambda_X}_{\lambda \Lambda}(k, p),  
\end{eqnarray}
with vertex functions defined as,
\begin{eqnarray}
\phi^{\Lambda_X}_{\lambda \Lambda}(k,p) &= & 
\Gamma(k) \,  \frac{\bar{U}_{\Lambda_X}(p-k)\, U_\Lambda(p)   }{k^2-m_g^2} \, \slashed{\epsilon}^*_{\lambda_g}(k) 
\end{eqnarray}
where a gluon mass, $m_g$, can also be  present since the struck gluon is off-shell  and,
\begin{eqnarray}
\Gamma(k) &\equiv& g \frac{k^2-m_g^2}{(k^2-M_\Lambda^2)^2},\hspace{10pt} 
\end{eqnarray}
so that $k^2-m_g^2$, cancels out. This coupling is used here as an ultraviolet regulator in the transverse momentum, ${\bf k}_T$ (more details of the model can be found in Appendix B of Ref.\cite{Kriesten:2021sqc}).
  
We assume that the hard-photon parton scattering happens in the frame where the initial proton is longitudinal,
\footnote{We use the notation: $a_\mu \equiv(a^+,a^-,{\bf a}_T)$, where $a^\pm$ are the LC components.}
\begin{eqnarray}
    p \equiv \left(p^+,\frac{M^2}{2p^+}, 0 \right)
\end{eqnarray}
and the scattered proton shifts its momentum to,
\begin{eqnarray}
    p' \equiv \left((1-\zeta) p^+,\frac{M^2+ {\bf \Delta}_T^2 }{2(1-\zeta)p^+}, -{\bf \Delta}_T \right)
\end{eqnarray}
In the DGLAP region ($X \geq \zeta$ and $1-\zeta <X<0$ )the the spectator is on-mass-shell,
\begin{eqnarray}
    k_X \equiv \left((1-X) p^+,\frac{M_X^2+ {\bf k}_T^2 }{2(1-X)p^+}, -{\bf k_T} \right)
    \end{eqnarray}
and the incoming ($k)$, and outgoing $(k')$, quarks four-momenta are defined as follows, 
\begin{eqnarray}
\label{eq:k}
    k &\equiv &\left(Xp^+,\frac{k^2+ {\bf k}_T^2}{2Xp^+}, 0 \right) \\
      k' &\equiv& \left((X-\zeta) p^+,\frac{k'^2+ ({\bf k}_T - {\bf \Delta}_T)^2 }{2(X-\zeta)p^+}, {\bf k}_T -{\bf \Delta}_T \right)
  \end{eqnarray}
with the off-shellness defined as,
the denominators are evaluated with the spectator mass, $M_X$ on shell,
\begin{subequations}
\label{denom}
\begin{eqnarray}
k^2-m^2 & = & X M^2 - \frac{X}{1-X} M_X^2 - m^2  - \frac{k_\perp^2}{1-X}  \\ 
k^{\prime \, 2}-m^2 & = & \frac{X-\zeta}{1-\zeta} M^2 - \frac{X-\zeta}{1-X} M_X^2 - m^2  - \frac{1-\zeta}{1-X} 
\left( {\bf k}_\perp - \frac{1-X}{1-\zeta} {\bf \Delta}_\perp \right)^2.
\end{eqnarray}
\end{subequations}
%
%
The various GPDs are then obtained as distinct, linear combinations of the $A$ amplitudes (see Refs.\cite{Goldstein:2010gu,Kriesten:2021sqc} for details), reflecting their polarization configurations, for both the quark and gluon sector, 
\begin{subequations}
\begin{eqnarray}
H_{q,g}(X,\zeta,t)
&=& \frac{1}{2} \left( A_{++,++} + A_{+-,+-} + A_{--,--} + A_{-+,-+} \right)  \\
-   \frac{\Delta_1 E_{q,g}(X,\zeta,t)}{ M}
&=& A_{++,-+}  + A_{+-,--} - A_{--,+-}  - A_{-+,++}  \\
\end{eqnarray}  
\end{subequations}
Evaluating the amplitudes for the quark GPDs, $H_{q}$ and $E_{q}$, using Eq.\eqref{eq:As}, one obtains,
\begin{eqnarray}
\label{eq:GPDHpar} 
H_q & = &  2\pi  \mathcal{N} \left(1-\displaystyle\frac{\zeta}{2}\right) 
 \int_0^\infty  \frac{d k_\perp k_\perp}{1-X} 
\frac{ 
 a \,  \left[(m+M X)  \left(m + M X' \right) +  k_\perp^2 \right]   - b \,  (1-X') \, k_\perp \Delta_\perp }%
{D^2 \,  \left(a^2 - b^2\right)^{3/2}  }       +   \frac{\zeta^2}{4(1-\zeta)} E_q, \nonumber  \\ \\
\label{eq:GPDEpar}
E_q & = &  2\pi  \mathcal{N}  \left(1-\displaystyle\frac{\zeta}{2}\right) \int_0^\infty    \frac{d k_\perp k_\perp}{1-X}
 \frac{-4M k_\perp^2 [(m + M X) - (m+ M X')] + a \, [2 M(m + M X) ] (1-X') }
{(1-\zeta) \,  D^2 \,  \left(a^2 - b^2\right)^{3/2}  }  
\end{eqnarray}
The set of mass parameters, $m, M_X, M_\Lambda$, can be determined using both LQCD form factors data and PDFs (Section \ref{sec:fitparameters}).
The functions $D$, $a$, and $b$, in Eqs.\ref{eq:GPDHpar},\ref{eq:GPDEpar}, are given by,
\begin{eqnarray}
    \label{eq:param_x_1}
D =  {\cal M}_X - \frac{k_T^2}{1-X}   , \quad\quad  a =  {\cal M}_{X'} -\frac{k_T^2+ \Delta_T^2(1-X')^2}{1-X'} , \quad\quad  b = 2k_T\Delta_T     , 
\end{eqnarray}
where,
\begin{subequations}
\begin{eqnarray}
\quad\quad X' &  = &   \frac{X-\zeta}{1-\zeta} , \qquad
    \mathcal{M}_{X} =  X M^2 -\frac{X}{1-X}M_X^2 -M_{\Lambda}^2, \quad \mathcal{M}_{X'} =  X' M^2 -\frac{X'}{1-X'}M_X^2 -M_{\Lambda}^2
\\
\Delta_T^2 & = & -(t-t_{min})(1-\zeta) , \qquad  t  =  \Delta^2 = -\frac{M^2 \zeta^2}{1-\zeta} - \frac{{\Delta}_T^2}{1-\zeta} = t_{min}- \frac{{\Delta}_T^2}{1-\zeta} \quad 
\end{eqnarray}
\end{subequations}

Similar parametric forms can be derived  in the antiquark and gluon sectors Ref.\cite{Kriesten:2021sqc},
\begin{eqnarray}
\label{eq:Hg} 
H_{g} & = & 2 \pi \mathcal{N} \int d k_{\perp} \frac{k_{\perp}}{1-X} \,  
 \frac{1}{D^2 \,  \left(a^2 - b^2\right)^{3/2}  } 
 \nonumber \\
 & \times & \left\{ a\left[ \frac{X X'}{1-X'} \Big((1-X) M - M_X \Big) \Big((1-X') M - M_X \Big) + \left(\displaystyle\frac{1}{1-X'} + (1-X) \right)k_\perp^2 \right]  - b \, \left(\displaystyle\frac{1}{1-X} + (1-X') \right) k_\perp \Delta_\perp \right\} 
 \nonumber \\
 & + & \frac{\zeta^2}{4(1-\zeta)} \, E_g  
\\
E_{g} & = & 2 \pi \mathcal{N} \int d k_{\perp} \frac{k_\perp}{1-X}  
\frac{1}{D^2 \,  \left(a^2 - b^2\right)^{3/2}  } 
\frac{-2M(1-\zeta)}{1-\frac{\zeta}{2}}
 \nonumber \\ 
 &\times & 
 \left\{ 2 k_T^2 \left[ X((1-X)M-M_X) - X'(1-\zeta)((1-X') M - M_X)  \right] - a (1-X') X\left[(1-X)M - M_X  \right]\right\} ,\nonumber \\ 
\label{eq:Eg}
\end{eqnarray}
\subsubsection{Regge Term}
As illustrated in Ref.\cite{Goldstein:2014aja}, the low $X$ behavior of GPDs is not fully captured by implementing a ``fixed mass" spectator: however, by letting the spectator invariant mass vary following a distribution function that extends to large values of $M_X$, one ensures a realistic  Regge behavior.  
The parametric form for this term is given by,  
\begin{equation}
\label{eq:regge}
    R_p^{\alpha,\alpha'}(X, t) = X^{-(\alpha + \alpha' (1-X)^p t)} .
\end{equation}
This term is given in a multiplicative form, {\it i.e.} it multiplies directly each of the expressions in the DGLAP region, Eqs.\eqref{eq:GPDHpar},\eqref{eq:GPDEpar},\eqref{eq:Hg},\eqref{eq:Eg}.

\subsubsection{Analytic expressions }
\label{subsec:quarks}

The  parametric forms for $H$ and $E$, presented as integrals in ${\bf k}_T$ in Eqs.\eqref{eq:GPDHpar} and \eqref{eq:GPDEpar}, and similar expressions for $\bar{q}$ and $g$, can be cast in the form,
\begin{align}
\label{eq:param_genform}
    H_{q,\bar{q}}(X,\zeta,t) &= \mathcal{A}_H\;R_p^{\alpha,\alpha'}\;(a_0 I_0 + a_{1}^{-}I_1 + I_2) \nonumber\\
    E_{q,\bar{q}}(X,\zeta,t) &= \mathcal{A}_E\;R_p^{\alpha,\alpha'}\;(b_0I_0 + b_{1}^{-}I_1) \nonumber\\
    H_g(X,\zeta,t) &= \mathcal{A}_H^g\;R_p^{\alpha,\alpha'}\;(a_{0}^g I_0 + a_{1}^{g+} I_1 + a_{2}^{g+} I_2) \nonumber\\
    E_g(X,\zeta,t) &= \mathcal{A}_E^g\;R_p^{\alpha,\alpha'}\;(b_{0}^g I_0 + b_{1}^{g} I_1)
\end{align}
where $\mathcal{A}_{H,E}$ and $\mathcal{A}^g_{H,E}$ are normalization factors given by,
\begin{subequations}
    \begin{eqnarray}
    \mathcal{A}_H^q &= & 2 \pi \mathcal{N} \left(1-\frac{\zeta}{2}\right) (1-X) (1-X')^2,  \\
    \mathcal{A}_E^q  & = &2 \pi \mathcal{N}  \left(1-\frac{\zeta}{2}\right) (1-\zeta)^2 (1-X')^4,\\
    \mathcal{A}_{H}^g &= & 2 \pi \mathcal{N}\left(1-\frac{\zeta}{2}\right)(1-X)^2 (1-X')^2, \\
    \mathcal{A}_E^g &  = & -4 \pi M\mathcal{N} \left(\frac{1-\zeta}{1-\zeta/2}\right)(1-X) (1-X')^3  .
\end{eqnarray}
\end{subequations}
${\cal N}$ is in GeV$^4$, and the mass parameters are in GeV. The integrals appearing in Eqs.\eqref{eq:param_genform} are of the form,
\begin{equation}
\label{eq:In}
   {I}_{n}= \int_{0}^{\infty}dk_T \frac{k_T^{2n+1}}{D^2 (a^2-b^2)^{\frac{3}{2}}}, \qquad\qquad n=0,1,2
\end{equation}
In what follows we present analytic solutions of the integral forms that are more readily usable in practical computations. 

\vspace{0.3cm}
\noindent 
\underline{{\it Forward limit: $\zeta=0, t=0$}}

\vspace{0.3cm}
\noindent 
In the forward limit, for $\zeta=0, t=0$, we obtain the following expressions consistent with \cite{Goldstein:2010gu,Kriesten:2021sqc} 

\begin{equation}
    \mathcal{A}_{H_q} =   2\pi  \mathcal{N}  \,  (1-X)^3 ,
    \quad \mathcal{A}_{E_q} = 2\pi \, \mathcal{N} (1-X)^4, 
    \quad  \mathcal{A}_{H_g} = 2\pi{\cal N} (1-X)^4 ,
    \quad \mathcal{A}_{E_g} = -4\pi\;M\mathcal{N} (1-X)^4 ,
    \quad R^{\alpha,\alpha'}_p \rightarrow X^{-\alpha} 
\end{equation}
\begin{eqnarray}
(a_0I_0 + a_{1}^{-}I_1 + I_2) &\rightarrow &  \frac{1}{12} \, \frac{(1-X)\mathcal{M}_X-2(m+MX)^2}{\mathcal{M}_{X}^3} \\
(b_0I_0 + b_{1}^{-}I_1) &\rightarrow & \frac{1}{3} \bigg(\frac{(1-X) M (m+MX)}{\mathcal{M}_{X}^3}\bigg) \\
(a_{0}^g I_0 + a_{1}^{g+} I_1 + a_{2}^{g+} I_2) &\rightarrow& \frac{1}{12\mathcal{M}_X^3} \, \Bigg[ \mathcal{M}_X\bigg(1 + \frac{1}{(1-X)^2}\bigg) - 2\bigg(\frac{X}{(1-X)}\bigg)^2 \bigg( M - \frac{M_X}{(1-X)} \bigg)^2 \Bigg] \\
(b_{0}^g I_0 + b_{1}^{g} I_1) &\rightarrow &  \frac{1}{6} \, \frac{X M^2 (1-X)^2}{\mathcal{M}_{X}^3} \; \bigg( 1 - \frac{M_X}{M(1-X)} \bigg)
\end{eqnarray}
whereas the Regge term is given by $R_\alpha = X^{-\alpha}$ for all parton flavors, with the exception of the ${\bar{u}}$ component where we introduced an extra term in Eq.\eqref{eq:GPDHpar}, leading to $ R_\alpha  \rightarrow  X^{-\alpha+ \alpha_2 \sqrt{c}} $.
Therefore, one has, 
\footnote{Notice that the subscripts $q,g$ on the mass parameters have been omitted in these expressions for simplicity.}
\begin{subequations}
\label{eq:forward}
\begin{eqnarray}
    H_{q,\bar{q}}(X,0,0) &=& \frac{\pi}{6} \mathcal{N}  \, X^{-\alpha}  \,\, \frac{(1-X)\mathcal{M}_X-2(m+MX)^2}{\mathcal{M}_{X}^3} \\
    E_{q,\bar{q}}(X,0,0) &=& -\frac{2\pi}{3} \, \mathcal{N}  \,\, X^{-\alpha}\bigg(\frac{(1-X) M (m+MX)}{\mathcal{M}_{X}^3}\bigg)  \\
    H_g(X,0,0) &=&  \frac{\pi}{6} \, \mathcal{N} \,  X^{-\alpha}\,\, \frac{(1-X)^2}{\mathcal{M}_X^3} \Bigg[ \mathcal{M}_X\bigg(1 + (1-X)^2 \bigg) -\; 2X^2 \bigg( M - \frac{M_X}{(1-X)} \bigg)^2 \Bigg]  \\
    E_g(X,0,0) &=& \frac{2\pi}{3} \mathcal{N} \, X^{-\alpha} \,\,   \frac{X M^2 (1-X)^2}{\mathcal{M}_{X}^3} \; \bigg( 1 - \frac{M_X}{M(1-X)} \bigg)
\end{eqnarray}
    \end{subequations}

\vspace{0.3cm}
\noindent \underline{\it  In the Limit, $\zeta=0$, t$\neq0$}:
\vspace{0.3cm}

The GPD $H_q , E_q , H_g , E_g$ in the zero-skewness limit and expanded in $\omega= (1-X) t/{\cal M}_X$ up to second order are expressed as:


\begin{align}\label{Hq_analytic}
    H_{q,\bar{q}}(X,0,t) &= \frac{\pi\mathcal{N}R_p^{\alpha,\alpha'}}{\mathcal{M}_{X}^3}\Bigg[ \bigg( \frac{\gamma-2\alpha^2}{6} \bigg) + \bigg( \frac{3\alpha^2 - 2\gamma}{15} \bigg) \frac{(1-X)t}{\mathcal{M}_{X}}   \nonumber \\
    &\quad
    +\; \bigg( \frac{9\gamma - 12\alpha^2}{140}\bigg) \frac{(1-X)^2t^2}{\mathcal{M}_{X}^2} - \bigg( \frac{65\alpha^2 + 3\gamma}{140}\bigg)\frac{(1-X)^3t^3}{\mathcal{M}_{X}^3} \Bigg] 
\end{align}
\\
\begin{align}\label{Eq_analytic}
    E_{q,\bar{q}}(X,0,t) &= \bigg( \frac{4\pi\;M\;(1-X)(m+MX)\;\mathcal{N}R_p^{\alpha,\alpha'}}{\mathcal{M}_{X}^3} \bigg) \Bigg[ \bigg( \frac{-1}{6} \bigg) + \bigg( \frac{1}{10} \bigg)\frac{(1-X)t }{\mathcal{M}_{X}}  \nonumber \\
    &\quad
    \;-\; \bigg( \frac{3}{70} \bigg)\frac{(1-X)^2t^2}{\mathcal{M}_{X}^2} \;-\; \bigg( \frac{13}{56} \bigg)\frac{(1-X)^3t^3}{\mathcal{M}_{X}^3} \Bigg] 
\end{align}
\\
\begin{align}\label{Hg_analytic}
    H_{g}(X,0,t) &= \frac{\pi\mathcal{N}R_p^{\alpha,\alpha'}}{\mathcal{M}_{X}^4}\Bigg[ \bigg( \frac{\gamma^2 (\eta+1) - 2 \beta^2 X^2 \mathcal{M}_{X}}{6} \bigg) - \bigg( \frac{2\gamma^2 (\eta+1) - 3\beta^2 X^2 \mathcal{M}_{X}}{15} \bigg) \frac{(1-X)t}{\mathcal{M}_{X}}   \nonumber \\
    &\quad
    +\; \bigg( \frac{9\gamma^2 (\eta+1) - 12\beta^2 X^2 \mathcal{M}_{X}}{140}\bigg) \frac{(1-X)^2t^2}{\mathcal{M}_{X}^2} - \bigg( \frac{3 \gamma^2 (\eta+1) + 65\beta^2 X^2 \mathcal{M}_{X}}{140}\bigg)\frac{(1-X)^3t^3}{\mathcal{M}_{X}^3} \Bigg] 
\end{align}
\\
\\
\begin{align}\label{Eg_analytic}
    E_{g}(X,0,t) &= \bigg(\frac{4\pi M^2 \mathcal{N}\;X\;(1-X)^2\;R_p^{\alpha,\alpha'}}{\mathcal{M}_{X}^3} \bigg)\; \bigg( 1 - \frac{M_X}{M(1-X)} \bigg) \Bigg[ \frac{1}{6} \nonumber \\
    &\quad
    \;-\; \frac{(1-X)t}{10\mathcal{M}_{X}} \;+\; \frac{3(1-X)^2t^2}{70\mathcal{M}_{X}^2} \;+\; \frac{13 (1-X)^3t^3}{56\mathcal{M}_{X}^3} \Bigg]
\end{align}
where, $\gamma = (1-X)\mathcal{M}_X$ , $\beta = [(1-X)M - M_X]$ and $\eta = (1-X)^2$ \\



\vspace{0.3cm}
\noindent 
\underline{{\it Full parametrization $\zeta \neq 0, \, t < 0$}}:
\vspace{0.3cm}

\noindent 
For non zero $(\zeta,t)$, the integrals in Eqs.\eqref{eq:param_genform} result in more complex expressions that are no longer easy to interpret or intuitive.
We, nevertheless, provide parametric forms which are relatively straightforward to evaluate numerically. Calculating these expressions directly allows us to streamline the parametrization usage in various regression applications since we avoid the numerical integrations in $k_T$.
The analytic solutions for  $I_n$, $n=0,1,2$, Eq.\eqref{eq:In}, are written in terms of recurrent expressions, $\chi$, $\omega$, $\gamma$, $\delta$, 
 \begin{eqnarray} 
    \chi = \frac{(1-\zeta)\mathcal{M}_{X} - (1-X^\prime)\Delta_\perp ^2}{\mathcal{M}_{X^\prime}}  , \quad
    \omega = -\frac{(1-X^\prime)\Delta_\perp ^2}{\mathcal{M}_{X^\prime}} , \quad
    \gamma = (1-X^\prime)\mathcal{M}_{X^\prime} , \quad 
    \delta = \sinh^{-1}\Bigg[\frac{1}{2}\frac{(\chi + 1)}{(\chi-\omega)}\sqrt{(\chi-1)^2+4\omega} \Bigg]   , \nonumber \\
\end{eqnarray}
that allow us to express the integrals in a compact form,
\begin{eqnarray}\label{Exact_integrals}
    I_0 &=& \frac{1}{\gamma^4[(\chi-1)^2+4\omega]^2} \Bigg[ \bigg(\frac{(\chi-3)^2}{4(1+\omega)} +\frac{(\chi+1)}{2(\chi-\omega)} - 2\bigg) + \bigg(\frac{3}{2}\frac{(\chi-1)}{\sqrt{(\chi-1)^2 + 4\omega}}\bigg)\delta \Bigg]
\\
    I_1 &=& \frac{-1}{\gamma^3[(\chi-1)^2+4\omega]^2}\Bigg[ \bigg(\frac{(\chi-1)(\chi+5)-8\omega}{4}\bigg) + \bigg( \frac{(\chi+1)^2 -(\chi-\omega)(3\chi+1)}{2\sqrt{(\chi-1)^2 +4\omega}}\bigg)\delta \Bigg]
\\
    I_2 &=& \frac{1}{\gamma^2\;[(\chi-1)^2+4\omega]^2}\Bigg[ \bigg ( \frac{(\chi + 1)^3 - (\chi^2 -1)(\chi-\omega) - 8(\chi-\omega)^2}{4} \bigg) \nonumber\\
    & \qquad & \qquad\qquad\qquad\qquad\quad + \bigg(  \frac{(\chi-\omega)}{2\sqrt{(\chi-1)^2 +4\omega}}[3(\chi-1)(\chi-\omega) -2(\chi-1)^2 - 8\omega] \bigg) \delta \Bigg]
\end{eqnarray}
We finally list the  kinematic coefficients for the GPD $H_{q,g}$, $a_j^{q,g}$,
\begin{subequations}
\begin{eqnarray}
    a_0^q &= & -(m+MX)(m+MX')\big[(\omega + 1) (1-X')\m_{X'}\big]\\
    a_1^{q\pm} & = &  (m+MX)(m+MX') - (\pm\; \omega \mp 1) (1-X')\m_{X'}\\
    a_2^q & = & 1 \\
    a_{01}^g & = & \frac{X X'}{(1-X')}\; \big[ \big((1-X)M-M_X\big) \big((1-X')M-M_X\big) \big] \\
    a_0^g &= &X X'\big[(1-X)M-M_X\big] \big[(1-X')M-M_X\big] [-(\omega + 1) \m_{X'}]\\ 
    a_{1}^{g+} &=& a_{01}^g + (\omega - 1) (1-X'){\cal M}_{X'}  [1 \pm (1-X)(1-X')]  \\
    a_{2}^{g+} &= & 1 + (1-X)(1-X') ,
\end{eqnarray}
\end{subequations}

and for the GPD $E_{q,g}$, $b_j^{q,g}$,        
\begin{subequations}
\begin{eqnarray}
    b_0^q &  = & -2M (\omega + 1)\; (m+MX)(1-X')\m_{X'}  \\
    b_1^{q\pm} & = &  2M \big[2(m+MX')  \pm (m+MX)\big]  \\
    b_0^g & = &  -(\omega+1)X\;(1-X')\m_{X'}\;\big((1-X)M-M_X\big)\\
    b_1^g & = &  2X((1-X)M-M_X)-X'(1-\zeta)((1-X')M-M_X)  .
\end{eqnarray}
\end{subequations}


\begin{figure}[H]
\centering
\includegraphics[width=0.45\linewidth]{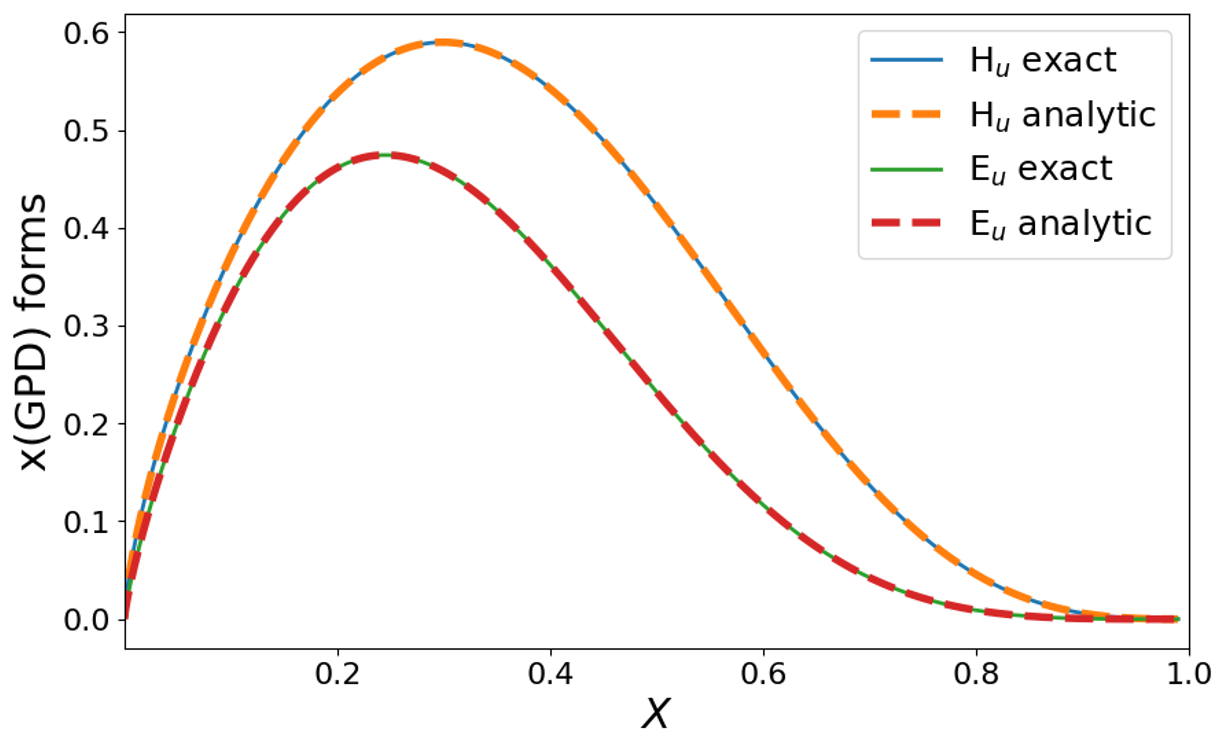}
\hspace{0.2cm}
\includegraphics[width=0.45\linewidth]{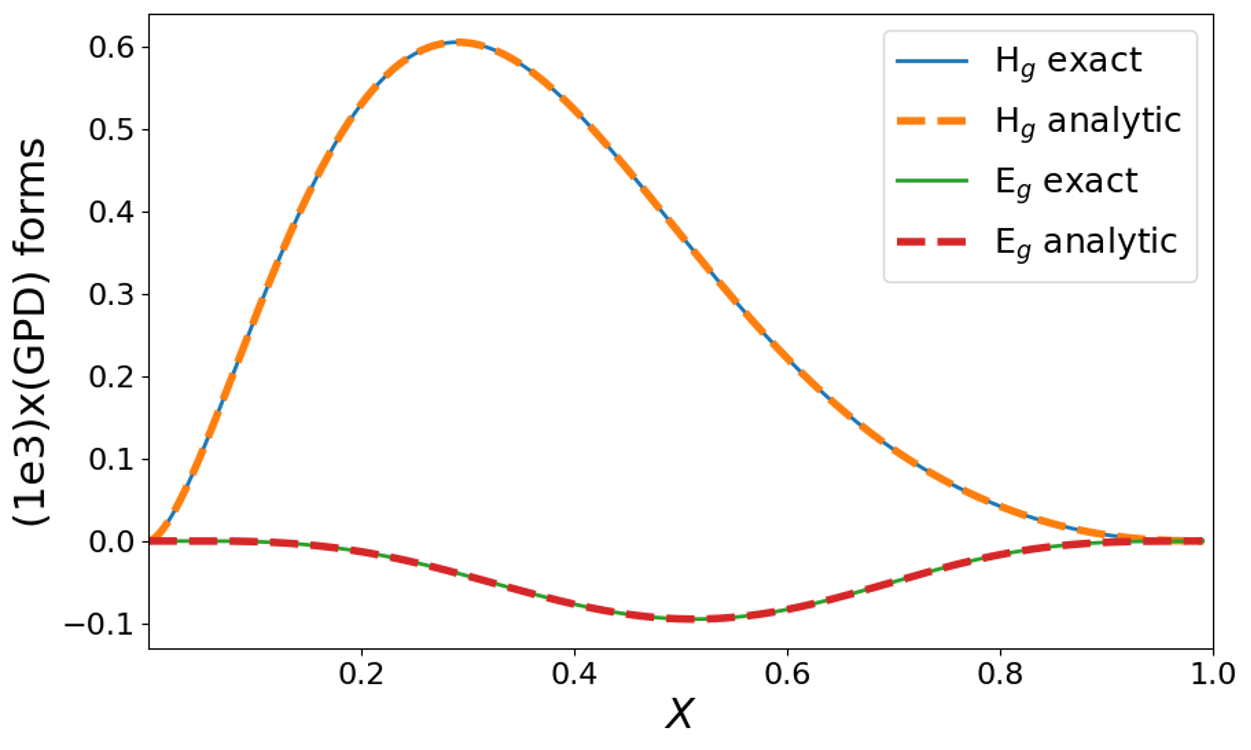}
\caption{The figure presents the comparison of exact (\ref{Exact_integrals}) and analytic expressions (\ref{Hq_analytic}, \ref{Eq_analytic}, \ref{Hg_analytic}, \ref{Eg_analytic}) for GPDs with parameter values from Eqs.\eqref{eq:paramERBL} at the initial scale of $Q^2 =0.58 \mathrm{GeV^2}$.}
\label{exact_analytic_GPDHu}
\end{figure}

More information about the solution of the integrals and associated parameters of the GPDs are provided in Appendix \ref{app:forward}. 
Insight details of the solutions for the integrals are elaborated in Appendix \ref{integral_solns}. Having both the exact (\ref{Exact_integrals}) and analytic expressions (\ref{Hq_analytic}, \ref{Eq_analytic}, \ref{Hg_analytic}, \ref{Eg_analytic}), we can see that the GPDs agree with each other. For example, in figure \ref{exact_analytic_GPDHu}, the exact and analytic expressions for GPD $\mathrm{H_{u_v}, E_{u_v}, H_g}$ and $\mathrm{E_g}$ show a good match with each other at the initial scale of $Q^2 =0.58 \mathrm{GeV^2}$.

\subsection{ERBL region}
\label{sec:ERBL}
\begin{figure}[H]
    \centering
    \includegraphics[width=0.5\linewidth]{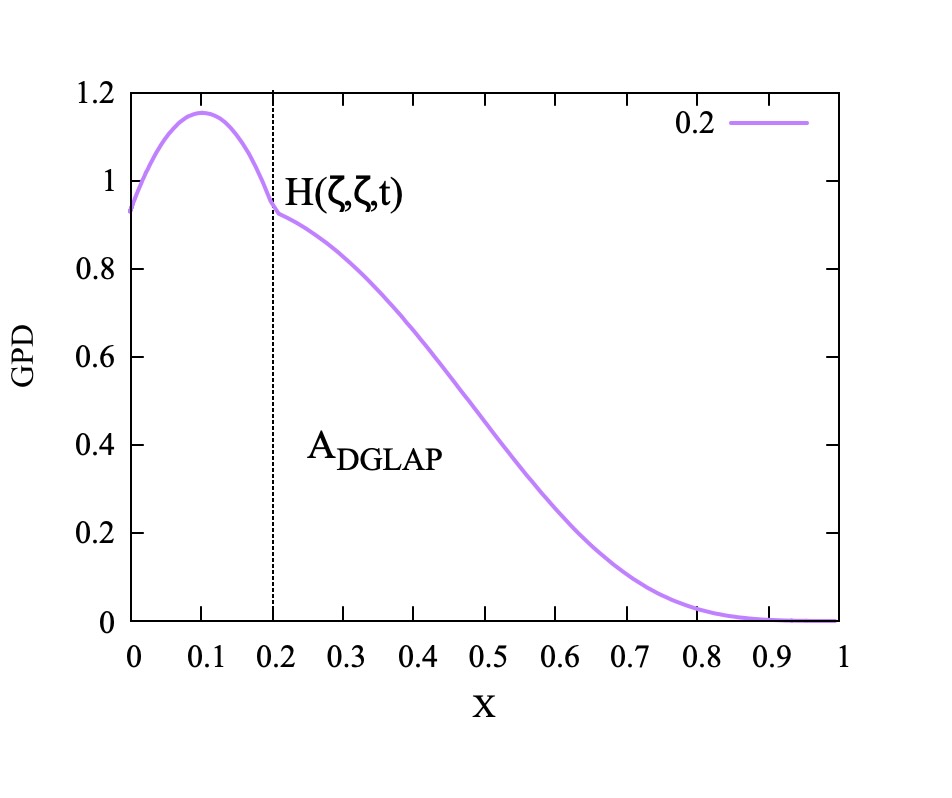}
    \caption{Illustration of the various quantities constraining the parameters of the valence and gluon distributions in the ERBL region: $A_\text{DGLAP}$, $\zeta$, and $H(\zeta,\zeta,t)$ (see discussion in text).}
    \label{fig:ERBLsimple}
\end{figure}

For $ 0< X  < \zeta$ (ERBL region) \cite{Belitsky:2005qn}, in Refs.\cite{Goldstein:2010gu,Kriesten:2021sqc} we introduced the following expressions obeying the quark parton symmetries detailed in Section \ref{sec:formalism}, where we chose the simplest polynomial forms in the variable $X$, that allow us to best satisfy the quark-antiquark symmetry relations (Section \ref{sec:sym}): a quadratic form 
for $F^-$, and a cubic form for $F^+$, respectively,   
\begin{align}
    H^-_{q}(X,\zeta,t) &= a^-(\zeta,t) X^2+b^-(\zeta,t) X + c^-(\zeta,t) \\
    H^+_q(X,\zeta,t) &= a^+(\zeta,t) X^3 + b^+(\zeta,t) X^2+c^+(\zeta,t) X + d^+(\zeta,t)   \\
    H_g(X,\zeta,t) &= a^-_g(\zeta,t) X^2+b^-_g(\zeta,t) X + c^-_g(\zeta,t) \\
    \end{align}
where the parameters $a^\pm, b^\pm, c^\pm, d^+$ are fixed from the area/form factors and symmetries constraints, and they are $\zeta$, $t$ and $Q^2$ dependent. At any given $Q^2$ scale the $\zeta$ and $t$ dependence of the $``-"$  and  gluon distributions is defined as,
\begin{subequations}
\label{eq:paramERBL}
\begin{eqnarray}
\underline{{H_q^-}, H_g}  \quad \quad\qquad &&  \nonumber \\
a^-_{(g)} &= & \frac{6}{\zeta^3}\{\zeta H^-_{q,g}(\zeta,\zeta,t) - 2[(1-\zeta/2)F_{q,g}(t)-A_{DGLAP}(\zeta)]\}\\
    b^-_{(g)} &= & -a^-\zeta\\
    c^-_{(g)} &= & H^-_{q,g}(\zeta, \zeta, t)
\end{eqnarray}
\end{subequations}
For the``+" distributions, we have,
\begin{subequations}
\label{eq:param_ERBL2}
\begin{eqnarray}
\underline{H_q^+}  && \nonumber \\
    b^+ &= & -\frac{3}{2} \, \zeta a^+\\
    c^+ &= & \left(2H^+_q(\zeta, \zeta, t) + \frac{1}{2}a^+ \zeta^3\right)\frac{1}{\zeta}\\
    d^+ &=&  -H^+_q(\zeta, \zeta, t)
\end{eqnarray}
\end{subequations}
where $a^+$ is let vary in our model as it can only be constrained upon the availability of  $\zeta$-dependent observables from experiment or from LQCD evaluations; $b^+, c^+, d^+$ are also $\zeta$ and $t$ dependent. 
The meaning of the various terms is illustrated in Figure \ref{fig:ERBLsimple} for the $``-"$ distribution: $A_{DGLAP}$  in Eqs.\eqref{eq:paramERBL}, is the area under the DGLAP region,
\[  A_{DGLAP} = \int_\zeta^1 dX H^-(X,\zeta,t) . \]
This area is calculated from the parametrization in Section \ref{sec:dglap};  
$(1-\zeta/2) F_{q}(t)$ is the total area subtended by the GPD curve, where $F_{q}$ is quark $q$ contribution to the proton form factors shown in Eqs.\eqref{eq:ud_formf}, and $1-\zeta/2$ is the Mellin moment jacobian (Section \ref{sec:formalism}); $H(\zeta,\zeta,t)$, is the value of the GPD at the ERBL-DGLAP crossover point. Therefore the difference $(1-\zeta/2) F_{q}(t) - A_{DGLAP}$ in Eq.\eqref{eq:paramERBL} is the area subtended by the GPD in the ERBL region. The gluon distribution, $H_g$, is described by a similar curve, having  the same symmetry.

In summary, the ERBL region is described by a parabolic parametric form for the valence and gluon distributions, centered in $X=\zeta/2$ (or $x=0$ in the symmetric frame), while the antisymmetric singlet term is described by a cubic form. These forms are built in the parametrization in a way that guarantees baryon number and momentum conservation. The user can easily code in these terms, which are constrained by our parametric form in the DGLAP region, using Eqs.(\ref{eq:paramERBL},\ref{eq:param_ERBL2}), or directly access our code, as detailed below.    

\vspace{0.3cm} 
\subsection{ERBL and DGLAP region}
\label{sec:ERBL_DGLAP}
The complete form of our parametrization is given below, 
\begin{eqnarray}
\label{eq:F_param_form}
\nonumber  \\ 
F^{q,g}(X,\zeta,t; Q^2)  =  F^{DGLAP}_{q,g}  \left[ \theta(X-\zeta) + \theta(-X) \right] 
+  F^{ERBL}_{q,g}  \left[ \theta(-X+\zeta) + \theta(X) \right] 
\\ \nonumber 
\end{eqnarray}
where we reiterate that although it is given for the ($X,\zeta$) choice of variables, this can be straightforwardly changed with the $(x,\xi)$ set. 
The parametric forms for the GPD $H$, are listed below. 

\vspace{0.2cm}
\noindent For $q=u_v, d_v$, the GPDs are defined as,
\begin{subequations}
\begin{eqnarray}
\label{eq:qvparam_D}
H^{DGLAP}_{q}  & = &  \mathcal{A}_H^q \; X^{-(\alpha_q+\alpha_q'(1-X)^{p_q}t)}\; \left[ a_0 I_0 + a_{1}^{-}I_1 + I_2\right]_{m_q,M_X^q,M_\Lambda^q}  \\
\label{eq:qvparam_E}
H^{ERBL\;\;}_{q} & = &  a^-X^2+b^-X+c^- 
\end{eqnarray}
\end{subequations}
for $q=\bar{u},\bar{d}$,  
\begin{subequations}
\begin{eqnarray}
\label{eq:qbparam_D}
H^{DGLAP}_{\bar{u}}  & = &   \mathcal{A}_H^{\bar{u}} \; X^{-\left[\alpha_{\bar{u}}'(1-X)^{[p_{\bar{u}}]}t  + \alpha_{\bar{u}} \sqrt{1 - \frac{t}{c}}+ \alpha_2 (1-\sqrt{1 - \frac{t}{c}} ) \right]}\; \left[ a_0 I_0 + a_{1}^{-}I_1 + I_2\right]_{m_{\bar{u}},M_X^{\bar{u}},M_\Lambda^{\bar{u}}} 
\\
\label{eq:dbparam_D}
H^{DGLAP}_{\bar{d}}  & = &   \mathcal{A}_H^{\bar{d}} \; X^{-(\alpha_{\bar{d}} + \alpha_{\bar{d}}'(1-X)^{[p_{\bar{d}}]}t) }\; \left[ a_0 I_0 + a_{1}^{-}I_1 + I_2\right]_{m_{\bar{d}},M_X^{\bar{d}},M_\Lambda^{\bar{d}}} 
\\
\label{eq:qbparam_E}
H^{ERBL\;\;}_{\bar{q}=\bar{u},\bar{d}} & = &  \frac{1}{2}\left[ (a^+X^3+b^+X^2+c^+X+d) - (a^-X^2+b^-X+c^-) \right]
\end{eqnarray}
\end{subequations}
and for gluons as,
\begin{subequations}
\begin{eqnarray}
\label{eq:gparam_D}
H^{DGLAP}_{g}  & = &  \mathcal{A}_H^g \;X^{-(\alpha_g + \alpha_{g}'(1-X)^{p_g}t)}\;(a_{0}^g I_0 + a_{1}^{g+} I_1 + a_{2}^{g+} I_2) 
\\
\label{eq:gparam_E}
H^{ERBL\;\;}_{g} & = &  a^-_g X^2+b^-_g X+c^-_g  
\end{eqnarray}
\end{subequations}
For the GPD $E$ one has for $q=u_v, d_v$ ,
\begin{subequations}
\begin{eqnarray}
E^{DGLAP}_{q}  & = &  \mathcal{A}_E^q \; X^{-(\alpha_q+\alpha_q'(1-X)^{p_q}t)}\; \left[ b_0 I_0 + b_{1}^{-}I_1 \right]_{m_q,M_X^q,M_\Lambda^q}  \\
E^{ERBL\;\;}_{q} & = &  a^-X^2+b^-X+c^- 
\end{eqnarray}
\end{subequations}
for $q=\bar{u},\bar{d}$,  
\begin{subequations}
\begin{eqnarray}
E^{DGLAP}_{\bar{q}}  & = &   \mathcal{A}_E^{\bar{q}} \; X^{-(\alpha_{\bar{q}} + \alpha_{\bar{q}}'(1-X)^{p_{\bar{q}}}t)}\; \left[ b_0 I_0 + b_{1}^{-}I_1 \right]_{m_{\bar{q}},M_X^{\bar{q}},M_\Lambda^{\bar{q}}} 
\\
E^{ERBL\;\;}_{\bar{q}} & = &  \frac{1}{2}\left[ (a^+X^3+b^+X^2+c^+X+d) - (a^-X^2+b^-X+c^-) \right]
\end{eqnarray}
\end{subequations}
and for gluons as,
\begin{subequations}
\begin{eqnarray}
E^{DGLAP}_{g}  & = &  \mathcal{A}_E^g \;R_p^{\alpha,\alpha'}\;(b_{0}^g I_0 + b_{1}^{g+} I_1 ) 
\\
E^{ERBL\;\;}_{g} & = &  a^-_g X^2+b^-_g X+c^-_g  ,
\end{eqnarray}
\end{subequations}
where the parametric expressions for all the coefficients are listed and explained above; the numerical values of the parameters from all the constraints satisfied by GPDs are provided for the initial scale, $Q_0^2 = 0.58 $ GeV$^2$, in the Tables listed in the following Section.    

\section{Results}
\label{sec:results}
\begin{figure}[H]
    \centering
    \includegraphics[width=0.75\linewidth]{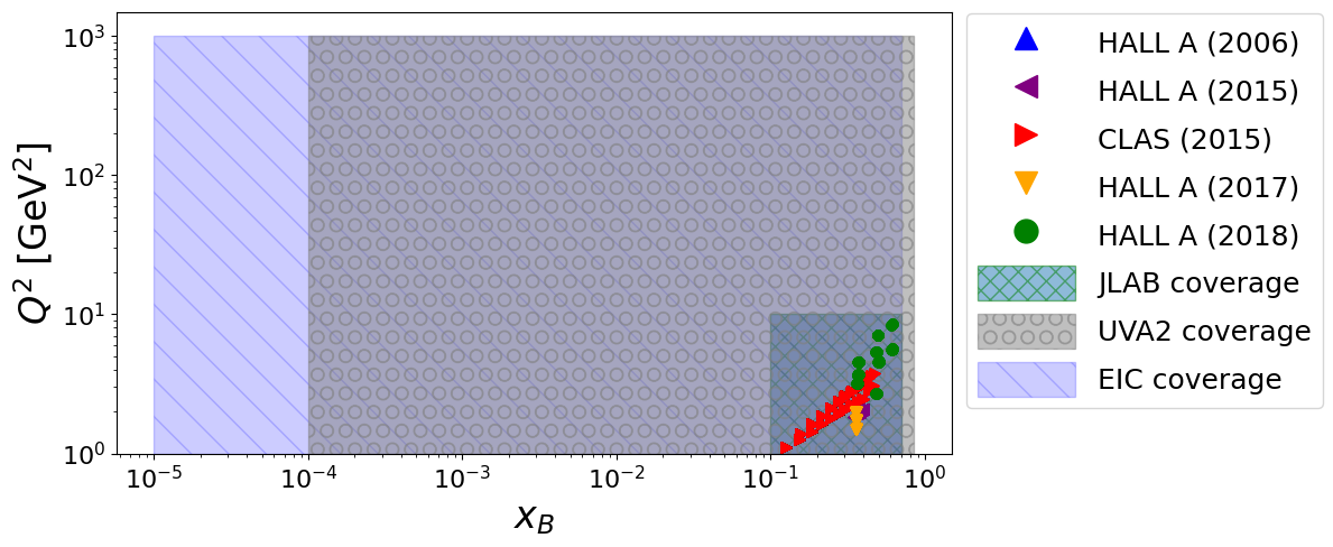}
    \includegraphics[width=0.75\linewidth]{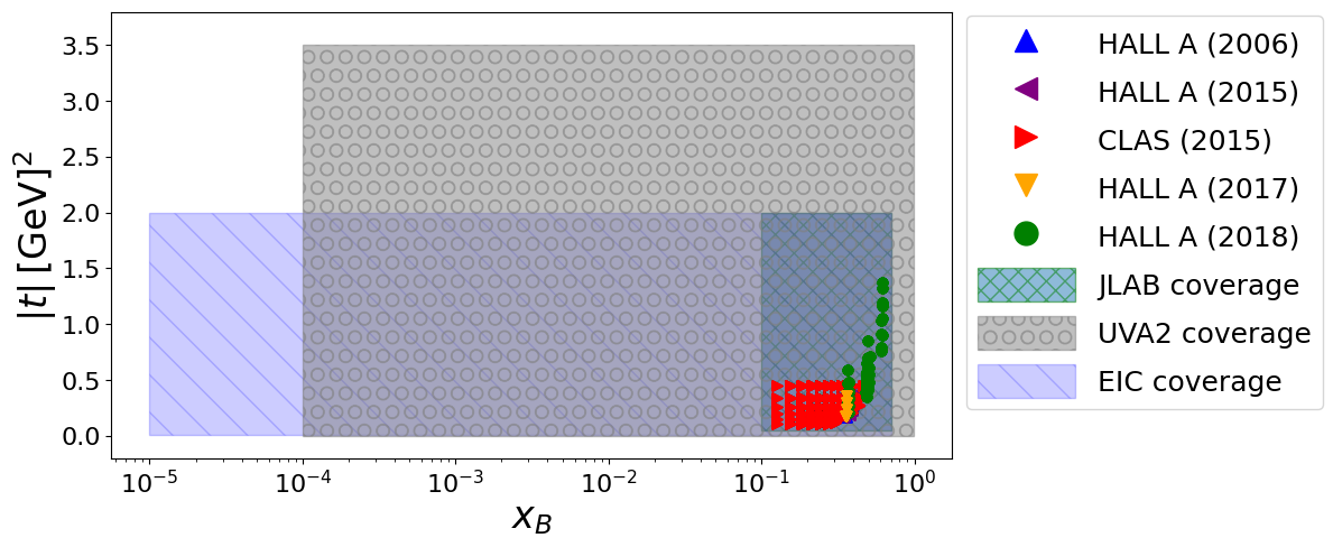}
    \caption{The above figure presents kinematic domains of various experimental facilities which provide experimental data and aim to cover kinematics in the future. We compare our model's kinematic coverage to the kinematic coverage of existing experimental data (JLAB) and data to be provided in the future (EIC).}
    \label{fig:kine}
\end{figure}

We present a global analysis of the GPDs $H_{q,g}$, $E_{q,g}$, and their corresponding CFFs, including the partons, $u_v, d_v, \bar{u}, \bar{d}, s$, and $g$. We first focus on the DGLAP region, where the values of our parameters at the initial QCD evolution scale, $Q_o^2=0.58 $ GeV$^2$, were found, as in the previous versions of our parametrization \cite{Goldstein:2013gra,GonzalezHernandez:2012jv,Kriesten:2021sqc}, using a standard curve fitting procedure with a Hessian-based method as our regression framework. 
The main features of the parametrization, as well as the differences and improvements with respect to the previous versions \cite{Ahmad:2007vw,Goldstein:2010gu,Goldstein:2012az,Kriesten:2021sqc} are described below.
The parametrizations estimated kinematic range of validity is illustrated in Figure \ref{fig:kine}, and it is determined by the current experimental ranges on nucleon form factors from Jefferson Lab experiments, as well as from the kinematic range in which the PDFs are known: we define as a conservative range of validity: $10^{-4} \leq X,\zeta < 0.90$, $| t | < 4$ GeV$^2$, $ 0.58 < Q^2 <10^3$ GeV$^2$.  In the figure we also show the kinematic coverage of present data at Jefferson Lab, and at the future EIC.

\begin{figure}[H]
    \centering
    \includegraphics[width=0.7\linewidth]{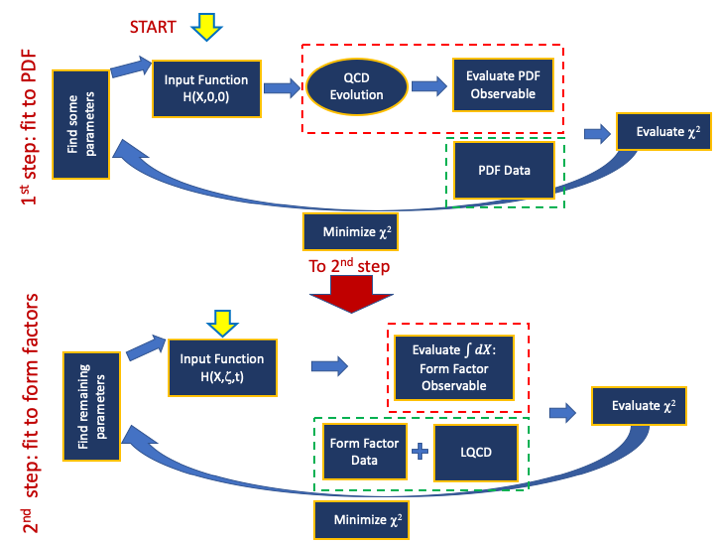}
    \caption{Flowchart describing our parametrization procedure: in the first step (upper panel) we obtain the parameters in the forward limit, where the GPD $H$ equals the PDFs for the corresponding flavors. In the second step (lower panel) we impose the constraint from the nucleon elastic form factors. PQCD evolution is applied throughout.}
    \label{fig:flowchart}
\end{figure}

Our recursive fitting procedure is illustrated in Figure \ref{fig:flowchart}: in a first step (upper panel) we obtain the parameters in the forward limit, for $\zeta=t=0$), by constraining the GPD $H$ with the PDF, for each flavor. 
\subsection{Fit parameters}
\label{sec:fitparameters}
Table \ref{tab:parameters} shows the parameters defining the DGLAP region, Eqs.\eqref{eq:qvparam_D},\eqref{eq:qbparam_D}, \eqref{eq:dbparam_D}, \eqref{eq:gparam_D}. The parameters in the ERBL region are constrained by the value of $H_{q,\bar{q},g}(\zeta,\zeta,t)$ and $E_{q,\bar{q},g}(\zeta,\zeta,t)$, at the crossover point and by the symmetries described in Section \ref{sec:formalism}: therefore, they are introduced in a variable form, as a function of $(\zeta,t)$.   

This choice of parameters guarantees that the GPDs observe the physical constraints from experimental data, as well as from the symmetries and conservation laws of the theory while adhering to a physical interpretation based on the reggeized spectator model.The expressions are valid in a wide kinematic range, including the low $x_{Bj}$ regime explored at the EIC. 
Compared to previous versions \cite{Kriesten:2021sqc}, the present formulation is more readily applicable to evaluate various observables, particularly for the fast implementations required to solve regression problems, a main improvement being that analytic forms are provided for all 
flavor components, $u_v$, $d_v$, $\bar{u}$, $\bar{d}$, $s$ and $g$, at the initial scale, $Q_o^2$, and evolved in PQCD.

The parameters defining GPDs in the forward limit, {\it i.e.} the $X$ dependence, are (${\cal N}, m,M_X,M_\Lambda,\alpha$), and they are fitted first to currently available PDF sets using the form in Eqs.\eqref{eq:forward}; the remaining $t$-dependent parameters  ($\alpha',p$) are fitted subsequently by integrating the GPDs to the flavor separated electromagnetic  form factors, $F_1^q$, $F_2^q$ \cite{Cates:2011pz,Arrington:2011kb,Qattan:2004ht}, in the valence quark sector, and to the nucleon gluon form factors, $A_{g}, B_{g}$, recently evaluated for vector operators in lattice QCD (LQCD) \cite{Hackett:2023rif,Pefkou:2021fni}. To evaluate the contribution of the sea quark GPDs, $H_{\bar{u}}$ and $H_{\bar{d}}$, we use LQCD results from Ref.\cite{Bhattacharya:2023ays}.

\begin{table}[H]
    \centering
    \begin{tabular}{lccccccccc}
\hline
\hline
    {\bf GPD} \quad \quad & $m \, ^\star$ \quad & $M_{X} \, ^\star $ \quad & $M_{\Lambda} \, ^\star $ \quad & $\alpha \, ^\star$ \quad & $\alpha^{\prime}$ \quad & $p$ & $\mathcal{N}$ \quad  & $\chi^2_{dof}$ \\
    \hline 
    \hline
      $H_{u_v}$ & 0.998 & 1.283 & 1.339 & 0.427 & 0.890(8) & 0.950(27) & $5.93^{\star}$ & 6.7\\
      $H_{d_v}$ & 0.8 & 1.45 & 1.08 & 0.33 & 0.869(33) & 0.15(14) & $2.2^{\star}$ & 0.7 \\
      $H_{\bar{u}}$ & 0.1 & 2.5 & 0.93 & 1.153 & 0.35(9) &  $0.25(18) \times 10$ & $0.19^{\star}$ & 2.8\\
      $H_{\bar{d}}$ & 0.9 & 2.94 & 1.6 & 1.150 & 0.14(11) & 2(1) & $1.1^{\star}$ & 0.5 \\
      $H_{g}$ & -- & 0.875 & 0.396 & $-0.54$ & 0.571(26) & 2.0(4) & $0.1^{\star}$ & 0.5 \\
      $E_{u_v}$ & 0.998 & 1.283 & 1.339 & 0.427 & 1.62(11) & 1.76(9) & 6.87(32) & 1.5\\
      $E_{d_v}$  & 0.8 & 1.45 & 1.08 & 0.33& 0.540(9) & 0.10(23) & $-3.55(5)$ & 0.8 \\
      $E_{\bar{u}}$  & 0.1 & 2.5 & 0.93 & 1.153 & 0.00(19) & 4(4) &  $-0.314(6)$ & 0.8\\
      $E_{\bar{d}}$ & 0.9 & 2.94 & 1.6 & 1.150 & 0.00(6) & 5(5) & $-0.46(20)$ & 0.9\\
      $E_{g}$  & -- & 0.875 & 0.396 & $-0.54$ & 0.00(18) & 5(5) & $-0.098(9) $ & 1.3\\
      \hline
      \hline
          \end{tabular}
    \caption{Fitted values of the UVA2 parameters in the {\bf DGLAP } region at $Q_{0}^{2} = 0.58 \: \mathrm{GeV}^{2}$. The quantities labeled with ``$\star$" are not the result of a parametric function to data fit, but were obtained by comparing parametrizations. The statistical error of our fit results, therefore, entirely from the fit to the $t$-dependence of the electromagnetic form factors, and moments $A_{20}$ and $B_{20}$. The $H^{\bar{u}}$ has two additional parameters, $\alpha_{2} = 0.80(4), c= 0.100(33)$ (see Eq.\eqref{eq:regge2}).}
    \label{tab:parameters}
\end{table}

PQCD evolution  is introduced in the first step, {\it i.e.} we solve (Eqs.\eqref{eq:evolNS},\eqref{eq:evolS})to determine the parameters (${\cal N}, m,M_X,M_\Lambda,\alpha$). 
To increase the speed of our fits we adopted the Adams method as an alternative to the commonly used ({\it e.g.} \cite{Freese:2024ypk}) Runge Kutta (RK) method.
The Adam's method is computationally faster than RK since it uses information from several previous points, {\it e.g.} two starting values, but only one slope evaluation (see also \cite{Maichum2022-sx}). RK, on the contrary, computes solutions at point ${n+1}$ using only data from one initial starting point, but it needs several slope evaluations to proceed. The Adams method is therefore ideal for the solution of DGLAP equations where efficiency matters since the {\it r.h.s.} of Eqs.\eqref{eq:evolNS},\eqref{eq:evolS} is a convolution integral, therefore notoriously expensive to evaluate. 

\begin{figure}[H]
    \centering
\includegraphics[width=0.4\linewidth]{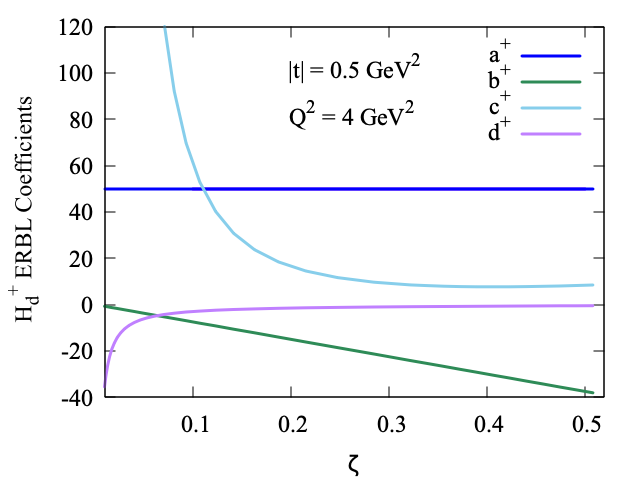}
    \caption{Kinematic dependence of the coefficents for the $H_d^+$ distribution, $a^+$, $b^+$, $c^+$, and $d^+$ Eqs.\eqref{eq:paramERBL}, plotted vs. $\zeta$ at $|t|=0.5$ GeV$^2$ and $Q^2 = 4$ GeV$^2$. } 
    \label{fig:erbl_d}
\end{figure}

An example illustrating the kinematic dependence of the coefficients, $a^+, b^+, c^+, d^+$, in the ERBL region is shown in Figure   \ref{fig:erbl_d}, for the GPD $H_d^+$.
A similar behavior is found for the remaining coefficients, as well as for different flavors. Notice that the coefficients also depend on $Q^2$, due to PQCD evolution. We reiterate that this behavior is obtained automatically, when using the parametrization, by evaluating Eqs.\eqref{eq:paramERBL}, once the parameters in the DGLAP region have been fixed.

As described in the lower panel in Fig.\ref{fig:flowchart}, after evaluating the forward limit, we then switch on the $t$ and $\zeta$ dependence and  we impose the constraint on the first moments of the GPDs using the nucleon elastic form factors (Eqs.\eqref{eq:dirac} and \eqref{eq:pauli}), and the second moments of GPDs (Eqs.\eqref{eq:A20}\eqref{eq:B20}) from LQCD calculations. 
The parameters constrained in the second step are from the Regge component of the parametrization  (Eq.\eqref{eq:regge}). 
The $\bar{u}$ parametrization uses a modified Regge term of the form,
\begin{equation}
\label{eq:regge2}
R_{\bar{u} } (X,t) = X^{-\left[\alpha_{\bar{u}} \sqrt{1 - \frac{t}{c}}+ \alpha_2 (1-\sqrt{1 - \frac{t}{c}} ) + \alpha_{\bar{u}}'(1-X)^{[p_{\bar{u}}]}t \right]}
\end{equation}
that is introduced in order to soften the slope in $t$, to reproduce the LQCD evaluation of $A_{20}$ and $B_{20}$ from Ref.\cite{Bhattacharya:2023ays}. We note that this step introduces a systematic error -- evaluated to be a $10\%$ effect -- stemming from the fact that the LQCD results on the $t$-dependent moments are for a pion mass still sensibly higher than the physical pion mass.   

With this recursive fitting procedure we are able to constrain a total number of 7 parameters per quark flavor, 8 for antiquarks, and 6 for the gluon. 
This number is the minimum required to guarantee enough flexibility of the multidimensional functional forms for $H_{q,g}(X,\zeta,t)$ and $E_{q,g}(X,\zeta ,t)$, to allow for a quantitative fit that preserves all the symmetry constraints (Sec.\ref{sec:param}). 
The rest of the parameters for the ERBL region, Eqs.\eqref{eq:paramERBL}, are automatically determined in the second  step (Fig.\ref{fig:flowchart}) since,  in this case, focusing on the $-$ distributions, $u_v$ and $d_v$, and $g$, we have the two remaining parameters  are fixed by  constraints from: 1) the value at the crossover point, $F(\zeta,\zeta,t;Q^2)$, $F=H,E$; 2) the GPDs first and second moments values which determine the areas subtended by the GPDs, and 3) imposing the symmetry around $X=\zeta/2$  for any value of $Q^2$; 
the $+$ distributions, $q+\bar{q}$ have 4 parameters per flavor ($u^+$, $d^+$), 2 of which are constrained in a similar way as for the $-$ distributions. The two remaining parameters which are free to vary even when all constraints from the symmetries and sum rules are satisfied, $a^+$, and $b^+$, can be fixed knowing the $\xi \, (\zeta)$ dependence of DVCS-type data. 
In our parametrization the matching between DGLAP and ERBL regions is done through the expressions in Eq.\eqref{eq:F_param_form}. As a result, the ERBL parameters will depend on the 
kinematics, as shown in what follows.

The details of the main changes with the previous versions of our parametrization are listed below:
\begin{enumerate}
\item \underline{Constraint from PDFs/forward limit}:
Compared to previous versions, we are now able to simultaneously evolve all relevant quark flavors as well as gluons, by providing their parametrization at the initial scale of $Q_0^2=0.58$ GeV$^2$ with a value of $\Lambda_{QCD}(N_f=4)$ = 0.257 GeV. Flavor thresholds are taken care of in the general-mass variable-flavor-number scheme (GM VFNS) \cite{Nadolsky:2008zw}. The parameters are fit to reproduce the LO NNPDF21 PDF set \cite{Ball:2011uy}. 
The fitting and parameter uncertainty estimation were performed using the python version of the MINUIT library \cite{iminuit}.
Comparisons of these curves and other parametrizations CJ15 \cite{Accardi:2016qay}, MMHT2014 \cite{Harland-Lang:2014zoa}, and MSHT20 \cite{Bailey:2020ooq} are shown in Figures \ref{fig:comparison_NS},\ref{fig:comparison_S}.
The set of parameters from this constraint is presented with no error bars because they are obtained by fitting PDF parametrizations, and not directly statistically distributed experimental data. 
For this case, we  show results of a Kolmogorov-Smirnov test which provides a measure of the similarity of the fitted and original PDF distributions.  

\item \underline{Analytic expressions}: The diquark/spectator model has been used ubiquitously to parametrize TMDs and fragmentation functions since it is sufficiently flexible to capture the essential physics features of the distributions while being cast in an easily calculable functional form. For GPDs, due to the dependence on the ``external" variables, $\zeta$ and $t$, the functional forms are more involved than for PDFs and TMDs. In this paper we present for the first time  analytic forms stemming directly 
from overlaps of lightcone wave
functions. Using these expressions, all parameters are fixed at a low scale by reproducing the phenomenological constraints described in Sec.\ref{sec:param}.

\item \underline{Constraint on $t$ dependence}: 
New LQCD results have become available since the previous version of this parametrization in Ref.\cite{Kriesten:2021sqc}, where the gluon GPDs were fitted to the $A_g, B_{g}$ moments from Ref.\cite{Pefkou:2021fni}. The new LQCD results from Ref.\cite{Hackett:2023rif}, better constrain the values of our parameters.  
The valence and gluon components are fitted following the standard procedure described in Refs.\cite{Goldstein:2013gra,Kriesten:2021sqc} while the antiquarks need to be extracted using $t$-dependent LQCD calculations of $A_{20}$ and $B_{20}$, recently evaluated in Ref. \cite{Bhattacharya:2023ays}. Since these moments are not at physical pion mass, we introduced a second $t$-dependent term in the Regge exponent of the $\bar{u}$ parametrization  (Eq.\eqref{eq:regge2}). 


\item \underline{QCD Evolution}: we present results using our parametrization evolved at LO with initial scale, $\mu_o^2= 0.58$ GeV$^2$. Using Eqs.\eqref{eq:SU6}, we will be able to eventually extract the gluon distributions exploiting the coupling of $H_g$ to the singlet component $H_S$, Eq.\eqref{eq:evolS}.

\item \underline{On the fitting procedure}:
we adopted a standard curve fitting procedure using MIGRAD -- Minuit’s principal optimization algorithm -- designed to efficiently minimize the chi-square through the evaluation of the inverse Hessian matrix (or the matrix of second derivatives) to find the minimum. MIGRAD is particularly effective for problems with many parameters such as the one at hand, providing a fast and stable method compared to simpler gradient-descent or fixed-metric methods. We view this as a first necessary step to gauge the main issues in defining a parametrization for our multidimensional problem. The parametrization is in a readily usable format for DVES observable predictions in a wide kinematic setting. In a second step, we will adopt more accurate approaches to a more thorough uncertainty quantification, as well as to estimate the size and impact of correlations in the data. The latter can be studied ideally in a  maximum likelihood analysis and Markov-Chain-Monte Carlo sampling procedure, as well as in neural network-based analyses, which we are in the process of developing. 
\end{enumerate}

Note that the parameters labeled with "$\star$" in Table \ref{tab:parameters} were obtained by performing a curve fit to match PDFs from DIS analyses. Therefore, we do not quote a statistical error on these parameters. The Kolmogorov-Smirnov test to the goodness of fit in a ``distribution-to-distribution" matching are shown in Section\ref{subsec:Kolmogorov}. As a result, the error on our GPDs originates entirely from the fit to the $t$-dependence of the electromagnetic form factors, and LQCD gluon form factors, as well as $A_{20}$ and $B_{20}$ moments of the quark distributions.

In what follows we show several examples illustrating the behavior of GPDs (Section \ref{sec:GPDs}), and CFFs (Section \ref{sec:CFF2}) obtained using our parametrization for $H_{q,g}$ and $E_{q,g}$, focusing on the different kinematic regimes at Jefferson Lab and EIC. 

\subsection{GPDs}
\label{sec:GPDs}
\begin{figure}[H]
\centering
\includegraphics[width=0.4\linewidth]{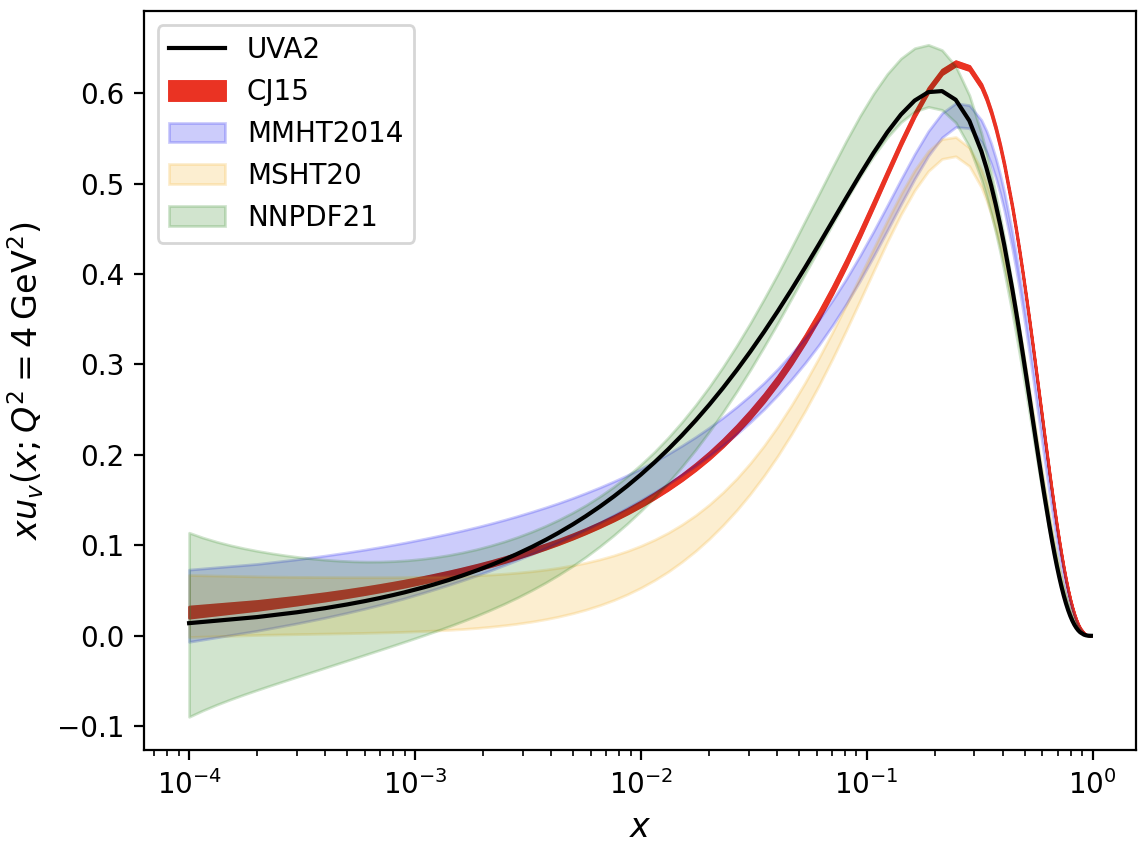}
\hspace{1.0cm}
\includegraphics[width=0.4\linewidth]{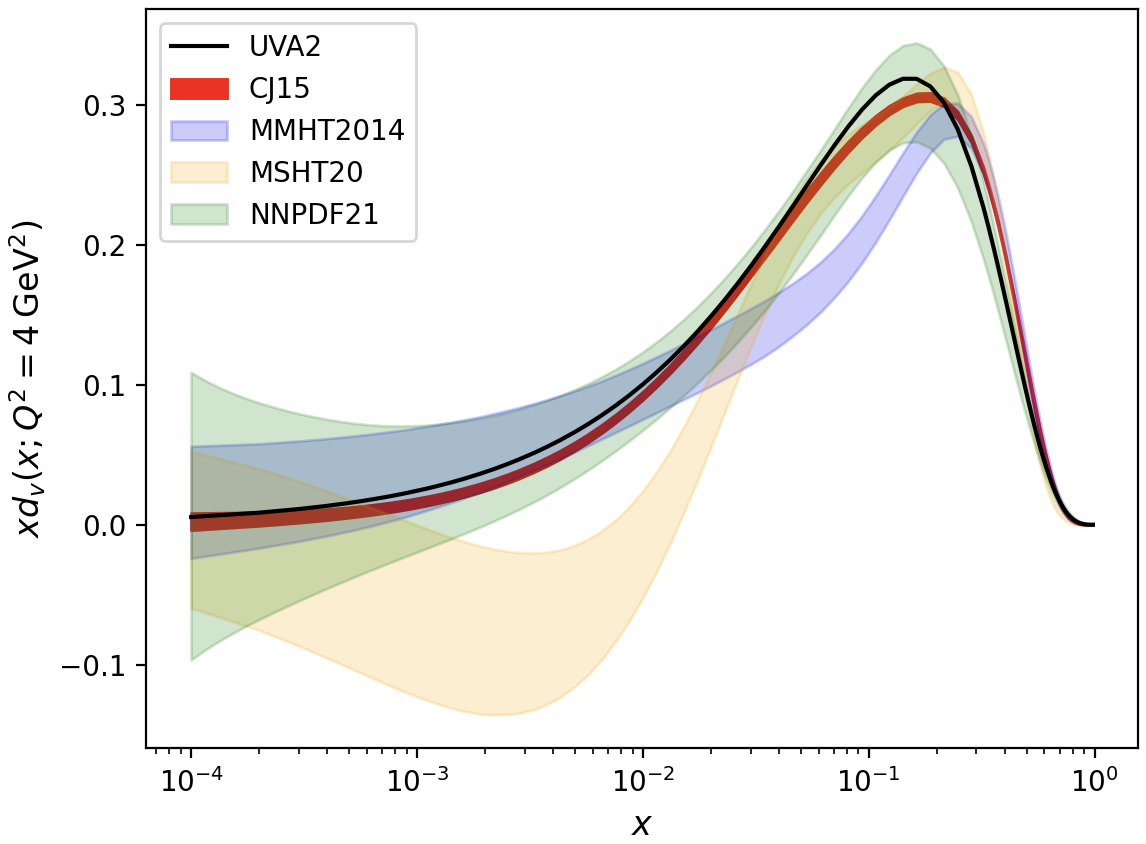}
\caption{Comparison of the $u_v$ and $d_v$ distributions in the forward limit with CJ15 \cite{Accardi:2016qay}, MMHT2014 \cite{Harland-Lang:2014zoa}, and MSHT20 \cite{Bailey:2020ooq} at $Q^2=4$ GeV$^2$.}
\label{fig:comparison_NS}
\end{figure}

\begin{figure}[H]
\centering
\includegraphics[width=0.4\linewidth]{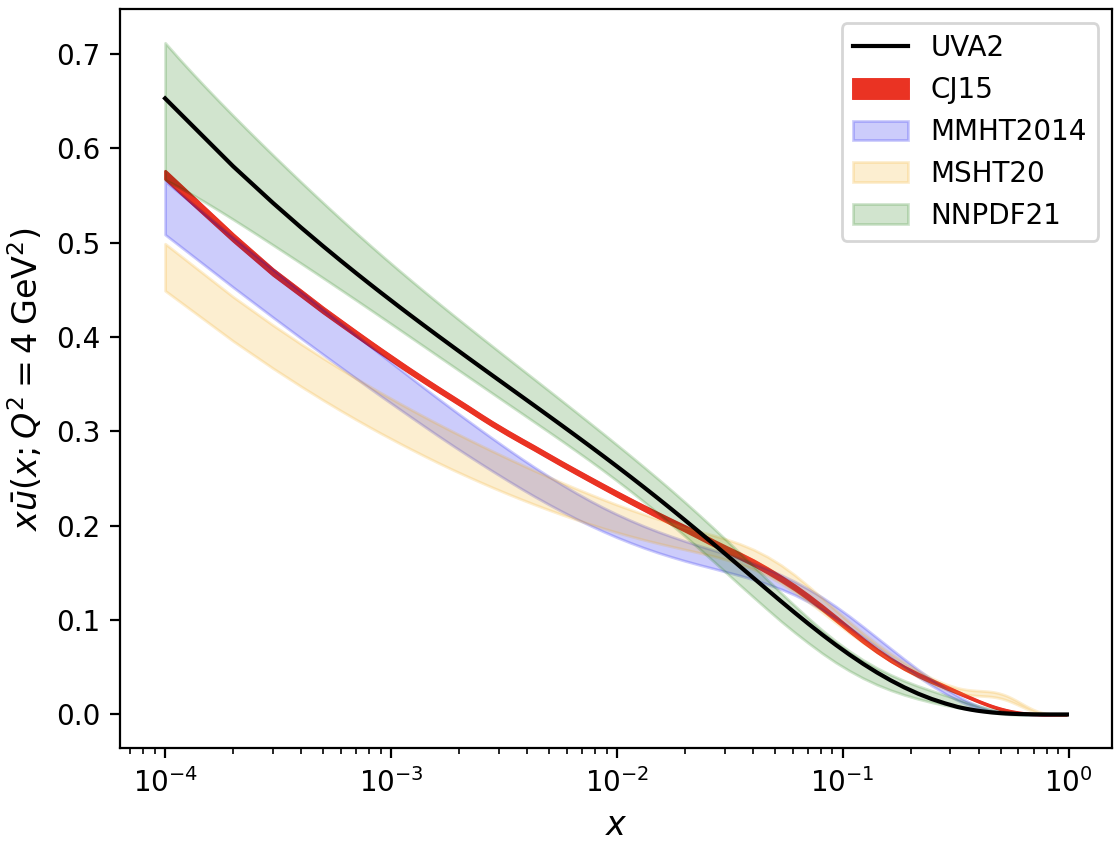}
\hspace{1.0cm}
\includegraphics[width=0.4\linewidth]{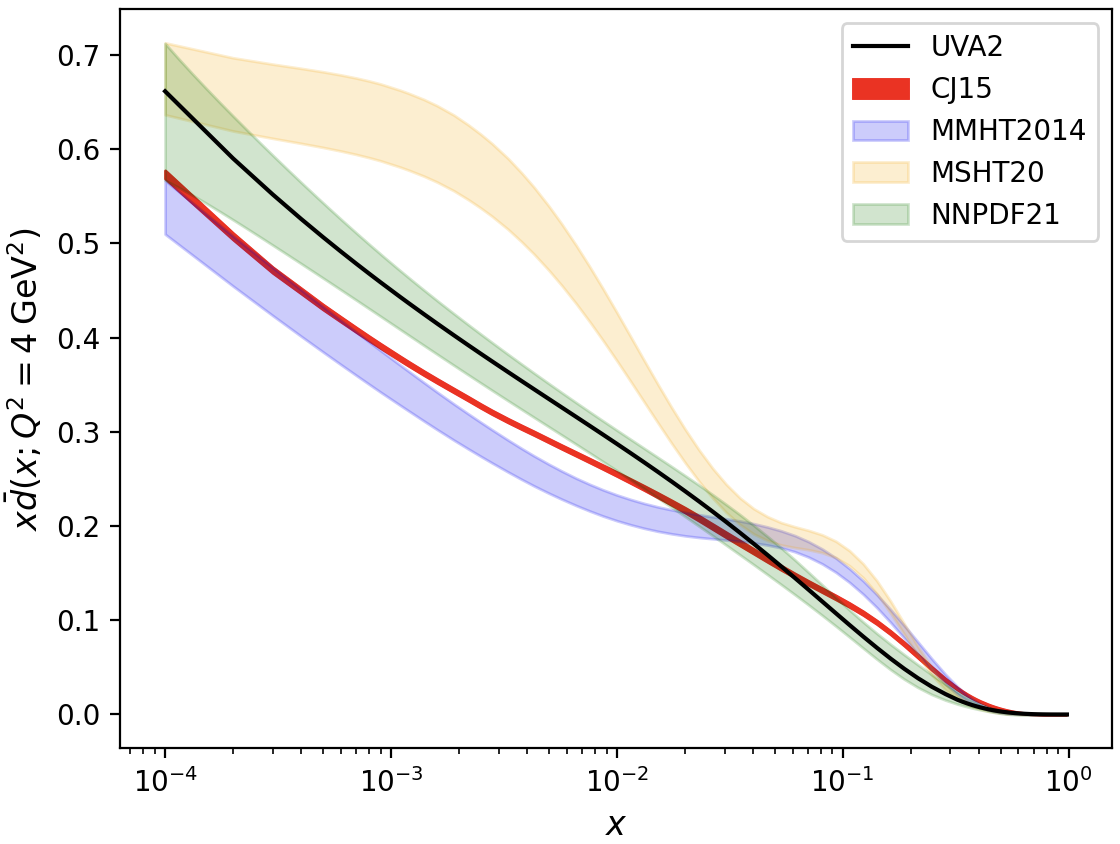}
\includegraphics[width=0.4\linewidth]{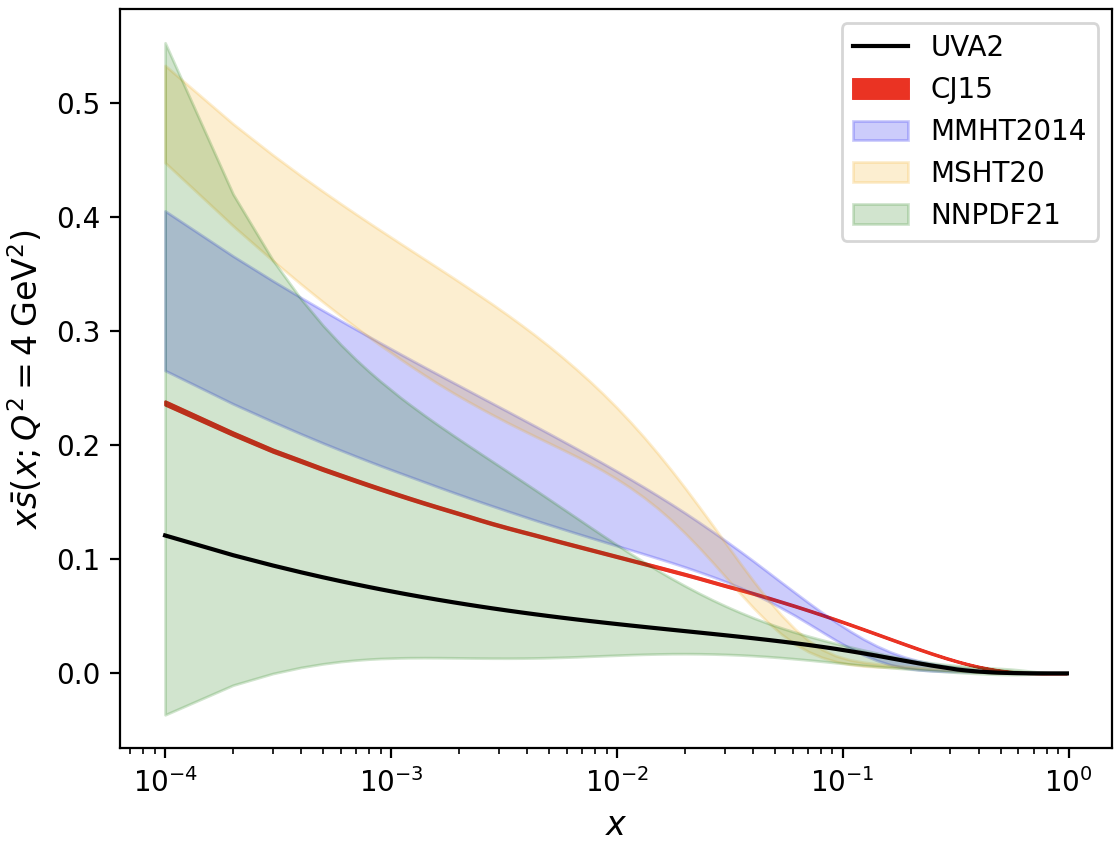}
\includegraphics[width=0.4\linewidth]{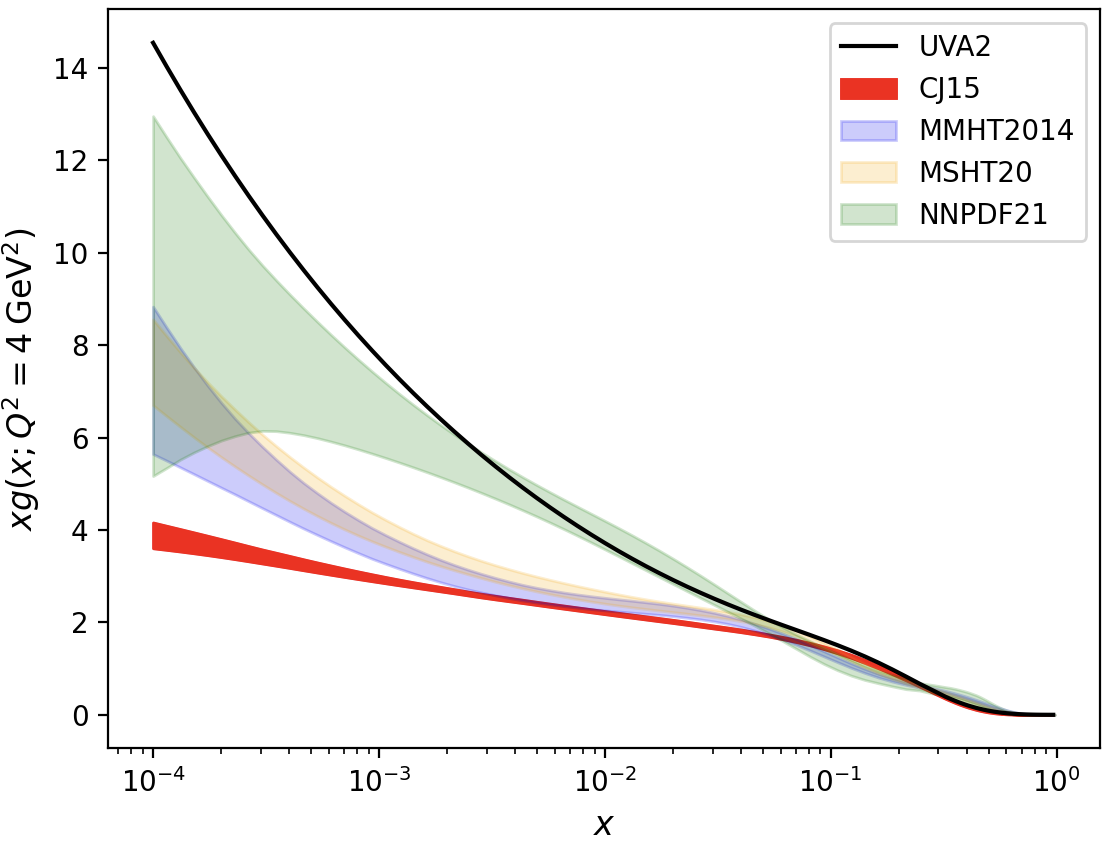}
\caption{Comparison of the $\bar{u}$, $\bar{d}, \bar{s}$, and gluon distributions in the forward limit with the parametrizations from CJ15 \cite{Accardi:2016qay}, MMHT2014 \cite{Harland-Lang:2014zoa}, and MSHT20 \cite{Bailey:2020ooq} at $Q^2=4$ GeV$^2$.}
\label{fig:comparison_S}
\end{figure}

We start by showing how the GPDs from the current UVA2 parametrization satisfy the constraints in the forward limit, {\it i.e.} from PDFs, from the $n=1$ moments from experiment, and the $n=2$ from LQCD. in Figures \ref{fig:comparison_NS}, \ref{fig:comparison_S}, all GPDs for $q=u_v, d_v, \bar{u}, \bar{d}, \bar{s}$ and $g$, in the limits $H_{q,g}(X,0,0;Q^2)$, are compared to available sets of PDFs from CJ15 \cite{Accardi:2016qay}, MMHT2014 \cite{Harland-Lang:2014zoa}, and MSHT20 \cite{Bailey:2020ooq}, at $Q^2= 4 $ GeV$^2$. Notice that in our model, the strange flavor GPD is generated entirely from PQCD evolution. 


\begin{figure}[H]
\centering
\includegraphics[width=0.4\linewidth]{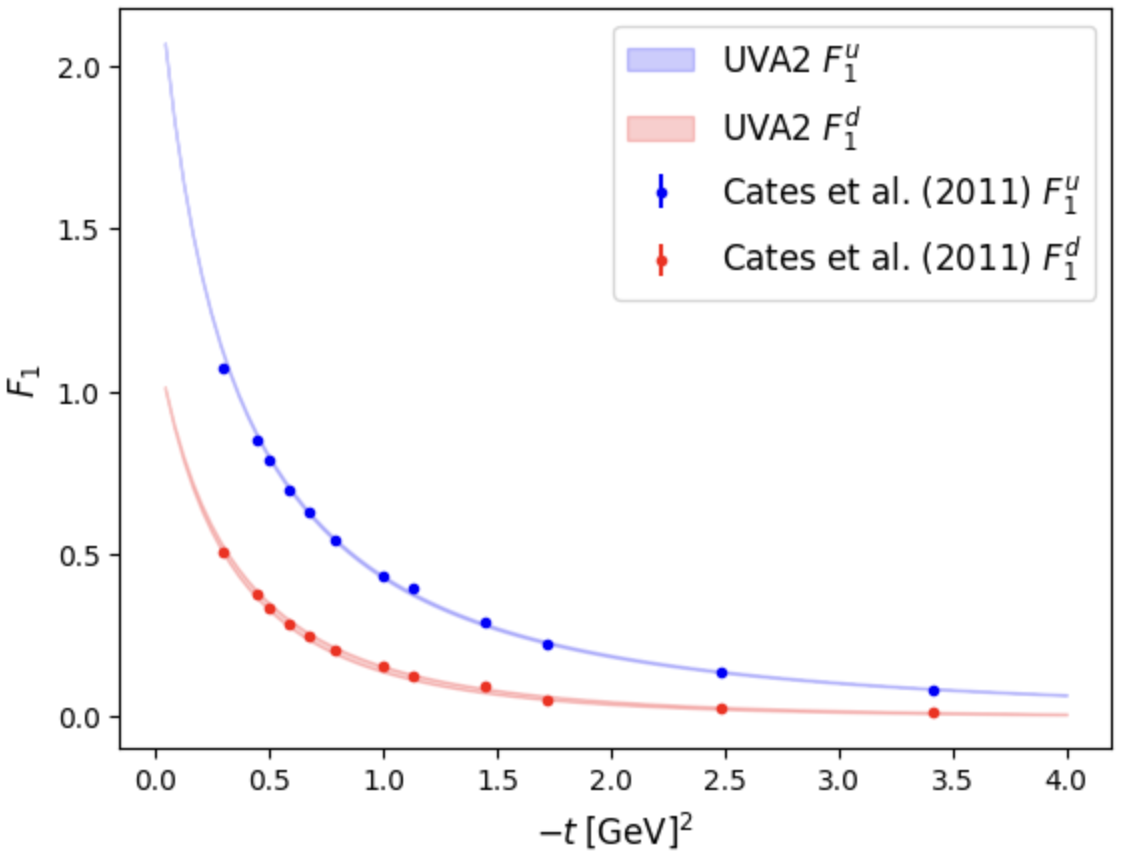}
\hspace{1.0cm}
\includegraphics[width=0.4\linewidth]{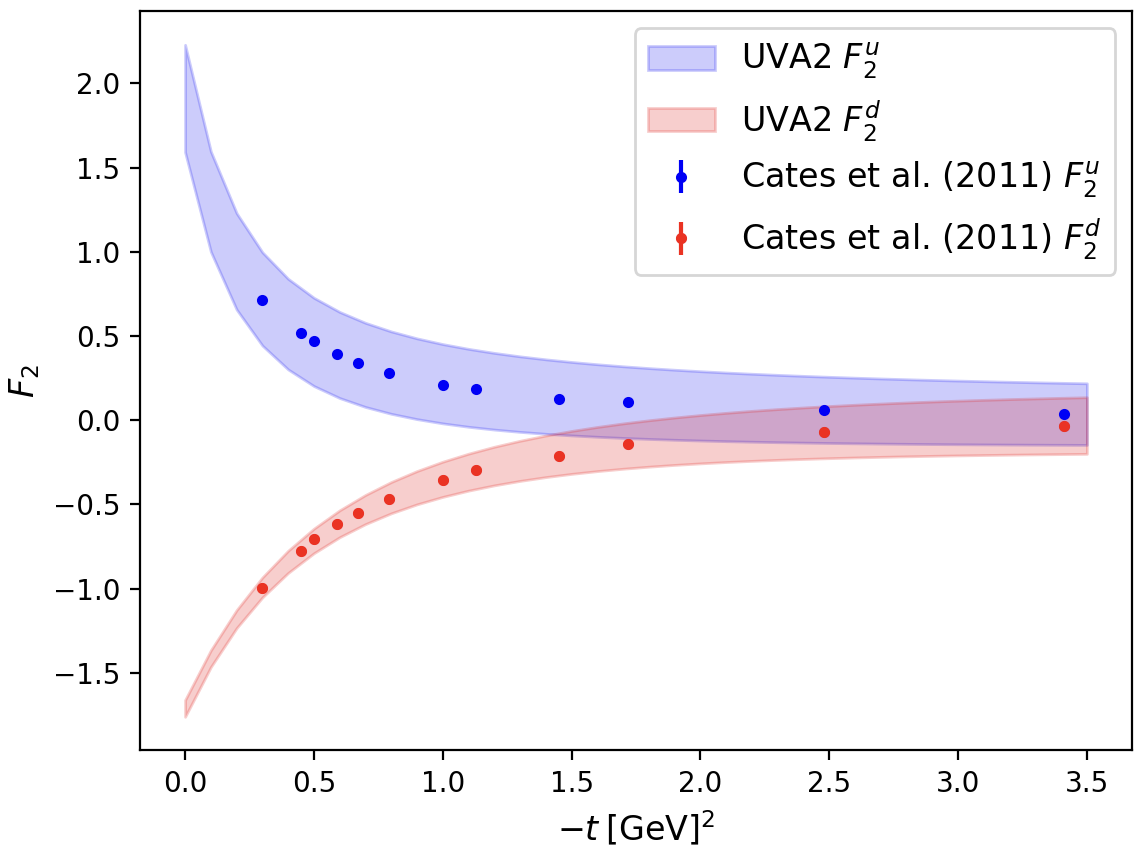}
\caption{Fit result of the t-dependence of the $u_v$ and $d_v$ distributions compared with data from Ref.\cite{Cates:2011pz}. The $u$ and $d$ contributions to the $F_1$ and $F_2$ form factors are shown on the the left and right panels, respectively.}
\label{fig:comparison_val_t}
\end{figure}

\begin{figure}[H]
\centering
\includegraphics[width=0.4\linewidth]{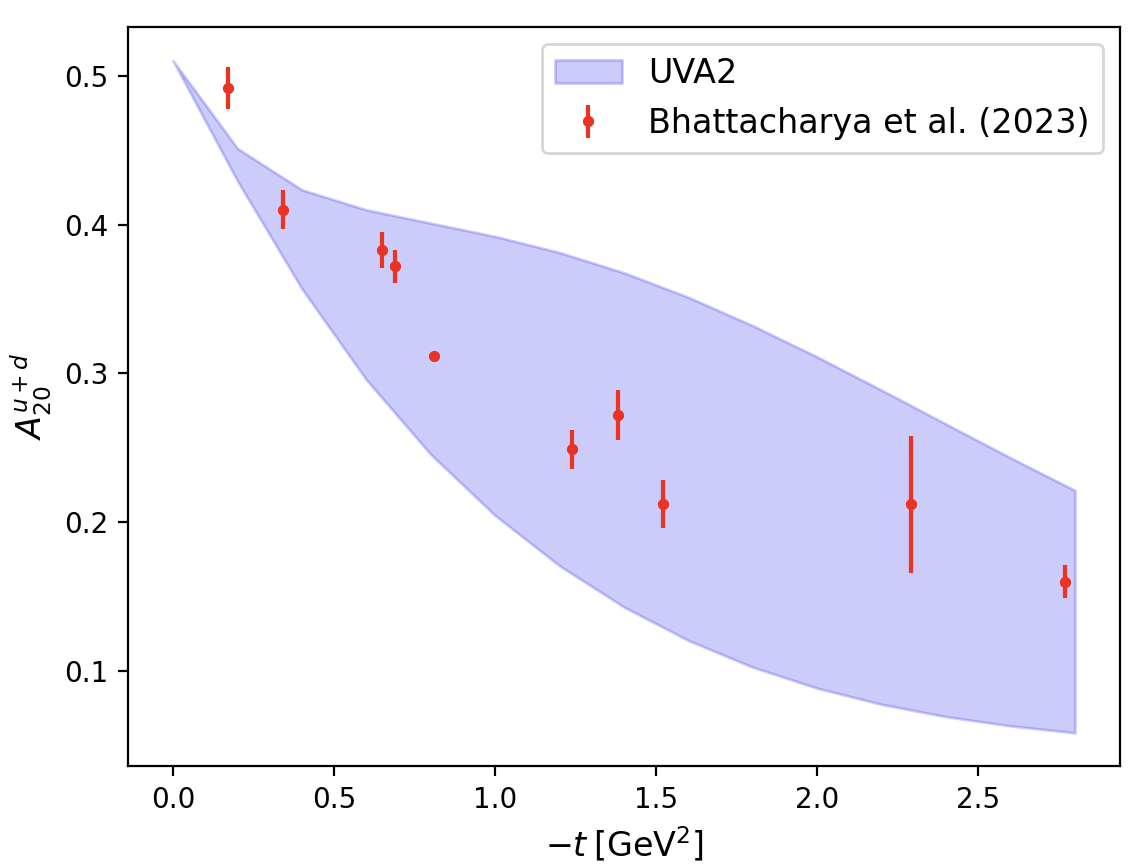}
\includegraphics[width=0.4\linewidth]{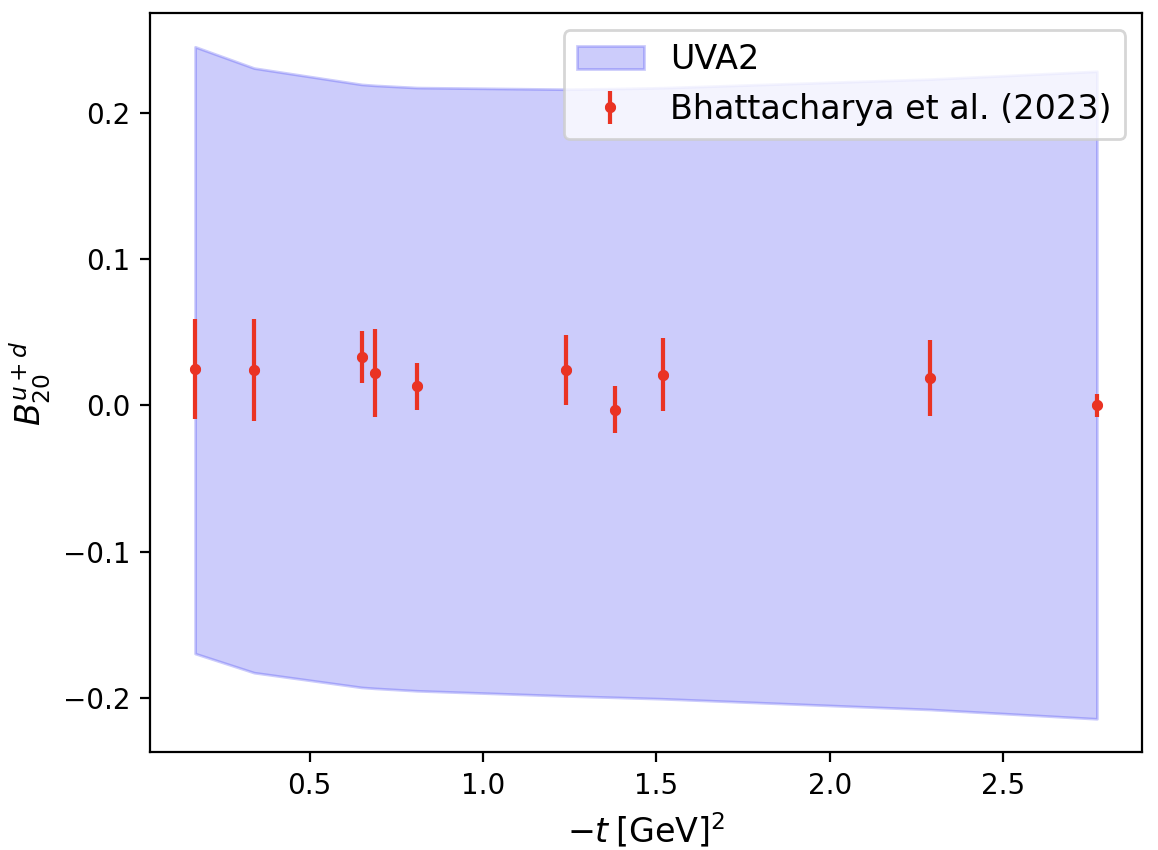}
\caption{Fit result of the $A_{20}^{u + d}, B_{20}^{u +d}$ form factors \cite{Bhattacharya:2023ays} constraining the antiquark $H_{\bar{q}}$ distributions.}
\label{fig:comparison_bar_Ht}
\end{figure}

\begin{figure}[H]
\centering
\includegraphics[width=0.4\linewidth]{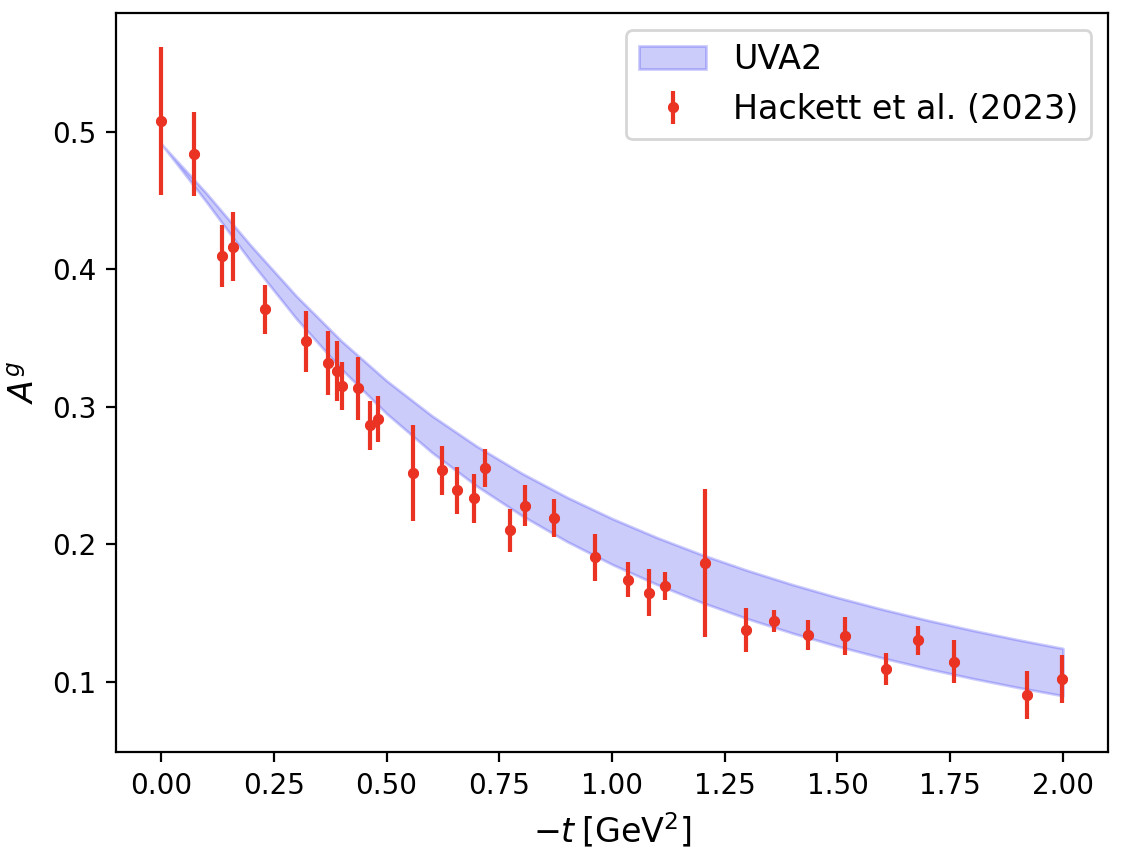}
\hspace{1.0cm}
\includegraphics[width=0.4\linewidth]{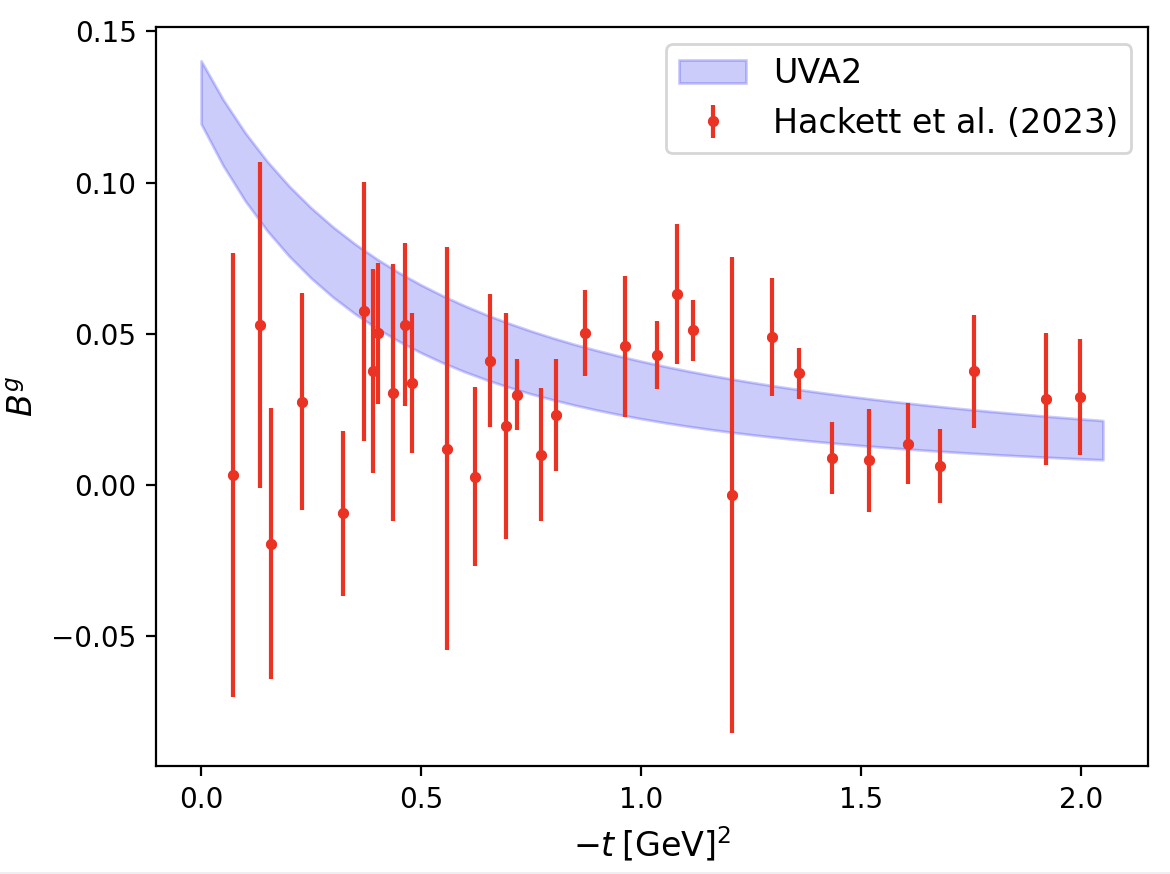}
\caption{Fit result of the t-dependence of the gluon distributions (LQCD results from Ref.\cite{Hackett:2023rif}).}
\label{fig:comparison_g_t}
\end{figure}

In Figures \ref{fig:comparison_val_t}, \ref{fig:comparison_bar_Ht}, \ref{fig:comparison_g_t} we show the results of the fit to the various form factors, namely the flavor separated electromagnetic form factors \cite{Cates:2011pz}, the $t$-dependent second moments \cite{Bhattacharya:2023ays}, and the gluon form factors \cite{Pefkou:2021fni}. The fit forms implements only two parameters for the $u_v$, $d_v$, and gluon components, consistently with a dipole form. We find a larger $\chi^2$ for $u_v$, that could result from either the presence of systematic/theoretical uncertainties in the flavor dependent form factor data, or from the need to insert correcting terms beyond the dipole form. Notice, also that the integral constraints for the $\bar{u}$ and $\bar{d}$  distributions were obtained in the Kuti-Weisskopf formulation of the second moments given in Eq.\eqref{eq:kutiW_moment}, as the difference  of the second Mellin moment \cite{Bhattacharya:2023ays}, and the second moment of the valence distribution from our model.  The fits for the GPD $E$ differ from $H$ in that the parameters for the forward limit are not constrained. Therefore we also used a third parameter, the normalization, ${\cal N}$, to fit the $t$-dependence.  

\begin{figure}[H]
\centering
\includegraphics[width=0.4\linewidth]{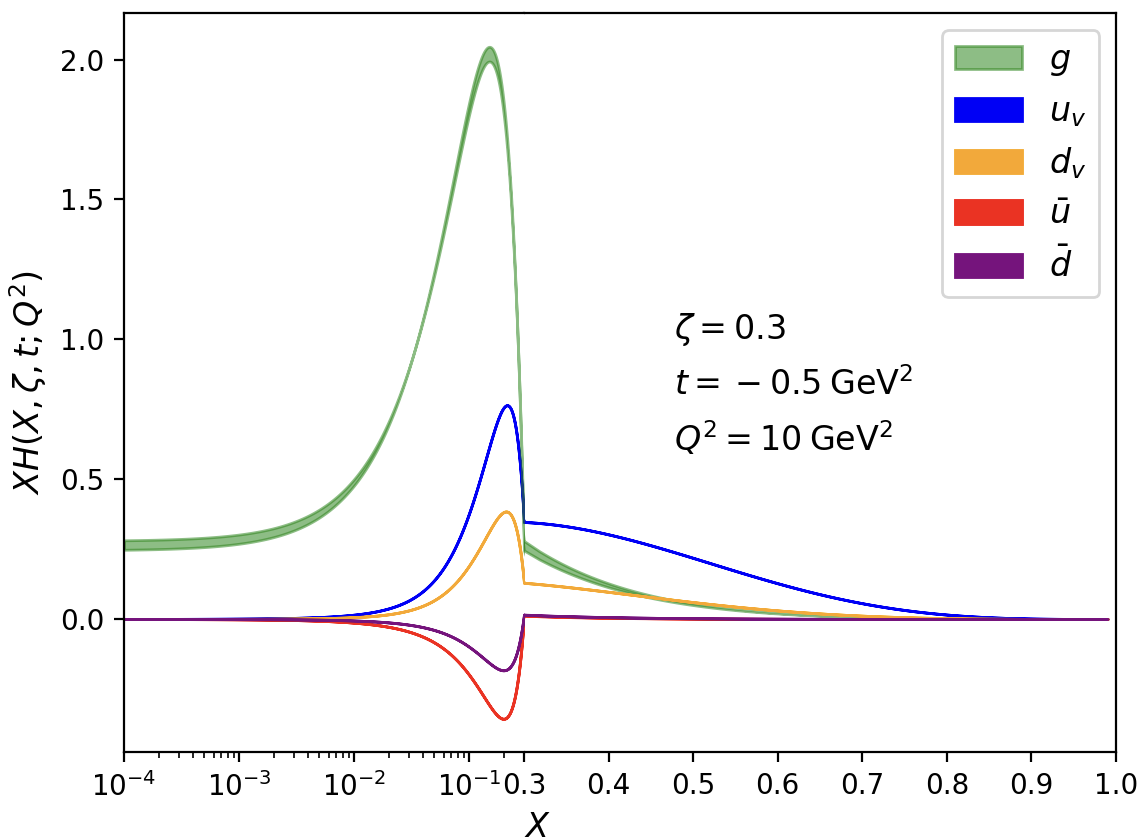}
\hspace{1.0cm}
\includegraphics[width=0.4\linewidth]{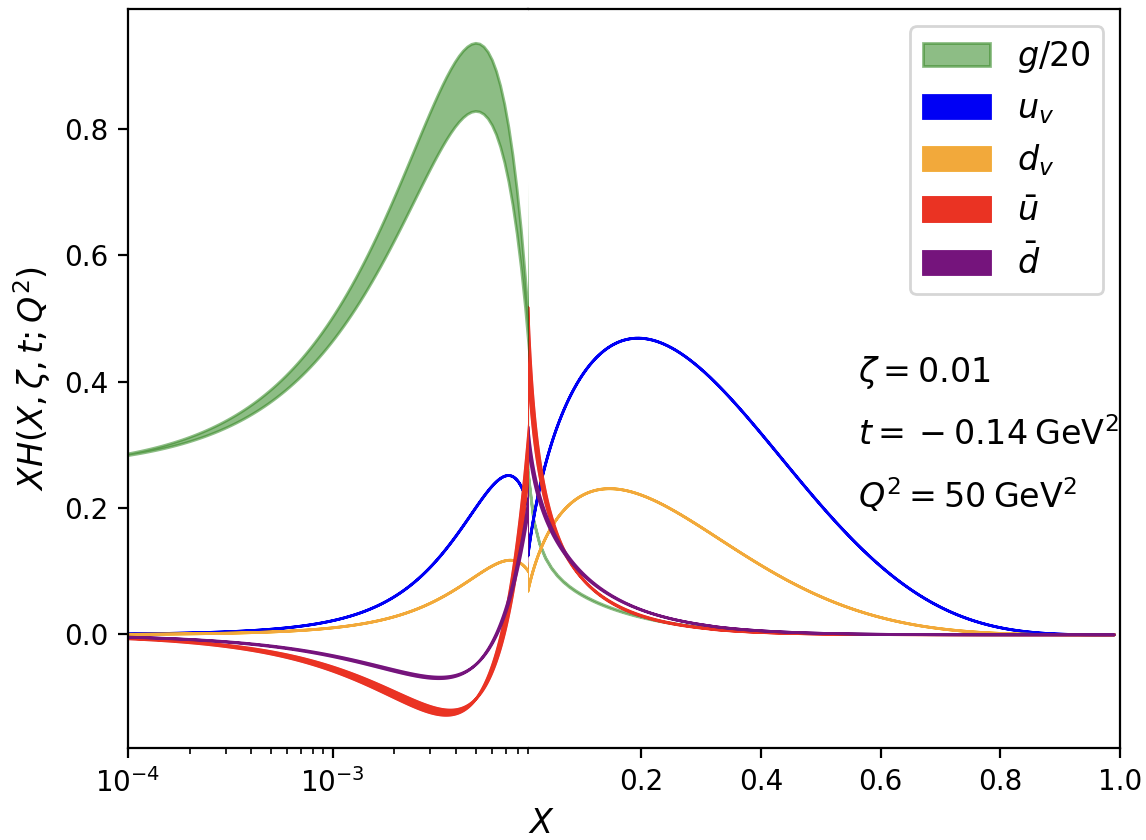}
\caption{The $H$ GPDs for representative JLab (left) and EIC (right) kinematics, $\zeta = 0.3, t = -0.5 \: \mathrm{GeV}^{2}, Q^{2} = 10 \: \mathrm{GeV^{2}}$ and $\zeta = 0.01, t = -0.14 \: \mathrm{GeV}^{2}, Q^{2} = 50 \: \mathrm{GeV^{2}}$, respectively.}
\label{fig:H_GPDs}
\end{figure}

\begin{figure}[H]
\centering
\includegraphics[width=0.4\linewidth]{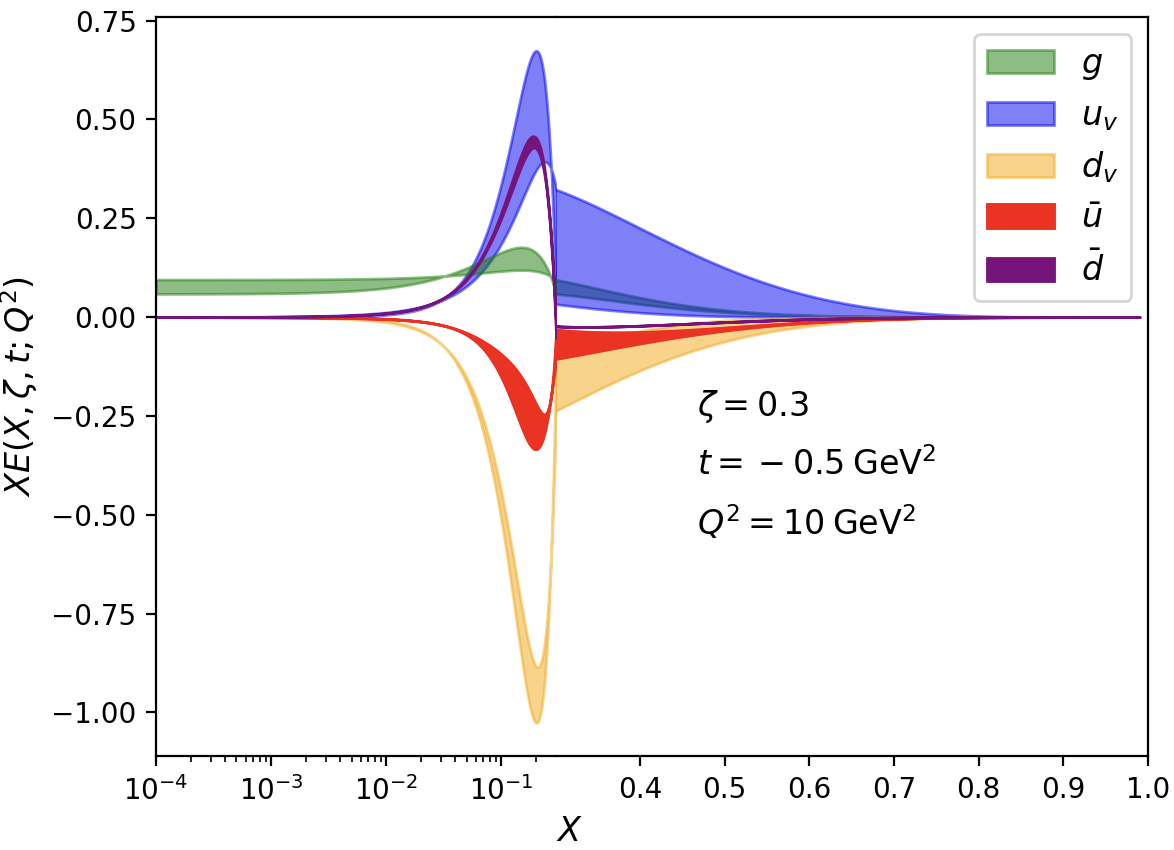}
\hspace{1.0cm}
\includegraphics[width=0.4\linewidth]{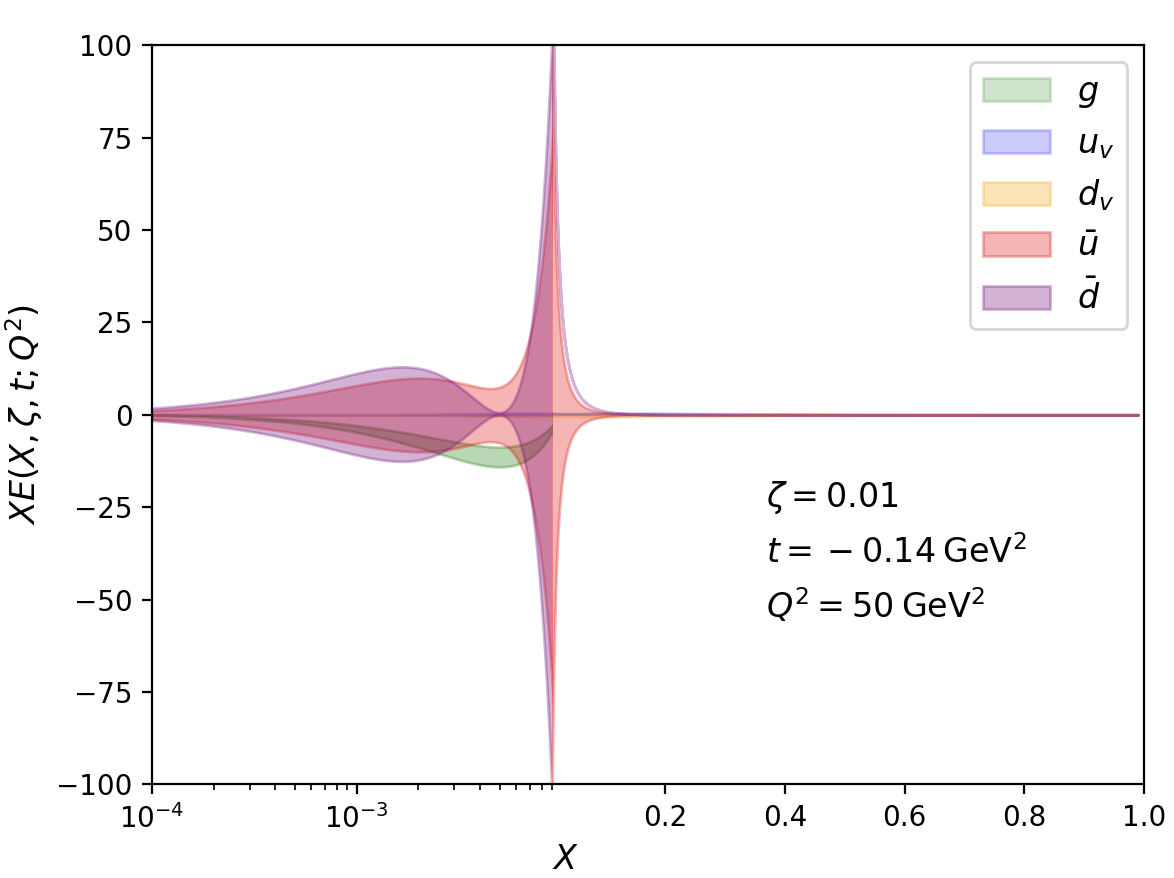}
\caption{The $E$ GPDs for representative JLab (left) and EIC (right) kinematics, $\zeta = 0.3, t = -0.5 \: \mathrm{GeV}^{2}, Q^{2} = 10 \: \mathrm{GeV^{2}}$ and $\zeta = 0.01, t = -0.14 \: \mathrm{GeV}^{2}, Q^{2} = 50 \: \mathrm{GeV^{2}}$, respectively. Notice that the $u_{v}$ and $d_{v}$ are highly suppressed compared to the sea quark and gluon distributions at EIC kinematics.}
\label{fig:E_GPDs}
\end{figure}

\begin{figure}[H]
\centering
\includegraphics[width=0.4\linewidth]{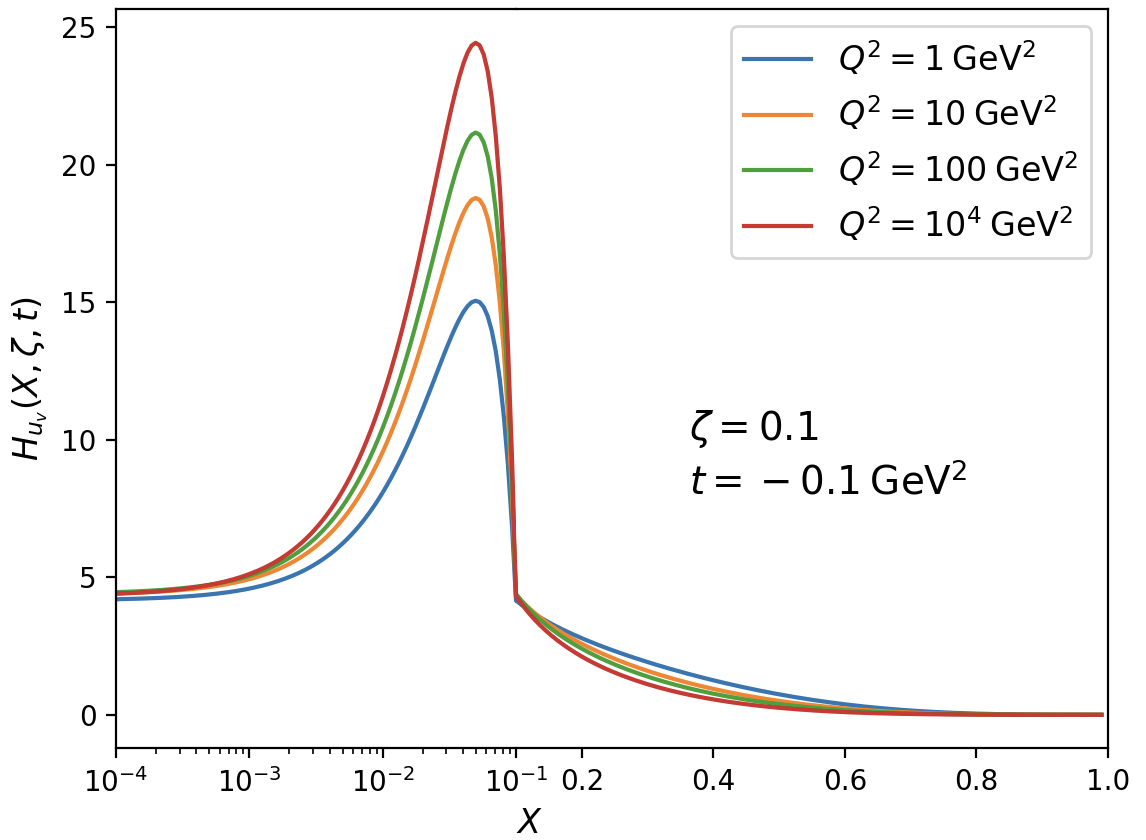}
\hspace{1.0cm}
\includegraphics[width=0.4\linewidth]{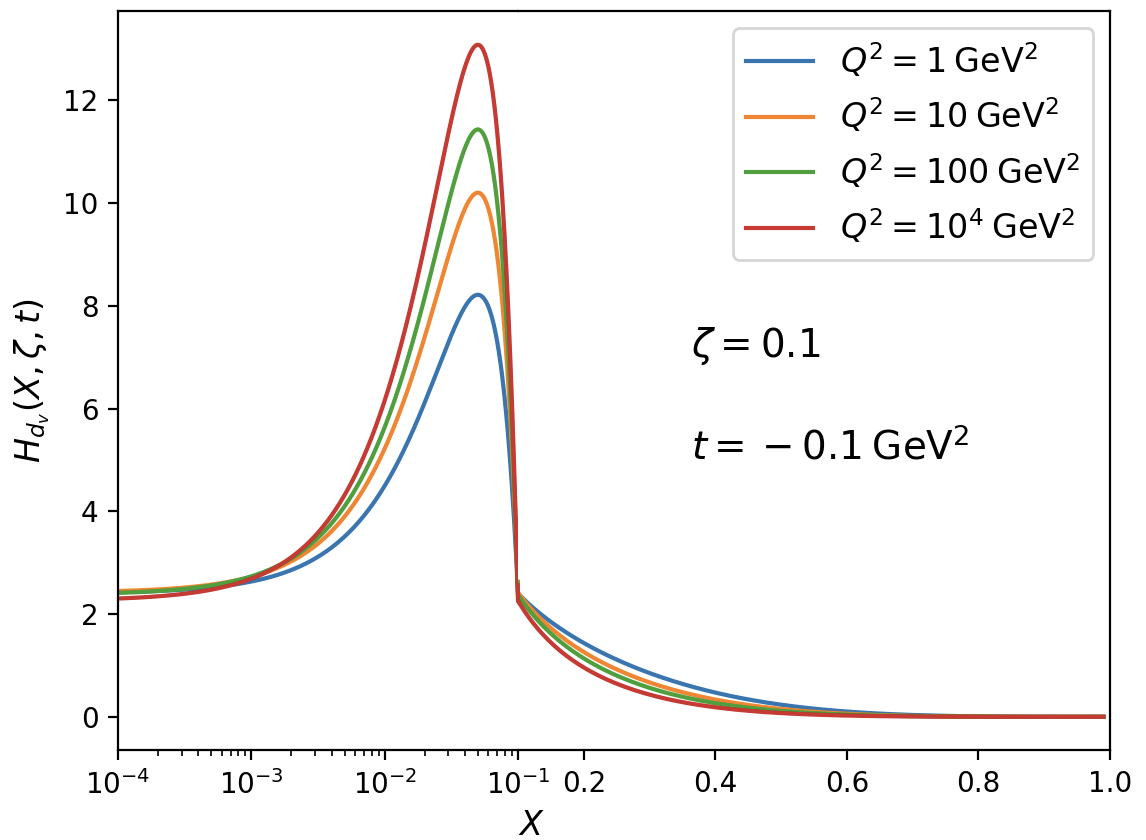}
\caption{$H_{u_v}$ ({\it left}), and $H_{d_v}$, ({\it right}) plotted vs. $X$, at kinematics: $|t| = 0.1$ GeV$^2$, $\zeta=0.1$, and varying $Q^2$ from 1 to $10^4$ GeV$^2$. Notice the change of scale from logarithmic for $X<0.1$, to linear for $X>0.1$. The error is not plotted.}
\label{fig:Hval_ERBL}
\end{figure}

\begin{figure}[H]
\centering
\includegraphics[width=0.4\linewidth]{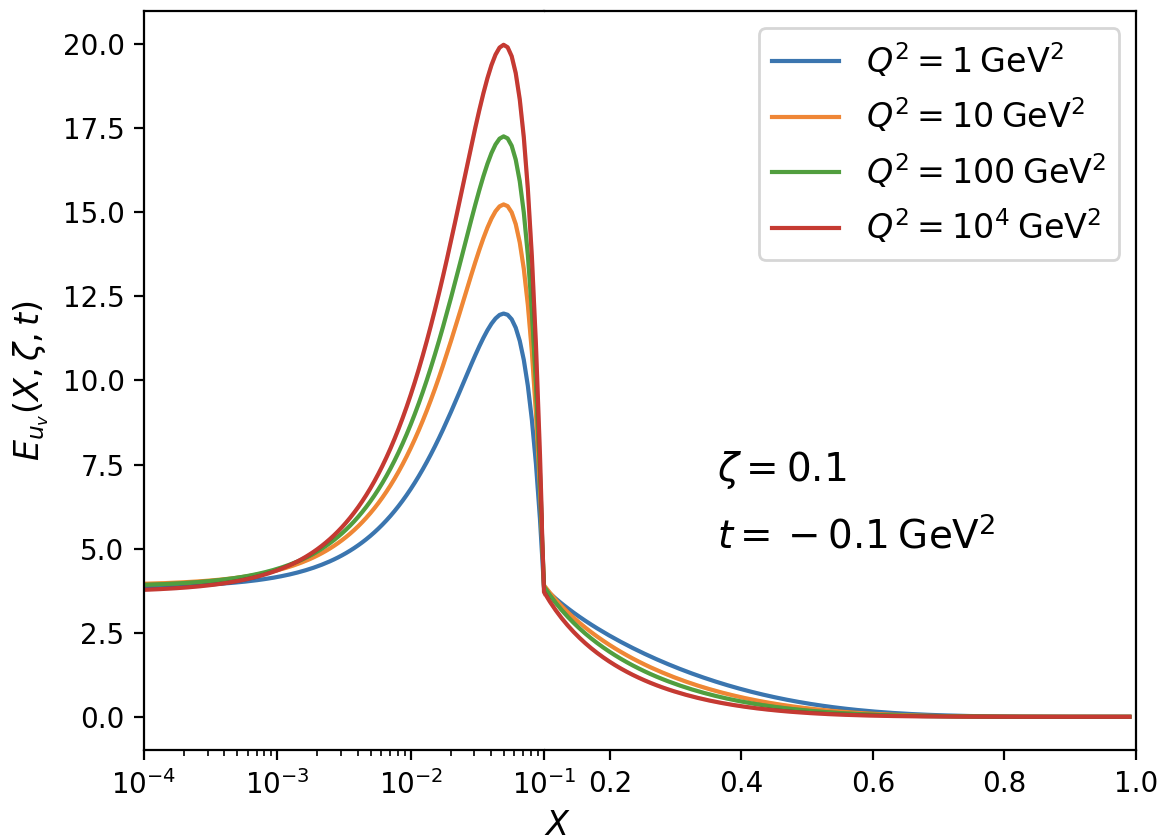}
\hspace{1.0cm}
\includegraphics[width=0.4\linewidth]{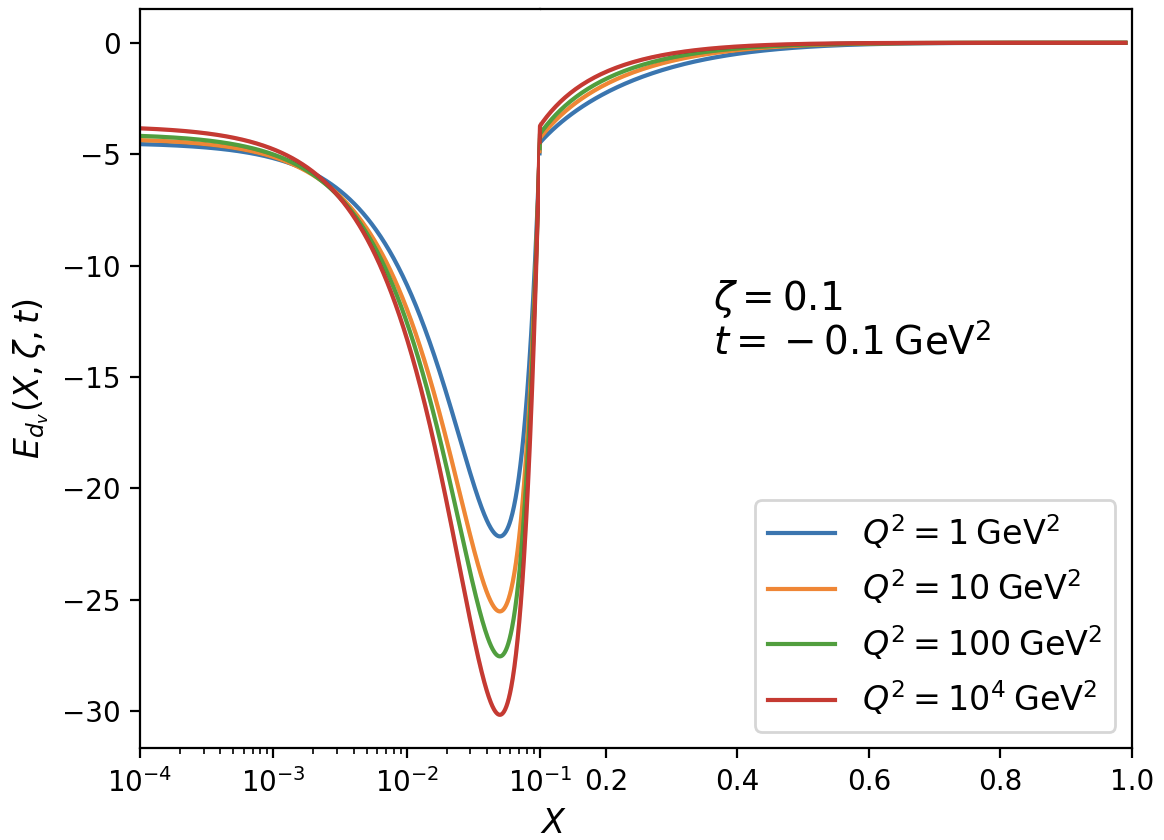}
\caption{$E_{u_v}$ ({\it left}), and $E_{d_v}$, ({\it right}) plotted vs. $X$, at kinematics: $|t| = 0.1$ GeV$^2$, $\zeta=0.1$, and varying $Q^2$ from 1 to $10^4$ GeV$^2$. Notice the change of scale from logarithmic for $X<0.1$, to linear for $X>0.1$. The error is not plotted.}
\label{fig:Eval_ERBL}
\end{figure}

\begin{figure}[H]
\centering
\includegraphics[width=0.4\linewidth]{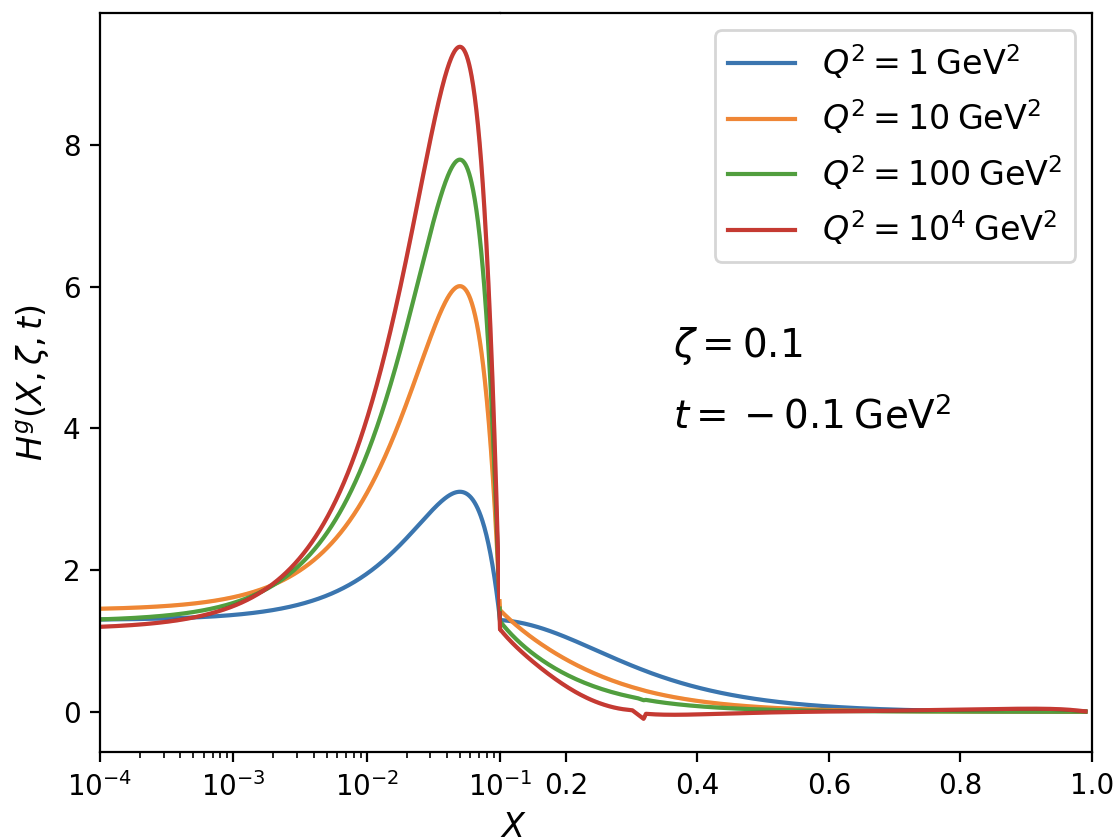}
\hspace{1.0cm}
\includegraphics[width=0.4\linewidth]{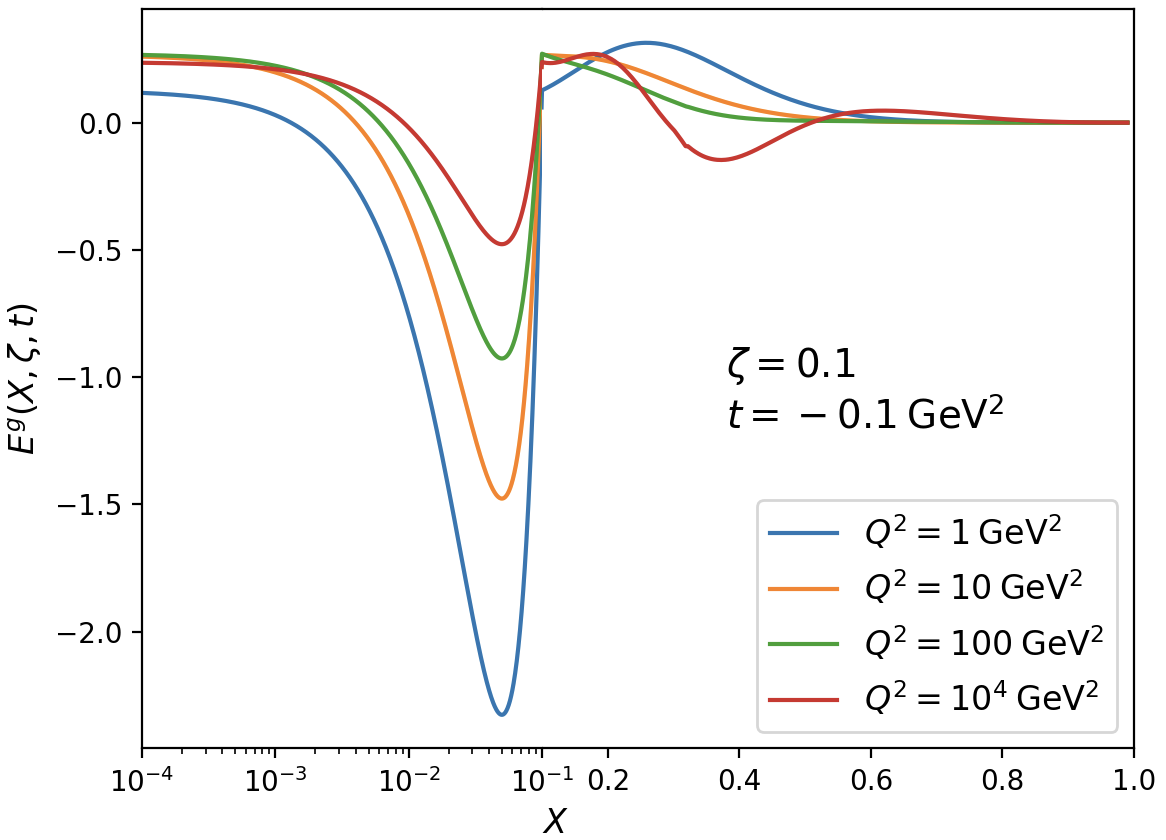}
\caption{$H_{g}$ ({\it left}), and $E_{g}$, ({\it right}) plotted vs. $X$, at kinematics: $|t| = 0.1$ GeV$^2$, $\zeta=0.1$, and varying $Q^2$ from 1 to $10^4$ GeV$^2$. Notice the change of scale from logarithmic for $X<0.1$, to linear for $X>0.1$. The error is not plotted.}
\label{fig:g_gpds_ERBL}
\end{figure}

Examples of the GPDs $H$ and $E$ are given in Figures \ref{fig:H_GPDs}, \ref{fig:E_GPDs}, \ref{fig:Hval_ERBL}, \ref{fig:Eval_ERBL} and \ref{fig:g_gpds_ERBL}. Figures \ref{fig:H_GPDs} and \ref{fig:E_GPDs} show the GPDs $H$ and $E$ for the various quark flavors and the gluon. The error on the curves reflects the statistical error from the form factors fit. On the {\it lhs} we show a typical Jlab kinematics ($x_{Bj} \approx \zeta = 0.3, |t|= 0.5$ GeV$^2$, $Q^2=10$ GeV$^2$), and on the {\it rhs} a typical EIC kinematics  ($x_{Bj} \approx \zeta = 0.01, |t|= 0.14$ GeV$^2$, $Q^2= 50$ GeV$^2$). In Figures \ref{fig:Hval_ERBL}, \ref{fig:Eval_ERBL} and \ref{fig:g_gpds_ERBL} we show the effects of PQCD evolution on the GPDs $H_{q_v}$, $E_{q_v}$, and ($H_g, E_g$) respectively, in the range $Q^2 \approx 1 - 10^3$ GeV$^2$.  

%
\subsubsection{Kolmogorov-Smirnov (KS) test}
\label{subsec:Kolmogorov}
The Kolmogorov-Smirnov (KS) test is a non-parametric method for comparing distributions, essential for various applications in diverse fields. Kolmogorov–Smirnov Test is a completely efficient manner to determine if two samples are significantly one of a kind from each other. It serves as an effective tool to determine the statistical significance of differences between two sets of data. The main idea behind using this Kolmogorov-Smirnov Test is to check whether the two samples that we are dealing with follow the same type of distribution or if the shape of the distribution is the same or not. This test is particularly valuable in various fields, including statistics, data analysis, and quality control, where the uniformity of random numbers or the distributional differences between datasets need to be rigorously examined. \\
One way the Kolmogorov-Smirnov Test works is by determining the Critical Value or p-value. For a one-sample Kolmogorov-Smirnov test, the test statistic (D) represents the maximum vertical deviation between the empirical distribution function (EDF) of the sample and the cumulative distribution function (CDF) of the reference distribution. For a two-sample Kolmogorov-Smirnov test, the test statistic compares the EDFs of two independent samples. The test statistic (D) is compared to a critical value from the Kolmogorov-Smirnov distribution table or, more commonly, a p-value is calculated. If the p-value is less than the significance level (commonly 0.05), the null hypothesis is rejected, suggesting that the sample distribution does not match the specified distribution. \\
However, KS test has some limitations too. One of them is the sensitivity to sample length. KS check may additionally have confined energy with small sample sizes and may yield statistically sizeable results with large sample sizes even for small versions. Another limitation is the lack of sensitivity to precise distributional properties; it assesses fashionable differences among distributions and might not be sensitive to variations, especially distributional houses. 

\begin{figure}[H]
    \centering
    \includegraphics[width=0.50\linewidth]{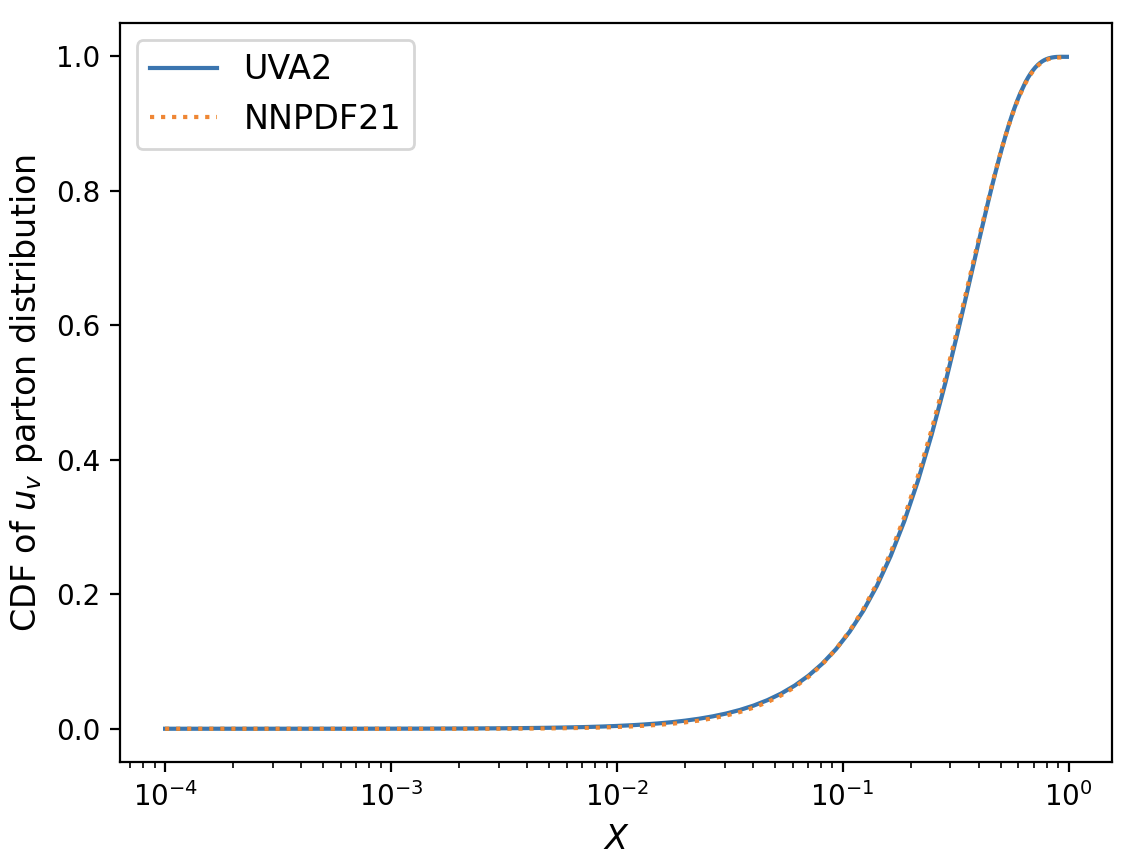}
    \caption{KS test applied to the $u_{v}$ distribution. The KS test quantifies the difference between the empirical cumulative distribution function that corresponds to UVA2's prediction for the parton distribution function and the underlying cumulative distribution function of the reference distribution, NNDPF21's PDF extraction. $p = 0.0074 < 0.05$ in this case, so the null hypothesis is rejected.}
    \label{fig:CDF_example}
\end{figure}

We provide an example of the KS test applied to our fit to the PDF in Figure \ref{fig:CDF_example}. Note that the integral of the $u_{v}$ was normalized to unity so that the CDF would range from 0 to 1; this differs from the physical normalization that is applied to the sum of the proton's degrees of freedom to ensure momentum conservation.


\subsection{Compton Form Factors}
\label{sec:CFF2}
\begin{figure}[H]
    \centering
    \includegraphics[width=0.5\linewidth]{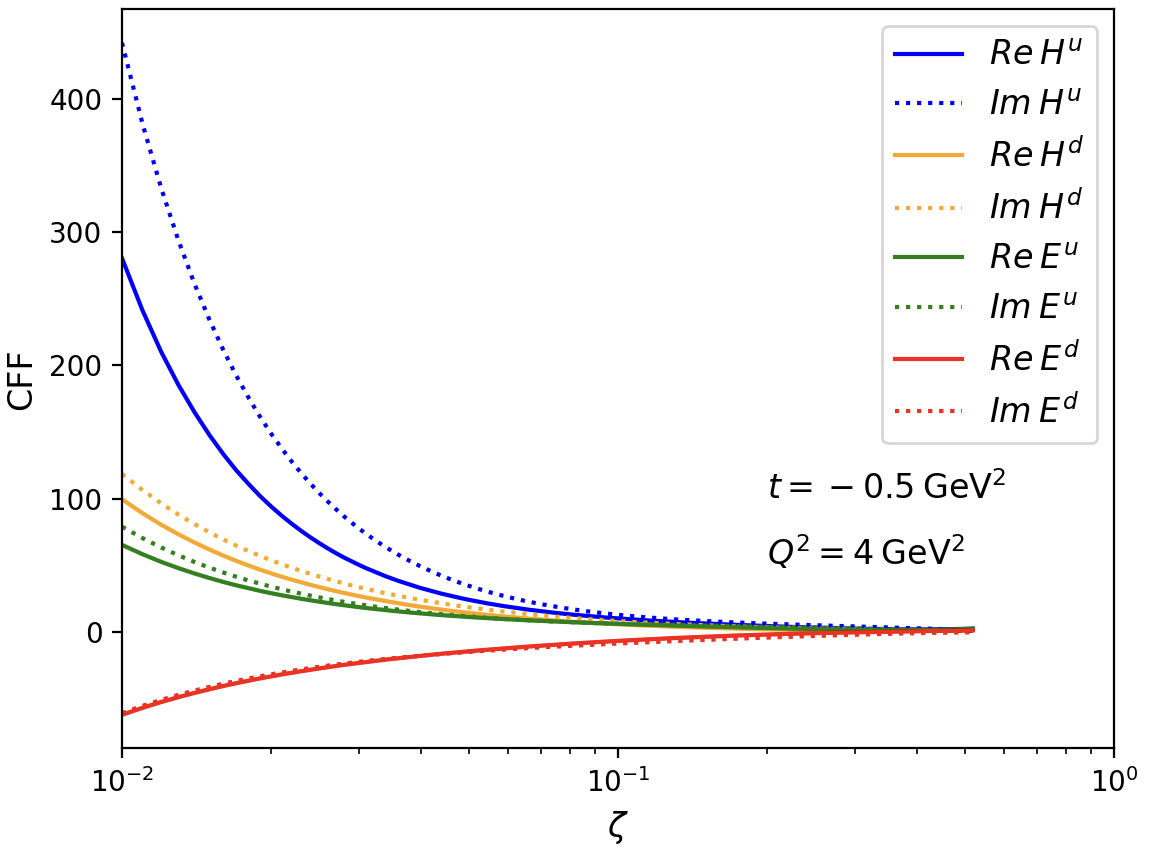}
    \caption{$\Re e H_q$ and $\Im m H_q$,  plotted vs. $\zeta \equiv x_{Bj}$ for $Q^2 = 4$ GeV$^2$ and $|t| = 0.5$ GeV$^2$.}
    \label{fig:cff_Real_Im}
\end{figure}

\begin{figure}[H]
    \centering
    \includegraphics[width=7.5cm]{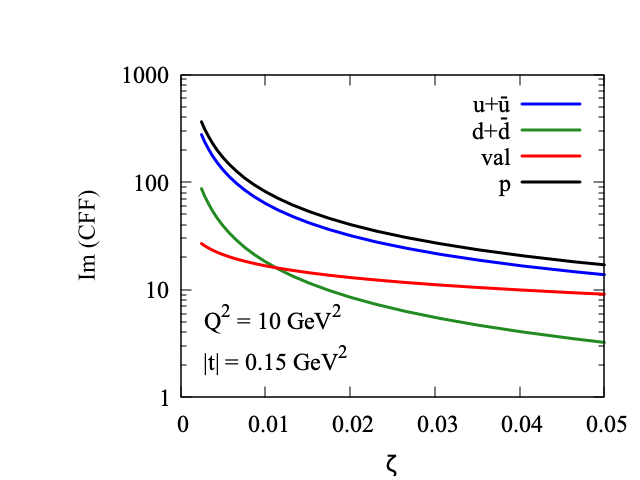}
    \includegraphics[width=7.5cm]{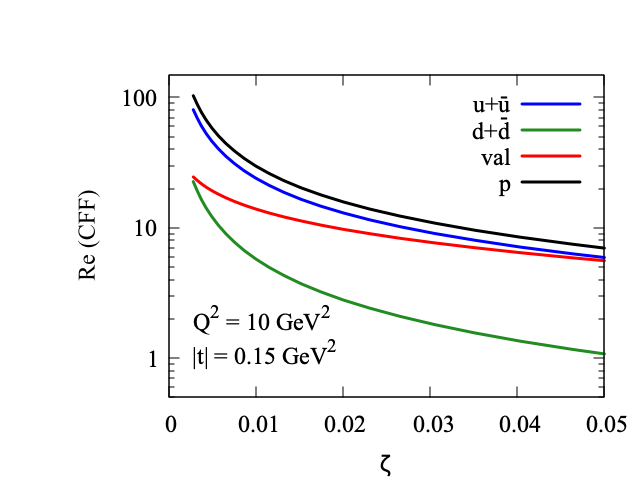}
    \caption{Contributions to the CFF $H$, plotted as a function of $x_{Bj}\equiv \zeta$ at EIC kinematics, for $|t|=0.15$ GeV$^2$, $Q^2= 10$ GeV$^2$ and $E_e \times E_p = 10\times 100$ GeV$^2$, with rapidity range, $0.05 <y< 0.95$. The imaginary part is plotted on the left, and the real part is plotted on the right. The various contributions correspond to $e_u^2(u+\bar{u}($ (blue), $e_d^2(d+\bar{d})$ (green), $e_u^2 u_v + e_d^2 d_v$ (red), denoted as ``val" in the legend, and $e_u^2 (u+\bar{u}) + e_d^2 (d+ \bar{d})$, denoted as $p$ in the figure. The small contribution from strange quark distribution is omitted in the figure. }
    \label{fig:cff_vxbj_eic}
\end{figure}

\begin{figure}[H]
    \centering
    \includegraphics[width=7.5cm]{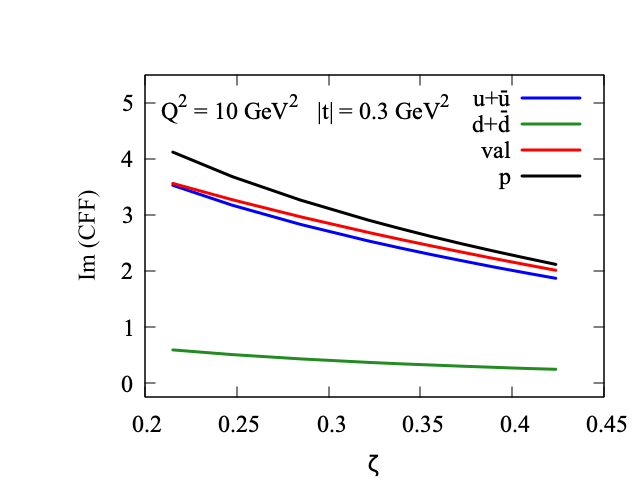}
    \includegraphics[width=7.5cm]{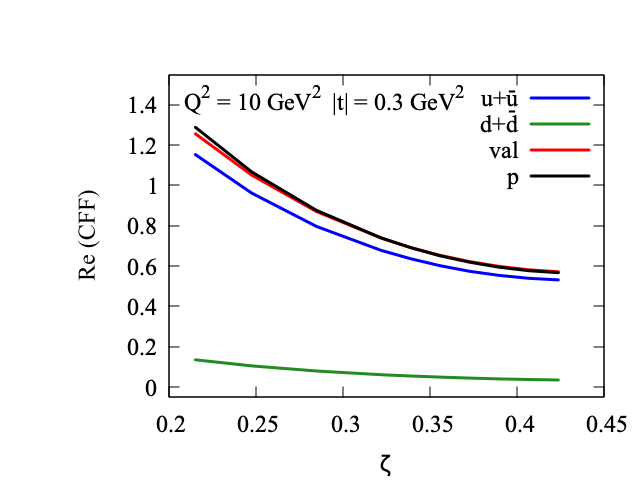}
    \caption{Contributions to the CFF plotted as a function of $x_{Bj} \equiv \zeta$ at JLab kinematics typical of a $20+$ GeV upgrade, with $|t|=0.3$ GeV$^2$, $Q^2=10$ GeV$^2$, and $0.05 < y < 0.95$. Notice the role of the valence quarks in the evaluation of the CFF given that $q + \bar{q} = q_{v} + 2\bar{q}$. Notation as in Fig.\ref{fig:cff_vxbj_eic}.}
    \label{fig:cff_vxbj_jlab}
\end{figure}

Results for the CFFs are shown in Figures \ref{fig:cff_Real_Im}, \ref{fig:cff_vxbj_eic} and \ref{fig:cff_vxbj_jlab}. In Fig.\ref{fig:cff_Real_Im} we show the Real and Imaginary parts of the different flavor contributions plotted vs. $x_{Bj} \approx \zeta$, at two different values of $|t|=0.5$ and $|t|=0.1$ GeV$^2$, respectively.
In Figs.\ref{fig:cff_vxbj_eic}  and \ref{fig:cff_vxbj_jlab}, we show the Im and Real parts of the proton CFF components from the quark flavor combinations, $e_u^2(u+\bar{u})$, $e_d^2(d + \bar{d})$, and $(e_u^2 u_v+ e_d^2 d_v)$, in two different kinematics settings attainable at the upcoming EIC (Fig.\ref{fig:cff_vxbj_eic}) and at a fixed target type experiments such as Jlab (Fig.\ref{fig:cff_vxbj_jlab}), respectively. The kinematics are defined as: $E_e E_p = (10 \times 100)$ for the EIC, and $E_e =22$ GeV for the fixed target setting. Both settings are such that the rapidity, $y$, varies in the interval, $0.05 < y < 0.95$. The fixed target scenario clearly shows sensitivity to the valence component, which appears instead to br suppressed for the CFFs evaluated in a similar range of rapidity, $y$, in a collider setting (Fig.\ref{fig:cff_vxbj_eic}). 
Notice that the gluon contribution will not be measured directly, but it can be inferred from the PQCD evolution equations, Eq.\eqref{eq:evolS}, from the coupling to the flavor singlet component. Another important feature highlighted in our graphs is the magnitude the CFFs measurable which in the low $x$ region accessible at the EIC are predicted to be several orders of magnitude larger than for the fixed target case, as expected from the forward limit behavior of the GPDs, inferred from our parametrization.         

%

\section{Conclusions and Outlook}
\label{sec:conclusions}
In summary, we presented an updated version of the UVA parametrization of GPDs, named UVA2, for the GPDs $H$ and $E$, and quark flavors, $u_v$, $d_v$, $\bar{u}$, $\bar{d}$, and $g$. The gist of our method is to provide the most stringent constraints on GPDs from the proton elastic form factors and second Mellin moments sum rules, from the forward limit given by PDFs, and from the crossing symmetries of the theory, without implementing DVES data. 
The parametrization is based on the reggeized spectator model developed in Refs.\cite{Ahmad:2006gn,Ahmad:2007vw,Goldstein:2010gu,Goldstein:2012az,Kriesten:2021sqc}. 

The main improvements compared to the previous versions include the availability of analytic expressions in $X,\zeta,t$ (straightforwardly translatable into $x,\xi,t$) at a common initial scale, $Q_o^{2}$, for all flavor components. We also introduced a more stringent constraint on the antiquark and gluon densities based on recent LQCD evaluations of the $t$-dependent form factors. 
The parametrization, which is evolved using off-forward evolution kernels/splitting functions at LO, can be readily used in the evaluation 
of the Compton form factors for DVCS and DVMP cross sections and asymmetries as well as other deeply virtual exclusive processes, for kinematic frameworks from Jefferson Lab (fixed target), to COMPASS, EIC and LHC kinematics. 
On the technical side, we will also provide an improved study of the implementation of Adam's method to solve the QCD evolution equations, compared to other methods that feature an improved speed in the $x$ integration, specifically defining a benchmarking process of the proposed numerical methods. 

The results presented here provide a first quantitative global framework for analyzing DVES data, ensuring a consistent treatment of all quark flavors and gluons.
An important outcome of our work is that, with a sufficiently large range in $Q^2$ such as the one accessible at the EIC, the gluon component can be extracted directly from the analysis of DVCS data.  

Our approach can be seamlessly extended to other aspects of the analysis, including NLO $Q^2$ evolution with target mass corrections, and evaluating dynamical higher twist matrix elements. The latter are likely to contribute in the low/intermediate $Q^2$ range, and they are important especially to determine the orbital component of angular momentum (\cite{Raja:2017xlo} and references therein). Being  based on the spectator model, we can extend our model to evaluate twist-three terms since it provides a natural framework  to introduce the interacting components between the struck quark and the spectator partons (this was already studied in QCD-based models of TMDs, see {\it e.g.} \cite{Kang:2010hg}).  
Finally, the statistical methods we adopted have been, so far, limited to determining the multidimensional space where a working sets of parameters can vary, within a Hessian framework. Having these results as a backdrop, a future detailed study using a maximum likelihood analysis of nucleon form factor data, will address the correlations among the parameters defining the dependence on $X, \zeta$ and $t$ of GPDs. Such a study would include using electromagnetic form factor data from multiple sources, including \cite{Qattan:2004ht, Qattan:2024pco} as well as the forthcoming data for large $t$ at Jefferson Lab \cite{2023APS..APRL01006D}. 

As a final point, we re-emphasize that the updated version of the UVA parametrization of GPDs, named UVA2, for the GPDs $H$ and $E$, and quark flavors, $u_v$, $d_v$, $\bar{u}$, $\bar{d}$, and $g$ in this paper is publicly available at 

\vspace{0.3cm}
\noindent\textbf{\textcolor{blue}{\url{https://github.com/Exclaim-Collaboration/UVA2} }}.

\acknowledgements
This work was completed under DOE grants DE-SC0016286 and DE-SC0024644. We thank Joshua Bautista and Brandon Kriesten for their contribution in the early stages of this work, and our colleagues from the EXCLAIM collaboration,  Jang (Jason) Ho, Gia-Wei Chern and Yaohang Li, for accurate tests of our model, as well as Michael Engelhardt for help on understanding the status of LQCD calculations of $t$-dependent nucleon Mellin moments.

\appendix
\section{Connection to the asymmetric frame}
\label{app:asymmetric}
In this section we will refer to a generic GPD in the asymmetric frame as $\hat{F}_q(\bar{X}, \zeta)$. The $t$ dependence is suppressed for brevity. We start with the transformation equations:
\begin{align}
    x = \frac{\bar{X}- \zeta/2}{1 - \zeta/2} \quad \quad \quad \xi = \frac{\zeta}{2 - \zeta}
\end{align}
We see that the interval $x \in [-1, 1]$ is transformed to $\bar{X} \in [-1 + \zeta, 1]$ and the midpoint of the interval is shifted from $0$ to $\zeta/2$. The quark and anti-quark distributions are defined as follows:
\begin{align}
    F_q(X, \zeta) &\equiv \hat{F}_q(X, \zeta) \quad\quad \quad \quad X\in [0,1]\\
    F_{\bar{q}}(X, \zeta) &\equiv -\hat{F}_q(\zeta - X, \zeta) 
\end{align}
Note that the quark and the anti-quark distributions have been defined only for positive $X$. Also note that they are not independent in the region $[0, \zeta]$.
We now define the plus and minus distributions:
\begin{align}
    F_q^{\pm}(\bar{X}, \zeta) = \hat{F}_q(\bar{X}, \zeta) \mp \hat{F}_{q}( \zeta - \bar{X}, \zeta)
\end{align}
From the definition above it follows that:
\begin{align*}
    F_{q}^+(\zeta/2 + \bar{X}, \zeta) &= \hat{F}_q(\zeta/2 + \bar{X} , \zeta) - \hat{F}_{q}( \zeta/2 - \bar{X}, \zeta)\\
    F_{q}^+(\zeta/2 - \bar{X}, \zeta)&= \hat{F}_q(\zeta/2 - \bar{X} , \zeta) - \hat{F}_{q}( \zeta/2 + \bar{X}, \zeta)\\
    F_{q}^+(\zeta/2 - \bar{X}, \zeta) &= - F_{q}^+(\zeta/2 + \bar{X}, \zeta)
\end{align*}
i.e. the plus distribution is anti-symmetric about $\zeta/2$. Similarly for the minus distribution we find that it is symmetric about $\zeta/2$.
\begin{align*}
    F_{q}^-(\zeta/2 + \bar{X}, \zeta) &= \hat{F}_q(\zeta/2 + \bar{X} , \zeta) + \hat{F}_{q}( \zeta/2 - \bar{X}, \zeta)\\
    F_{q}^-(\zeta/2 - \bar{X}, \zeta)&= \hat{F}_q(\zeta/2 - \bar{X} , \zeta) + \hat{F}_{q}( \zeta/2 + \bar{X}, \zeta)\\
    F_{q}^-(\zeta/2 - \bar{X}, \zeta) &=  F_{q}^-(\zeta/2 + \bar{X}, \zeta)
\end{align*}
From the definition of the plus and the minus distributions, we can write:
\begin{align}
    \hat{F}_q(\bar{X}, \zeta)  = \frac{1}{2}\big(F^+_q(\bar{X}, \zeta) + F^-_q(\bar{X}, \zeta)\big)
\end{align}
From the transformation equations, we see that:
\begin{align*}
    \int_{-1}^{1} dx \rightarrow \int_{-1+\zeta}^1 \frac{d\bar{X}}{(1-\zeta/2)}
\end{align*}
Inegrals over GPDs are usually of the form:
\begin{align}
    I = \int_{-1}^{1}dx K^{\pm} \hat{F}_q(x, \xi) \rightarrow \int_{-1+\zeta}^1 \frac{d\bar{X}}{(1-\zeta/2)} K^{\pm} \hat{F}_q(\bar{X}, \zeta)
\end{align}
where $K^{\pm}$ is a kernel that is antisymmetric/symmetric about the midpoint of the integration interval. Using the symmetries of the plus/minus distributions, the integrals can be simplified:
\begin{align}
   I &= \frac{1}{2} \int_{-1+\zeta}^1 \frac{d\bar{X}}{(1-\zeta/2)} K^{\pm} \big(F^+_q(\bar{X}, \zeta) + F^-_q(\bar{X}, \zeta)\big)\\
   &= \int_{\zeta/2}^1 \frac{dX}{(1-\zeta/2)} K^{\pm} F^{\pm}_q(X, \zeta)
\end{align}   
Note that in the last step we changed the integration variable from $\bar{X}$ to $X$ since we are only integrating over positive values. We can now write:
\begin{align}
   I &= \int_{\zeta/2}^1 \frac{dX}{(1-\zeta/2)} K^{\pm} F^{\pm}_q(X, \zeta)\\
   &=\int_{\zeta/2}^1 \frac{dX}{(1-\zeta/2)} K^{\pm} \big(\hat{F}_q(X, \zeta) \mp \hat{F}_{q}( \zeta - X, \zeta\big)\\
   &=\int_{\zeta/2}^1 \frac{dX}{(1-\zeta/2)} K^{\pm} \big(F_q(X, \zeta) \pm F_{\bar{q}}(X, \zeta)\big)
\end{align}

\section{Specifics of off forward PQCD evolution equations}
\label{app:evol}
\begin{eqnarray}
\mu \frac{\partial }{\partial \mu} H_{NS}(X,\zeta) &= &\frac{\alpha_S(\mu)}{2\pi}  \int_X^1 \frac{dZ}{Z}  \, 
C_F \left[  \left(1+ \frac{X}{Z} \frac{X-\zeta}{Z-\zeta} \right) H_{NS}(Z,\zeta) - \left( 1+ \frac{X-Z}{Z-\zeta} \frac{Z}{X} \right)H_{NS}(X,\zeta) \right]   \frac{X/Z}{1-{X}/{Z} }   \nonumber \\
&+ & \frac{\alpha_S}{2 \pi} C_F \left(\frac{3}{2} + \frac{1}{2} \ln\big[(1-X)^2/(1-\zeta) \big]\right)  H_{NS}(X,\zeta) 
\end{eqnarray}

\section{Forward Limit}
\label{app:forward}
\begin{figure}[H]
    \centering
    \includegraphics[width=0.55\linewidth]{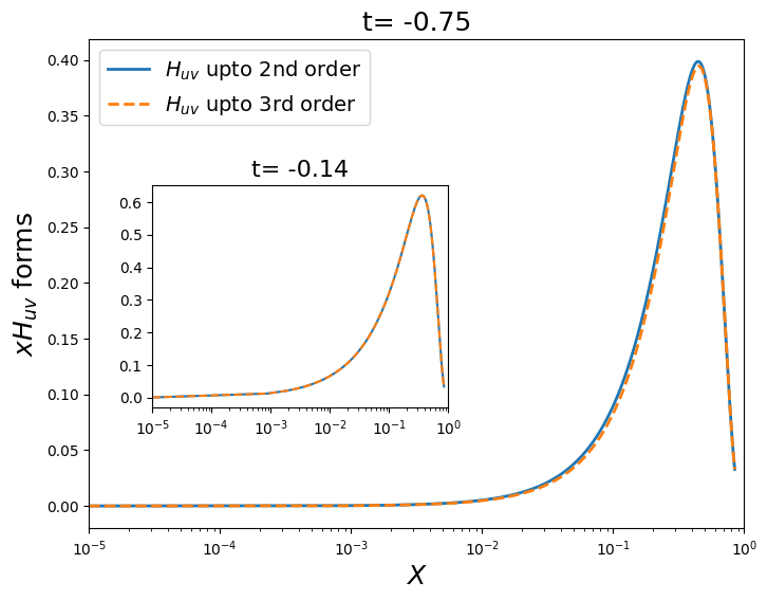}
    \caption{The figure shows a comparison of the GPD $H_{uv}$ at two extreme values of t to illustrate the saturation of integrals in \eqref{integrals_omega} at second order.}
    \label{orders_compare}
\end{figure}

\textbf{Forward limit for the integrals}\\
\vspace{0.3cm}
\noindent \underline{\it  In the Limit, $\zeta=0$, t$<0$}:

\begin{align}
    \omega &= \frac{(1-X)t}{\mathcal{M}_{X}} \label{variables} \nonumber\\
    \chi &= (1 + \omega ) \nonumber \\
    \delta &= \sinh^{-1}\Bigg[\frac{1}{2}(\omega+2)\sqrt{\omega(\omega + 4)} \Bigg]
\end{align}
In order to set $t=0$ which amounts $\omega \rightarrow0$, we expand $\delta$ around $\omega = 0$, giving $\delta$ as:

\begin{equation}
    \delta = 2\sqrt{\omega} - \frac{\omega \sqrt{\omega}}{12} + \cdots
\end{equation}
With this expansion, we obtain the following results for $I_0, I_1, \text{and } I_2$ up to the second order in $\omega$ in the zero-skewness limit:
\begin{align}\label{integrals_omega}
    I_0 &= \frac{1}{\gamma^4} \bigg[ \frac{1}{8} - \frac{1}{5} \omega + \frac{13}{56} \omega^2 - \cdots \bigg] \\
    I_1 &= \frac{1}{\gamma^3} \bigg[ -\frac{1}{24} + \frac{1}{40} \omega - \frac{3}{280} \omega^2 + \cdots \bigg] \\
    I_2 &= \frac{1}{\gamma^2} \bigg[ \frac{1}{24} - \frac{1}{280} \omega^2 + \cdots \bigg] \, .
\end{align}
It is to be noted that one can easily go to higher orders in $\omega$, but even up to two orders, the expressions are precise, as there is (almost) no difference in the GPDs if we go to higher orders in $\omega$. A very feeble difference does arise, but at large t ($t>-0.7$). This can be evidently seen from Figure~\ref{orders_compare}. \\
In the limit of zero-skewness and $t<0$, using the terms of the integrals up to third order, the integrals are: \\
\begin{align}
    I_0 &= \frac{1}{\mathcal{M}_{X}^3\;\gamma^4} \bigg[ \frac{\mathcal{M}_{X}^3}{8} - \frac{\mathcal{M}_{X}^2}{5}(1-X)t + \frac{13\mathcal{M}_{X}}{56}(1-X)^2 t^2 \bigg] \\
    I_1 &= \frac{1}{\mathcal{M}_{X}^3\;\gamma^3} \bigg[ -\frac{\mathcal{M}_{X}^3}{24} + \frac{\mathcal{M}_{X}^2}{40}(1-X)t - \frac{3\mathcal{M}_{X}}{280}(1-X)^2 t^2  \bigg] \\
    I_2 &= \frac{1}{\mathcal{M}_{X}^3\;\gamma^2} \bigg[ \frac{\mathcal{M}_{X}^3}{24} - \frac{\mathcal{M}_{X}}{280}(1-X)^2 t^2 \bigg] \, .
\end{align}
where $\gamma = (1-X)\mathcal{M}_{X}$.\\
With these expansions, it becomes clear that, in the forward limit, i.e., as $t \to 0$, and hence $\omega \to 0$, we obtain
\begin{align}
    I_0 &= \frac{1}{8\gamma^4} = \frac{1}{8(1-X)^4\mathcal{M}_{X}^4} \\
    I_1 &= -\frac{1}{24\gamma^3} = \frac{-1}{24(1-X)^3\mathcal{M}_{X}^{3}} \\
    I_2 &= \frac{1}{24 \gamma^2} = \frac{1}{24(1-X)^2\mathcal{M}_{X}^{2}} .
\end{align}

In the forward limit, $\zeta = 0, t<0$ we obtain,

\text{The normalization parameters in the forward limits:}

\begin{equation}
    \mathcal{A}_H = 2\pi\mathcal{N}(1-X)^3
\end{equation}
\begin{equation}
    \mathcal{A}_E = 2\pi\mathcal{N}(1-X)^4
\end{equation}
\begin{equation}
    \mathcal{A}^G_H = 2\pi\mathcal{N}(1-X)^4
\end{equation}
\begin{equation}
    \mathcal{A}^G_E = -4\pi M \mathcal{N}(1-X)^4
\end{equation}

Lets us suppose $\alpha = (m+MX)$, $\beta = [(1-X)M - M_X]$ , and $\eta = (1-X)^2$. (It is to be noted that these parameters here $\alpha, \beta, \eta$ are not the ones in the Regge term $R_p^{\alpha,\alpha'}$). These are been used here just as notations for easier understanding.
\vspace{0.3cm}
The analytic expressions for the numerical parameters of the GPDs in the limit $\zeta=0$ (zero skewness) and t$<$0 are as follows:\\

\begin{align}
    a_0 &= -\gamma \alpha^2 (1+\omega) \\
    a_1^{-} &= \gamma(\omega-1) + \alpha^2\\
    a_1^{+} &= \gamma(1 - \omega) + \alpha^2\\
    b_0 &= -2M\gamma\alpha(1+\omega)\\
    b_1^{+} &= 6M\alpha \\
    b_1^{-} &= 2M\alpha \\
    a_{0}^G &= -\mathcal{M}_{X}\;X^2\;(\omega+1)\;\beta^2 \\
    a_{1}^{G+} &= \frac{X^2\;\beta^2}{(1-X)} + \gamma(\omega-1)\;(1 + \eta) \\
    a_{2}^{G+} &= (1 + \eta) \\
    b_0^{G} &= -X\gamma(\omega+1)\beta \\
    b_1^{G} &= X\beta
\end{align}
 where $\omega$ is given by equation \eqref{variables}. \\
 
These expressions above in the forward limit, are expressed as (excluding the Regge term):

\begin{align}
    a_0 &= -\gamma \alpha^2  \\
    a_1^{-} &= -\gamma + \alpha^2\\
    a_1^{+} &= \gamma + \alpha^2\\
    b_0 &= -2M\gamma\alpha\\
    b_1^{+} &= 6M\alpha \\
    b_1^{-} &= 2M\alpha \\
    a_{0}^G &= -\mathcal{M}_{X}\;X^2\;\beta^2 \\
    a_{1}^{G+} &= \frac{X^2\;\beta^2}{(1-X)} - \gamma\;(1 + \eta) \\
    a_{2}^{G+} &= (1 + \eta) \\
    b_0^{G} &= -X\gamma\beta \\
    b_1^{G} &= X\beta
\end{align}


\section{Evaluating the integrals}
\label{integral_solns}
The $k_\perp$ integrals in the GGL parameterization of the GPD's are of the following form:
\begin{equation}
   \Tilde{I}_{n}= \int_{0}^{\infty}dk \frac{k^{2n+1}}{D^2 (a^2-b^2)^{\frac{3}{2}}}
\end{equation}
Where $k = k_{\perp}$ and $n=0,1,2$. The functions $D$, $a$, and $b$ are given by:
\begin{align}
    D &= X M^2 -M_{\Lambda}^{2} - \frac{X}{1-X}M_{X}^{2}-\frac{k^2}{1-X}\\
    a &=  X' M^2 -M_{\Lambda}^{2} - \frac{X'}{1-X'}M_{X}^{2}-\frac{k^2}{1-X'} -\Delta_{\perp}^2(1-X')\\
    b &= 2k\Delta_{\perp}
\end{align}
Written out in full:
\begin{align}
    \Tilde{I}_n=\int_{0}^{\infty} dk \frac{k^{2n+1}}{\left( X M^2 -M_{\Lambda}^{2} - M_{X}^{2}\frac{X}{1-X}-\frac{k^2}{1-X}\right)^2\left(\left(X' M^2 -M_{\Lambda}^{2} - M_{X}^{2}\frac{X'}{1-X'}-\frac{k^2}{1-X'} -\Delta_{\perp}^2(1-X')\right)^2 -4k^2 \Delta_{\perp}^{2}\right)^{\frac{3}{2}}}
\end{align}
We can pull out factors of $1-X$ and $1-X'$ to isolate the integration variable as much as possible. The integral then becomes:
\begin{equation}
\Tilde{I}_{n}  = (1-X)^2 (1-X')^3\int_{0}^{\infty} dk
\frac{k^{2n+1}}{\left( k^2 + x\right)^2 \left[(k^2+(\xp+\d))^2 -4 k^2 \d \right]^{\frac{3}{2}}}
\end{equation}
where:
\begin{align}
    x &= XM_{X}^2 +(1-X)M_{\Lambda}^2-X(1-X)M^2\\
    \xp  &=X'M_{X}^2 +(1-X')M_{\Lambda}^2-X'(1-X')M^2\\
    \d &= \Delta_{\perp}^2(1-X')^2
\end{align}
We define:
\begin{equation}
     I_n=\frac{\tilde{I}_n}{(1-X)^2(1-X')^3}
\end{equation}
Expanding out the square in the second term in the denominator in (6), we get:
\begin{align}
    I_n= \int_{0}^{\infty} dk \frac{k^{2n+1}}{(k^2+x)^2(k^4+2k^2(\xp-\d)+(\xp+\d)^2)^{\frac{3}{2}}}
\end{align}
We now make the following substitution: $u = k^2$.
\begin{align}
    I_n &= \frac{1}{2}\int_{0}^{\infty} du\frac{u^n}{(u+x)^2(u^2+2u(\xp-\d)+(\xp+\d)^2)^{\frac{3}{2}}}\\
    &=\frac{1}{2}\int_{0}^{\infty} du\frac{u^n}{(u+x)^2\left[(u+\xp-\d +2i \sqrt{\xp \d})(u+\xp-\d -2i \sqrt{\xp \d}) \right]^{\frac{3}{2}}}
\end{align}
We make another substitution: r = u+x.
\begin{align}
    I_{n} &= \frac{1}{2} \int_{x}^{\infty}dr\frac{(r-x)^n}{r^2\left[(r-x+\xp-\d +2i\sqrt{\xp \d})(r-x+\xp-\d -2i\sqrt{\xp \d}) \right]^{\frac{3}{2}}}\\
    &= \frac{1}{2}\int_{x}^{\infty}dr\frac{(r-x)^n}{r^2(r^2+Br+A)^{\frac{3}{2}}}\\
    & = -\frac{\partial}{\partial A}\int_{x}^{\infty}dr\frac{(r-x)^n}{r^2\sqrt{r^2+Br+A}}
\end{align}
with:
\begin{align}
    B&= -2(x-\xp +\d)\\
    A&= (x-\xp+\d)^2+4\xp \d
\end{align}
The integral above can be looked up in a standard table of integrals. 
\bibliography{references}

\end{document}